\documentclass[12pt]{iopart}

\usepackage[british]{babel}
\usepackage{color}
\usepackage{mathrsfs}
\usepackage{setstack}
\usepackage{iopams}
\usepackage[T1]{fontenc}
\usepackage{graphicx}
\usepackage{array}
\usepackage{multirow}
\usepackage{adjustbox}
\usepackage{footnote}

\def\Xint#1{\mathchoice
{\XXint\displaystyle\textstyle{#1}}%
{\XXint\textstyle\scriptstyle{#1}}%
{\XXint\scriptstyle\scriptscriptstyle{#1}}%
{\XXint\scriptscriptstyle\scriptscriptstyle{#1}}%
\!\int}
\def\XXint#1#2#3{{\setbox0=\hbox{$#1{#2#3}{\int}$}
\vcenter{\hbox{$#2#3$}}\kern-.55\wd0}}

\def\dashint{\Xint-}

\begin{document}

\title{First passage properties of asymmetric L\'evy flights}
\author{Amin Padash$^{\dagger,\sharp}$, Aleksei V. Chechkin$^{\sharp,\ddagger}$,
Bartlomiej Dybiec$^{\flat}$, Ilya Pavlyukevich$^{\S}$, Babak Shokri$^{\dagger,
\P}$, \& Ralf Metzler$^{\sharp}$}
\address{$^\dagger$ Physics Department of Shahid Beheshti University, 19839-69411
Tehran, Iran\\
$^\sharp$ Institute for Physics \& Astronomy, University of Potsdam,
14476 Potsdam-Golm, Germany\\
$^\ddagger$ Akhiezer Institute for Theoretical Physics, 61108 Kharkov,
Ukraine\\
$^\flat$ Marian Smoluchowski Institute of Physics and Mark Kac Center
for Complex Systems Research, Jagiellonian University, ul. St. Lojasiewicza 11,
30-348 Krakow, Poland\\
$^\S$ Friedrich Schiller University Jena, Faculty of Mathematics and
Computer Science, Institute for Mathematics, 07737 Jena, Germany\\
$^\P$ Laser and Plasma Research Institute, Shahid Beheshti University,
19839-69411 Tehran, Iran}
\ead{rmetzler@uni-potsdam.de (Corresponding author)}

\begin{abstract}
L{\'e}vy Flights are paradigmatic generalised random walk processes, in which
the independent stationary increments---the "jump lengths"---are drawn from an
$\alpha$-stable jump length distribution
with long-tailed, power-law asymptote. As a result, the variance of L{\'e}vy
Flights diverges and the trajectory is characterised by occasional extremely
long jumps. Such long jumps significantly decrease the probability to revisit
previous points of visitation, rendering L{\'e}vy Flights efficient search
processes in one and two dimensions. To further quantify their precise
property as random search strategies we here study the first-passage time
properties of L{\'e}vy Flights in one-dimensional semi-infinite and bounded
domains for symmetric and asymmetric jump length distributions. To obtain the
full probability density function of first-passage times for these cases
we employ two complementary methods. One approach is based on the
space-fractional diffusion equation for the probability density function,
from which the survival probability is obtained for different values of
the stable index $\alpha$ and the skewness (asymmetry) parameter $\beta$.
The other approach is based on the stochastic Langevin equation with
$\alpha$-stable driving noise. Both methods have their advantages and
disadvantages for explicit calculations and numerical evaluation, and the
complementary approach involving both methods will be profitable for concrete
applications. We also make use of the Skorokhod theorem for processes with
independent increments and demonstrate that the numerical results are in
good agreement with the analytical expressions for the probability density
function of the first-passage times.
\end{abstract}

\section{Introduction}

Normal Brownian motion described by Fick's second law, the diffusion equation,
is characterised by the linear time dependence $\langle x^2(t)\rangle\simeq t$
of the mean squared displacement (MSD) \cite{vankampen}. Deviations from this
Fickean time dependence typically occur in the power-law form $\langle x^2(t)
\rangle\simeq t^{\kappa}$ of the MSD \cite{Bouchaud1990,RMetzler2000}. Depending
on the value of the anomalous diffusion exponent $\kappa$ we distinguish
between subdiffusion for $0<\kappa<1$, normal, Fickean diffusion for $\kappa
=1$, superdiffusion for $1<\kappa<2$, and ballistic, wave-like motion for
$\kappa=2$. The range $\kappa>2$ is sometimes referred to as hyperdiffusion.
The theoretical description of anomalous diffusion phenomena of physical
particles (passive or active) often requires a radical departure from the
classical formalism for Brownian motion. Namely, effects of energetic or
spatial disorder, collective dynamics, or non-equilibrium conditions need to
be addressed with more complex approaches \cite{Bouchaud1990,pccp}. For
instance, fractional Brownian motion \cite{fbm} is a process in
which the Langevin equation is driven by Gaussian yet power-law correlated
noise (fractional Gaussian noise) effecting both sub- and superdiffusion.
The generalised Langevin equation \cite{gle} includes a memory integral
with a kernel, that balances the input fractional Gaussian noise and effects
a thermalised process. Processes with explicitly time or position dependent
diffusion coefficients such as scaled Brownian motion \cite{sbm} or
heterogeneous diffusion processes \cite{hdp}, respectively, also lead to
sub- and superdiffusion. Diffusion on fractals \cite{fractals} due to the
fact that the particle in the highly ramified environment often has to
back-track its motion, has similar characteristics as subdiffusive fractional
Brownian motion. We finally mention the continuous time random walk
model, in which the standard Pearson walk was generalised to include
continuous waiting times \cite{ctrw}. When the distribution of waiting times
becomes scale-free, with diverging characteristic waiting time, the
continuous time random walk process is subdiffusive \cite{EWMontroll1969,
HScher1975}. Conversely, when the continuous time random walk has a finite
characteristic waiting time but is equipped with a scale-free distribution
of jump lengths
with power-law asymptote $\lambda(x)\sim|x|^{-1-\alpha}$ ($0<\alpha<2$)
the resulting process is a "L{\'e}vy flight" (LF). As in this case the variance
of the process diverges, diffusion can be characterised in terms of
rescaled fractional order moments ${\langle|x(t)|^{\eta}\rangle}^{2/\eta}
\simeq t^{2/\alpha}$ with $0<\eta<\alpha$ \cite{RMetzler2000, RMetzler2004,
HCFogedby1997,RMetzler2012,yanovsky,chechgon}. Mathematically, asymptotic
power-law forms of the jump length distribution can be explained by the
generalised central limit theorem \cite{Bouchaud1990,Samorodnitsky-Taqqu,
Khintchine1936,gnedenko}, which gives rise to the much higher likelihood for
extremely long jumps \cite{BMandelbrot1977,MFShlesinger1993,MFShlesinger2001}
in comparison to conventional Pearson random walks.

L{\'e}vy stable laws play a crucial role in the statistical description
of scale invariant stochastic processes \cite{BMandelbrot1977,PLevy1954}
not only in physical contexts but also in biological, chemical, geophysical,
sociological, economical or financial systems, among others. In physics
L{\'e}vy statistics were demonstrated to explain deviations of complex
systems from the Gaussian paradigm, inter alia, for the power-law blinking of
nano-scale light emitters \cite{FDStefani2009}, diffusive transport of light
\cite{PBarthelemy2008}, photons in hot atomic vapours \cite{NMercadier2009},
tracer particles in a rotating flow \cite{THSolomon1993}, passive scalars in
vortices in shear \cite{DCNegrete1998}, anomalous diffusion in disordered media
\cite{Bouchaud1990}, weakly chaotic and Hamiltonian systems \cite{TGeisel1985,
JKlafter1994}, in the divergence of kinetic energy fluctuations of a single
ion in an optical lattice \cite{HKatori1997}, fluctuations in the transition
energy of a single molecule embedded in a solid \cite{GZumofen1994}, in the
interaction of two level systems with single molecules \cite{EBarkai1999},
the distribution of random single molecule line shapes in low temperature
glasses \cite{EBarkai2000}, the diffusion of a collection of ultra-cold atoms
and single ions in optical lattices \cite{EBarkai2014}, but also in reaction
diffusion systems \cite{DCNegrete2003,DCNegrete2009}. L{\'e}vy statistics was
observed experimentally in tokamak and stellarator fusion devices
\cite{plasma1,plasma2,plasma3}. It was also shown that the phenomenon of L-H
transitions observed in the stellarator is accompanied by the crossover from
L{\'e}vy to Gaussian fluctuation statistics \cite{plasma4}.
Numerous examples for L{\'e}vy statistics exist in the
dynamics of plasmas, including the anomalous transport in magnetic confinement
\cite{DCNegrete2008,AKullberg2014,ABovet2014,ABovet20142,ABovet2015}, the dynamics
of a charged particle in a plasma \cite{plasma5,DCNegrete2016}, anomalous transport of
ions and electrons in solar winds \cite{SPerri2009}, nonlocal transport in plasma
turbulence \cite{DCNegrete2005,DCNegrete2006,DCNegrete2007, DCNegrete2010,
DCNegrete20132}, and heat transport in magnetised plasmas \cite{DCNegrete2011,
DCNegrete2013}.

In the biological sciences, many organisms from bacteria
to humans use L{\'e}vy stable relocation statistics in their search for
resources \cite{AMReynolds2009}, tracer motion in living cells \cite{ACaspi2000,
NGal2010,KJChen2015,SSMinho2018}, or the superdiffusive motion of bacteria within
a swarm \cite{AGil2015}. On long, fast-folding polymers the search process
of a binding molecule is based on L{\'e}vy motion \cite{IMSokolov1997,
DBrockmann2003,MALomholt2005,MMajka2015}. In geoscience paleoclimate time series
show signatures of L{\'e}vy noise \cite{PDitlevsen1999}, earthquake statistics
exhibit distinct power-law behaviour \cite{ACorral2006}, as well as tracer
plumes in heterogeneous aquifers \cite{RinaSchumer2001,RinaSchumer2003,brian},
and the transport of ensembles of particles on the Earth surface
\cite{RinaSchumer2009}. The mechanisms of the worldwide spread of infectious
diseases \cite{LHufnagel2004}, pollen dispersal by bees \cite{VVallaeys2017}
and human mobility patterns and social interactions revealed by tracing mobile
phones or following banknotes \cite{DBrockmann2006,MCGonzalez2008,CSong2010,
CSong20102,InjongRhee2011,PDeville2016} also reveal L{\'e}vy statistics.
Evidence of L{\'e}vy stable laws was also unveiled in the human cognition for
the retrieval dynamics of memory \cite{TRhodes2007} and in human mental search
\cite{FRadicchi2012,FRadicchi2012PRE,TCosta2016}, as well as search in
multiplex networks \cite{QGuo2016}. The optimal search patterns of robots were
shown to be based on L{\'e}vy stable laws \cite{MVDartel2004}. In finance and
economical contexts \cite{BMandelbrot1963,RNMantegna1996,JPBouchaud2001,
BPodobnik2011} L{\'e}vy statistics govern the distribution of trades.
A particular area in which L{\'e}vy relocation statistics has been widely
explored is movement ecology \cite{ran}. Search patterns of foraging animals
\cite{GMViswanathan2011} that follow L{\'e}vy statistics include marine predators
\cite{DWSims2008, NEHumphries2010}, albatross birds \cite{NEHumphries2013},
Agrotis segetum moths \cite{AMReynolds2007}, fruit flies (Drosophila
melanogaster) \cite{AMReynolds20072}, bumblebees \cite{MLihoreau2016},
jellyfish \cite{GCHays2011}, goats \cite{HJdeKnegt2007,SFocardi2009},
immune function of human T cells \cite{THHarris2012} and human hunter-gatherers
\cite{DARaichlen2014}. We hasten to note that in the context of animal
movement there exist some debates on the predominance of L{\'e}vy search
patterns, especially the disqualification of L{\'e}vy statistics for
albatrosses \cite{viswa1996,viswanathan} in \cite{albatros} became a strong
argument against the LF hypothesis. However, there is strong evidence that for
individual albatross birds LFs are indeed a real search pattern \cite{PNAS2012}.
Moreover, \cite{DBoyer2006} reported that for spider monkeys the foraging pattern
is deterministic, mussel movements are rather multimodal \cite{AMashanova2010,
VAAJansen2012} and black bean aphids individually move in a predominantly
diffusive manner \cite{SPetrovskii2011}.

The efficiency of search processes is typically benchmarked by the time it
takes the searcher to reach a certain region in space. One relevant measure
is therefore the first-passage time, quantifying the time it takes from the
original position to first cross a point located at a given distance away.
For instance, the first-passage time in a financial time series could be
defined by a preset increase or decrease in the price of a given stock. Once
this threshold value is reached, the stock is sold or bought. Similarly, we
could talk about the instant of time when
foraging animals first randomly locate a resource-rich area away from
their original location. Such first-passage times in a stochastic search
process will vary from realisation to realisation, and can be quantified by
the first-passage time density $\wp(t)$. While the mean first-passage time
$\langle t\rangle=\int_0^{\infty}t\wp(t)dt$ can capture some aspects of this
dynamics,\footnote{Often, a better choice is the mean inverse first-passage
time $\langle1/t\rangle$ \cite{vladimir}.} the full information encoded
in $\wp(t)$ provides significant additional insight \cite{aljaz,aljaz1,denis}.
Here we study the first-passage properties for a general class of $\alpha$-stable
L{\'e}vy laws. We go beyond previous approaches \cite{vladimir,jpa,vlad1,vlad2,
vlad3,vlad5,eli,Dybiec2016,Koren2007PhysicaA,Koren2007PRL,vinc,frisch,zukla95}
focusing on symmetric and one-sided $\alpha$-stable relocation distributions
and consider $\alpha$-stable
laws with \emph{arbitrary asymmetry\/} in semi-infinite and bounded
domains. Our approach is based on the convenient formulation of LFs in terms of
the space-fractional diffusion equation. We derive these integro-differential
equations for LFs based on general asymmetric $\alpha$-stable distributions of
relocation lengths in finite domains, and thus go beyond studies of the exit
time and escape probability in bounded domain for symmetric LFs
\cite{JDuan2014,JDuan2015,JDuan2018}. An important aspect of this study is
that we complement our results with numerical analyses of the (stochastic)
Langevin equation for LFs and show how both approaches complement each
other.

The paper is organised as follows. In section \ref{SFDE-Alpha} we define
L{\'e}vy stable laws and the associated fractional diffusion equation. In
section \ref{NumSch} we set up our numerical model for the fractional
diffusion equation and the associated Langevin equation. Moreover, a
comparison between the numerical method and $\alpha$-stable distributions
for symmetric and asymmetric density functions is presented. Section
\ref{surv-first} then presents the numerical results for the survival
probability and first-passage time density for both symmetric and asymmetric
probability density functions. Our findings are compared with results
derived from the Skorokhod theorem for symmetric, one-sided, and extremal
two-sided stable distributions. We draw our conclusion in section \ref{concl}.
In the Appendices, we present details of several derivations.

\section{$\alpha$-stable processes and space-fractional diffusion equations}
\label{SFDE-Alpha}

An $\alpha$-stable L{\'e}vy process $X(t)$ with $X(0)=0$ and probability density
function (PDF) $\ell_{\alpha,\beta}(x,t)$ is fully specified by its characteristic
function in the Fourier domain as \cite{Samorodnitsky-Taqqu,Gikhman-Skorokhod}
\begin{eqnarray}
\nonumber
\hat{\ell}_{\alpha,\beta}(k,t)&=&\int\limits_{-\infty}^{\infty}\ell_{\alpha,\beta}
(x,t)\exp(\mathrm{i}kx)\mathrm{d}x\\
&=&\exp(-tK_{\alpha}|k|^{\alpha}[1-\mathrm{i}\beta\mathrm{sgn}(k)\omega(k,\alpha)]
+\mathrm{i}\mu kt), 
\label{eq:charecA}
\end{eqnarray}
where the index $\alpha$ with $0<\alpha\leq2$ is called the index of stability
or L{\'e}vy index, and the skewness parameter $\beta$ is allowed to vary within
the limits $-1\leq\beta\leq1$. Further, the generalised diffusion coefficient
$K_\alpha>0$ is a scale parameter, the shift parameter $\mu$ is any real number,
and the phase factor $\omega$ is defined as
\begin{equation}
\omega(k,\alpha)=\left\{\begin{array}{lr}\tan(\frac{\pi\alpha}{2}),&\alpha\neq1\\
-\frac{2}{\pi}\ln|k|,&\alpha=1\end{array}\right..
\end{equation}
In the real space-time domain the PDF of the $\alpha$-stable distribution can be
expressed via elementary functions for the following three cases:

(a) Gaussian distribution, $\alpha=2$, $\beta$ irrelevant:
\numparts
\begin{equation}
\ell_{2}(x,t)=\frac{1}{\sqrt{4\pi K_2t}}\exp\left(-\frac{(x-\mu t)^2}{4 K_2t}
\right), \qquad -\infty<x<\infty.
\end{equation}

(b) Cauchy distribution, $\alpha=1$, $\beta=0$:
\begin{equation}
\ell_{1,0}(x,t)=\frac{1}{\pi}\frac{K_1 t}{(x-\mu t)^2+(K_1t)^2},\qquad
-\infty<x<\infty.
\end{equation}
In physics, the Cauchy distribution is also often called a Lorentz distribution.

(c) L{\'e}vy-Smirnov distribution, $\alpha=1/2$, $\beta=1$:
\begin{equation}
\ell_{1/2,1}(x,t)=\frac{K_{1/2}t}{\sqrt{2 \pi (x-\mu t)^3}}\exp\left(-\frac{(K_{1/2}
t)^2}{2 (x-\mu t)}\right),\qquad x\geq0.
\end{equation}
\endnumparts

Schneider reported the representation of general L{\'e}vy stable densities in terms
of Fox $H$-functions \cite{schneider,fox,fox1}. Somewhat simpler representations
for rational indices $\alpha$ and $\beta$ are given in \cite{gorska,gorska1}. More
information on L{\'e}vy stable densities and their parametrisation are
provided in \ref{Appendix Char}.

Physically, the parameter $\mu$ accounts for a constant drift in the system. In
this paper we consider the first-passage process in the absence of a drift, thus
in what follows we set $\mu=0$. Let us first consider the case $\alpha\neq1$ and
$-1\leq\beta\leq1$. The corresponding space-fractional diffusion equation for the
PDF $\ell_{\alpha,\beta}(x,t)$ then reads
\begin{equation}
\frac{\partial\ell_{\alpha,\beta}(x,t)}{\partial t}= K_{\alpha}\,D^{\alpha}_{x}
\ell_{\alpha,\beta}(x,t),\quad\ell_{\alpha,\beta}(x,0)=\delta(x),
\label{eq:ffpe}
\end{equation}
where $D_x^{\alpha}$ is the space-fractional derivative operator,
\begin{equation}
D^{\alpha}_x\ell_{\alpha,\beta}(x,t)=L_{\alpha,\beta}\,_{-\infty}D_x^{\alpha}
\ell_{\alpha,\beta}(x,t)+R_{\alpha,\beta}\,_xD_{\infty}^{\alpha}\ell_{\alpha,
\beta}(x,t),
\label{eq:fraccoef}
\end{equation}
that we compose of $_{-\infty}D_x^{\alpha}$ and $_xD_\infty^{\alpha}$, the left and
right hand side space-fractional operators, respectively. We use the Caputo form
of the operators defined by $(n-1<\alpha <n)$ \cite{Podlubny1999}
\numparts
\begin{eqnarray}
\label{eq:lcapu}
_{-\infty}D_x^{\alpha}f(x)&=&\frac{1}{\Gamma(n-\alpha)}\int\limits_{-\infty}^x
\frac{f^{(n)}(\zeta)}{(x-\zeta)^{\alpha-n+1}}\,\mathrm{d}\zeta\\
_xD_\infty^\alpha f(x)&=&\frac{(-1)^n}{\Gamma(n-\alpha)}\int\limits_x^{\infty}
\frac{f^{(n)}(\zeta)}{(\zeta-x)^{\alpha-n+1}}\,\mathrm{d}\zeta
\label{eq:rcapu}
\end{eqnarray}
\endnumparts
and $L_{\alpha,\beta}$ and $R_{\alpha,\beta}$ are the left and right weight
coefficients, defined as \cite{DCNegrete2006,DCNegrete2007} 
\begin{equation}
L_{\alpha,\beta}=-\frac{1+\beta}{2\cos(\frac{\alpha\pi}{2})},\qquad
R_{\alpha,\beta}=-\frac{1-\beta}{2\cos(\frac{\alpha\pi}{2})}. 
\end{equation}
The Fourier transforms of the operators (\ref{eq:lcapu}) and (\ref{eq:rcapu})
have the forms \cite{Podlubny1999}
\numparts
\begin{eqnarray}
\label{eq:Flcapu}
_{-\infty}D_x^{\alpha} f(x)\div(-\mathrm{i}k)^{\alpha}\hat{f}(k)=|k|^{\alpha}
\exp\left({-\frac{\alpha\pi\mathrm{i}}{2}\mathrm{sgn}(k)}\right)\hat{f}(k),\\
_xD_\infty^\alpha f(x)\div(\mathrm{i}k)^{\alpha}\hat{f}(k)=|k|^{\alpha}\exp\left(
{\frac{\alpha\pi\mathrm{i}}{2}\mathrm{sgn}(k)}\right)\hat{f}(k),
\label{eq:Frcapu}
\end{eqnarray}
\endnumparts
where $\div$ defines the Fourier transform pairs.

For the case $\alpha=1$ and $\beta=0$ we have $L_{1,0}=R_{1,0}=1/\pi$
\cite{samko}, and instead of equation (\ref{eq:fraccoef}) we find
\begin{equation}
\frac{\partial\ell_{1,0}(x,t)}{\partial t}=-K_{\alpha}\frac{\partial}{\partial
x}\mathscr{H}\Big\{\ell_{1,0}(x,t)\Big\},
\label{eq:hilbertdiffeq}
\end{equation}
where $\mathscr{H}$ is the Hilbert transform
\begin{equation}
\mathscr{H}\Big\{f(x)\Big\}=\frac{1}{\pi}\dashint\limits_{-\infty}^{\infty}
\frac{f(\zeta)}{x-\zeta}
\mathrm{d}\zeta,
\label{eq:Hilbe}
\end{equation}
in terms of the principle value integral $\dashint$. In Fourier space the
Hilbert transform has the simple form 
\begin{equation}
\mathscr{H}\Big\{f(x)\Big\}\div\mathrm{i}\mathrm{sgn}(k)\hat{f}(k).  
\label{eq:Fhilb}
\end{equation}
In what follows we do not consider the particular case $\alpha=1$, $\beta\neq0$
since it cannot be described in terms of a space-fractional operator. For all
other choices of the parameters by substitution of relations (\ref{eq:Flcapu}),
(\ref{eq:Frcapu}), and (\ref{eq:Fhilb}) into equation (\ref{eq:ffpe}) we recover
the characteristic function (\ref{eq:charecA}) of the $\alpha$-stable process
after Fourier transform.

\section{Numerical scheme}
\label{NumSch}

To determine the first-passage properties of $\alpha$-stable processes we will
employ different numerical schemes based on the space-fractional diffusion
equation and the Langevin equation for LFs. We here detail their implementation.

\subsection{Diffusion description}
\label{Diffdesc}

There are several numerical methods to solve space-fractional diffusion equations,
such as the finite difference \cite{Jia2015,Shimin2018} and finite element
\cite{Deng-Weihua2008,W. Melean2007,Fix2004} methods as well as the spectral
method \cite{Bhrawy2016,Li2009}. In this paper we use the finite difference
method that uses differential quotients to replace the derivatives in the
differential equations. The domain is partitioned in space and time, and
approximations of the solution are computed. Due to causality we use forward
differences in time on the left hand side of equation (\ref{eq:ffpe}),
\begin{equation}
\frac{\partial}{\partial t}f(x_i,t_j)\approx\frac{f^{j+1}_i-f^j_i}{\Delta t},
\label{eq:disctime}
\end{equation}
where $f^j_i=f(x_i,t_j)$, $x_i=(i-I/2)\Delta x$, and $t_j=j\Delta t$, where
$\Delta x$ and $\Delta t$ are step sizes in position and time, respectively.
The $i$ and $j$ are non-negative integers, $i=0,1,2,\ldots,I$, and further
$x_0=-L$, $x_I=L$, and $\Delta x=2L/I$. Similarly, $j=0,1,2,\ldots,J-1$, $t_0=0$,
$t_J=t$, and $\Delta t=t/J$. Absorbing boundary conditions for the
determination of the first-passage event imply $f_0^j=f_I^j=0$ for all $j$.
The integrals on the right hand side of equation (\ref{eq:ffpe}) are
discretised as follows. For $0<\alpha<1$,
\numparts
\begin{equation}
\int\limits_{-L}^{x_i}\frac{f^{(1)}(\zeta,t_j)}{(x_i-\zeta)^{\alpha}}\mathrm{d}
\zeta\approx\displaystyle\sum_{k=1}^i\frac{f^j_k-f^j_{k-1}}{ \Delta x}\int
\limits_{x_{k-1}}^{x_k}\frac{1}{(x_i-\zeta)^{\alpha}}\mathrm{d}\zeta,
\label{eq:discleft1}
\end{equation}
for the left side derivative, and
\begin{equation}
\int\limits_{x_i}^L\frac{f^{(1)}(\zeta,t_j)}{(\zeta-x_i)^{\alpha}}\mathrm{d}
\zeta\approx\displaystyle\sum_{k=i}^{I-1}\frac{f^j_{k+1}-f^j_k}{ \Delta x}
\int\limits_{x_k}^{x_{k+1}}\frac{1}{(\zeta-x_i)^{\alpha}}\mathrm{d}\zeta,
\label{eq:discright1}
\end{equation}
\endnumparts
for the right side derivative. \textcolor{black}{Thus the idea is to approximate
only the derivative by the differences. The integral kernel is then calculated
explicitly. For the estimation of the error in this scheme we refer to \ref{error}.}
For the case $1<\alpha<2$ we use the central
difference approximation, namely,
\numparts
\begin{equation}
\int\limits_{-L}^{x_i}\frac{f^{(2)}(\zeta,t_j)}{(x_i-\zeta)^{\alpha-1}}\mathrm{d}
\zeta\approx\displaystyle\sum_{k=1}^i\frac{f^j_{k+1}-2f^j_k+f^j_{k-1}}{(\Delta x
)^2}\int\limits_{x_{k-1}}^{x_k}\frac{1}{(x_i-\zeta)^{\alpha-1}}\mathrm{d}\zeta,
\label{eq:discleft2}
\end{equation}
for the left side, and
\begin{equation}
\int\limits_{x_i}^L\frac{f^{(2)}(\zeta,t_j)}{(\zeta-x_i)^{\alpha-1}}\mathrm{d}
\zeta\approx\displaystyle\sum_{k=i}^{I-1}\frac{f^j_{k+1}-2f^j_{k}+f^j_{k-1}}{
(\Delta x)^2}\int\limits_{x_k}^{x_{k+1}}\frac{1}{(\zeta-x_i)^{\alpha-1}}
\mathrm{d}\zeta,
\label{eq:discright2}
\end{equation}
\endnumparts
for the right side. For the special case $\alpha=1$, $\beta=0$ by using the discrete
Hilbert transform \cite{SKak1970} we approximate the derivative in space, namely
\begin{equation}
\fl-\frac{d}{d x}(\mathscr{H}\left\{f(x,t)\right\})\approx-\frac{2}{\pi}
\displaystyle\sum_{k=1}^i\frac{f^j_k
-f^j_{k-1}}{\Delta x}\frac{1}{2(i-k)+1}-\frac{2}{\pi}\displaystyle\sum_{k=i}^{I-1}
\frac{f^j_k-f^j_{k+1}}{\Delta x}\frac{1}{2(k-i)+1}.
 \label{eq:Hermitd1}
\end{equation}
For further details of the numerical scheme we refer the reader to \ref{Appendix Num}.
To improve the stability we use the Crank-Nicolson method. By substitution of
equations
(\ref{eq:disctime}) to (\ref{eq:Hermitd1}) into equation (\ref{eq:ffpe}) we obtain
\begin{equation}
\mathbf{A}f^{j+1}=\mathbf{B}f^{j},
\label{eq:matrixab}
\end{equation}
in which the coefficients $\mathbf{A}$ and $\mathbf{B}$ have matrix form with
dimension $(I+1)\times (I+1)$ and $j=0,1,2,\ldots,J-1$. In the numerical scheme
the initial condition $f(x,0)=\delta(x)$ is approximated as
\numparts
\begin{equation}
f(x_i,t_0)=\left\{\begin{array}{ll}\frac{1}{\Delta x},& i=L/\Delta x\\
0 ,& \mbox{otherwise} \end{array}\right..
\end{equation}
For the setup used in our numerical simulations, see section \ref{surv-first}
below, the initial point is $f(x,0)=\delta(x-x(0))$ at $x(0)=L-d$,
\textcolor{black}{where $d<
2L$ is the distance of $x(0)$ from the right interval boundary (see figure
\ref{fig:fig3})}. For this case we
implement the initial condition as
\begin{equation}
f(x_i,t_{0})=\left\{\begin{array}{ll}\frac{1}{\Delta x},& i=(2L-d)/\Delta x\\
0,& \mbox{otherwise}\end{array}\right..
\end{equation}
\endnumparts
In the next step the time evolution of the PDF is obtained by using the absorbing
boundary conditions $f_0^j=f_I^j=0$ for all $j$ and applying to the matrix
coefficients $\mathbf{A}$ and $\mathbf{B}$.

\subsection{Langevin dynamics}
\label{Lange}

The fractional diffusion equation (\ref{eq:ffpe}) can be related to the Langevin
equation \cite{HCFogedby1997,Dybiec2016,jespersen}
\begin{equation}
\frac{\mathrm{d}}{\mathrm{d}t}{x}(t)=K_\alpha^{1/\alpha}\zeta(t),
\label{eq:langevin}
\end{equation}
where $\zeta(t)$ is white L{\'e}vy noise characterised by the same $\alpha$ and
$\beta$ parameters as the space-fractional operator (\ref{eq:fraccoef}) and with unit
scale parameter. The Langevin equation (\ref{eq:langevin}) provides a microscopic
(trajectory-wise) representation of the space-fractional diffusion equation
(\ref{eq:ffpe}). Therefore, from an ensemble of trajectories generated from
equation (\ref{eq:langevin}) it is possible to estimate the time dependent PDF
whose evolution is described by equation (\ref{eq:ffpe}). In numerical simulations
L{\'e}vy flights can be described by the discretised form of the Langevin equation
\begin{equation}
x(t+\Delta t)=x(t)+K_\alpha^{1/\alpha}(\Delta t)^{1/\alpha}\zeta, 
\label{eq:discrete}
\end{equation}
where $\zeta$ stands for the $\alpha$-stable random variable with a unit scale
parameter \cite{Samorodnitsky-Taqqu,janicki1994} and the same index of stability
$\alpha$ and skewness $\beta$ parameters as in equation (\ref{eq:langevin}).
Relation (\ref{eq:discrete}) is exactly the Euler-Maruyama approximation
\cite{janicki1996,kloeden2011numerical,maruyama1955continuous} to the general
$\alpha$-stable L{\'e}vy process. From the trajectories $x(t)$, see equations
(\ref{eq:langevin}) and (\ref{eq:discrete}), it is also possible to estimate the
first-passage time $\tau$ as
\begin{equation}
\tau=\min\{t:|x(t)|\geq L\}.
\end{equation}
From the ensemble of first-passage times it is also
possible to estimate the survival probability $S(t)$, the complementary cumulative
density of first-passage times. More precisely, the initial condition is $S(0)=1$,
and at every recorded first-passage event at time $\tau_i$, $S(t)$ is decreased by
the amount $1/N$, where $N$ is the number of records of first-passage times. If
a given estimation of the first-passage time is recorded $k$ times the survival
probability is decreased by $k/N$. For a finite set of first-passage times there
exists a small fraction of very large first-passage times. Therefore, this
estimation becomes poorer with increasing time $t$. In the next section we present
a comparison between the numerical solution of equation (\ref{eq:ffpe}) and the
$\alpha$-stable probability laws with characteristic function (\ref{eq:charecA}).

\subsection{Comparison with $\alpha$-stable distributions}
\label{compar}

We now show that the difference scheme for the space-fractional diffusion equation
provides excellent agreement with the theoretical results for the shapes of
$\alpha$-stable probability densities. To this end we use a MATLAB code to obtain
the inverse Fourier transform of the characteristic function \cite{JPNolan1997}.
This programme employs Zolotarev's so-called M-form for the parametrisation of
$\alpha$-stable distributions with parameters $\alpha$, $\beta_M$, $\mu_M$, and
$K_{\alpha}^M$, while in the main text we use the A-form with parameters $\alpha$,
$\beta_A$, $\mu_A$ and $K_{\alpha}^A$ \cite{Zolotarev1986}. Thus, along with the
code we use the corresponding change of the distribution's parameters, see
\ref{Appendix Char} and, in particular, equation (\ref{eq:relparAandM}) for details.

\subsubsection{Symmetric $\alpha$-stable distributions.}

In this section we show a comparison between $\alpha$-stable distributions
obtained by inverse Fourier transform of the characteristic function with the
numerical solution of the space-fractional diffusion equation in a bounded
domain $[-L, L]$ for skewness parameter $\beta=0$.

\begin{figure}
\centering
\includegraphics[width=0.49\textwidth]{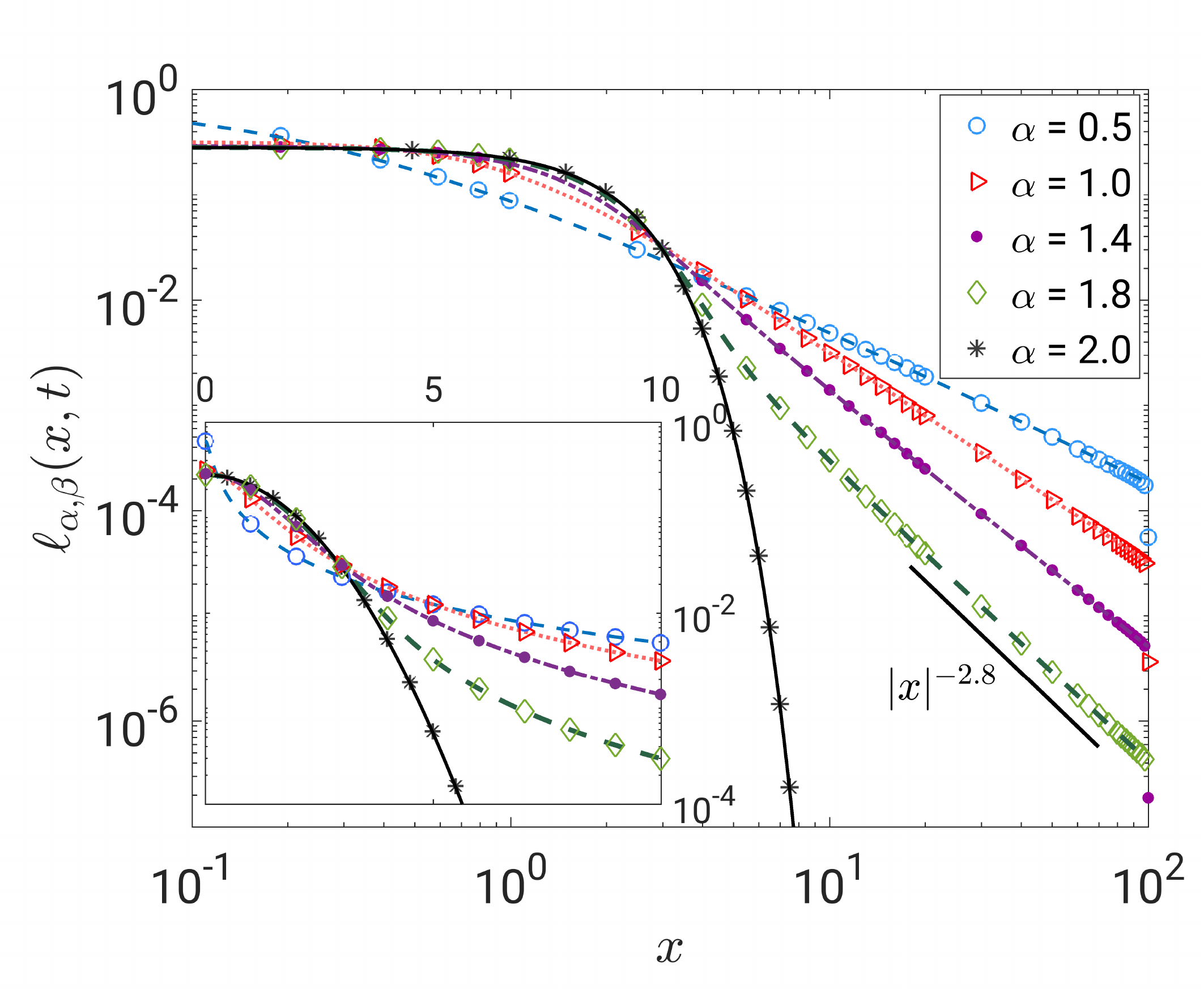}
\includegraphics[width=0.49\textwidth]{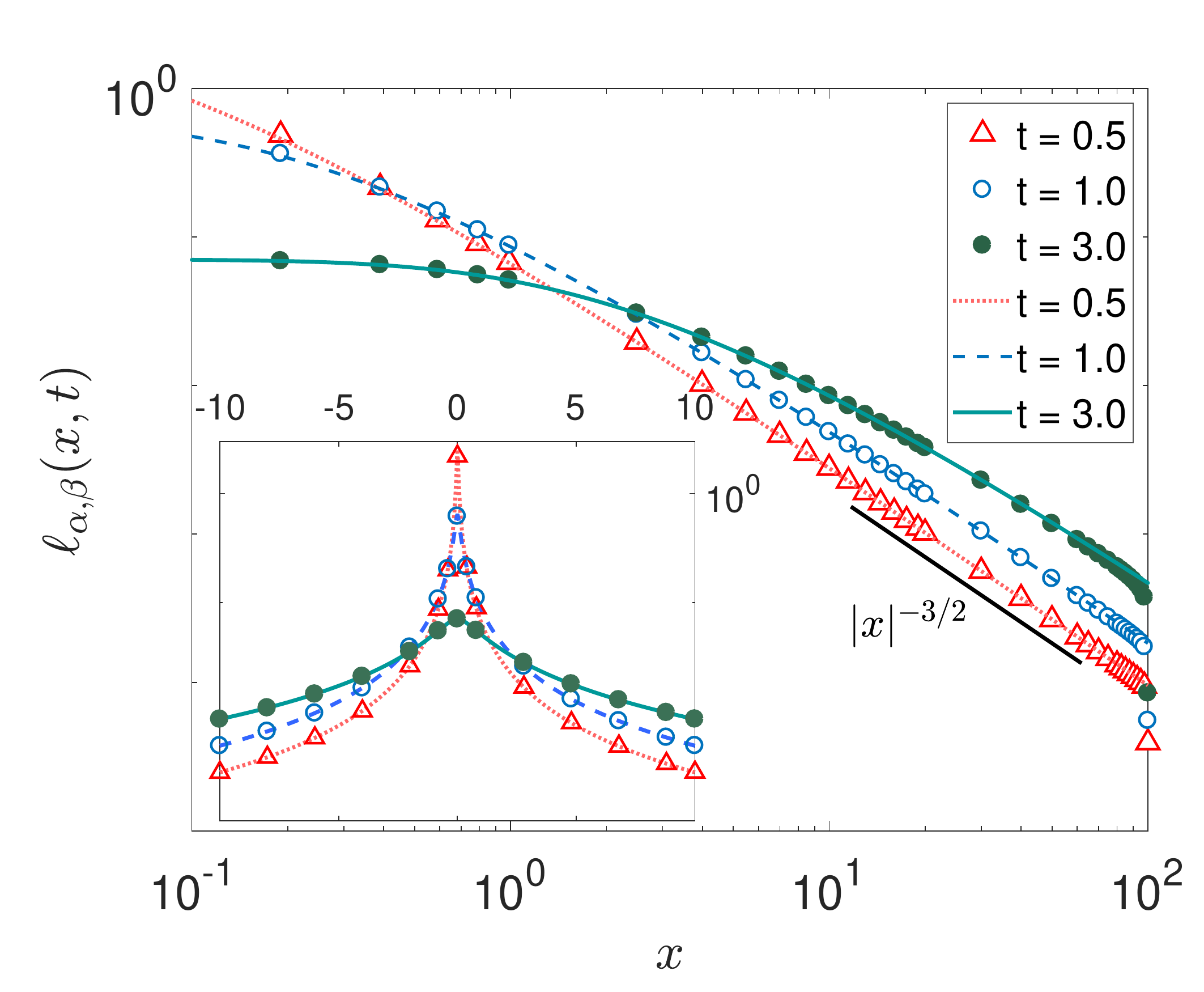}
\caption{Probability density function of symmetric $\alpha$-stable probability law
with $\beta=0$ and interval length $L=100$. Left: for different sets of $\alpha$
(see figure legend) for time $t=1$. Right: for $\alpha=0.5$ at different times. In
both figures we use $\Delta t=0.001$ as time step and $\Delta x=0.01$ for the step
length. In the insets we show a zoom into the central part of the PDF on a
log-linear scale. The symbols show the solution of the diffusion equation
(\ref{eq:ffpe}) while the lines show the $\alpha$-stable distributions obtained
from Fourier inversion of the characteristic function, displaying excellent
agreement. The black solid line shows the asymptotic behaviour of the PDF (main
panels). Effects of the absorbing boundary condition at $L=100$ can be seen as
fairly sudden drops of the PDF while at the times shown the central part of
the PDF remains hardly affected.}
\label{fig:fig1}
\end{figure}

We use absorbing boundary conditions and a finite domain with half length $L=100$
in one dimension, the initial condition is a Dirac $\delta$-function located at
$x(0)=0$. The probability density function for $\beta=0 $ and different sets of
the index of stability $\alpha$ at $t=1$ is displayed in figure \ref{fig:fig1}
on the left. The tails display the correct power-law scaling. In the right panel
of figure \ref{fig:fig1} we show the PDF for $\alpha=0.5$ and $\beta=0$ at
different times. The insets focus on the central part of the PDFs. In all cases
and over the entire plotted range the agreement between the numerical solutions
and the theoretical densities is excellent.

\subsubsection{Asymmetric $\alpha$-stable distributions.}

\begin{figure}
\centering
\includegraphics[width=0.49\textwidth]{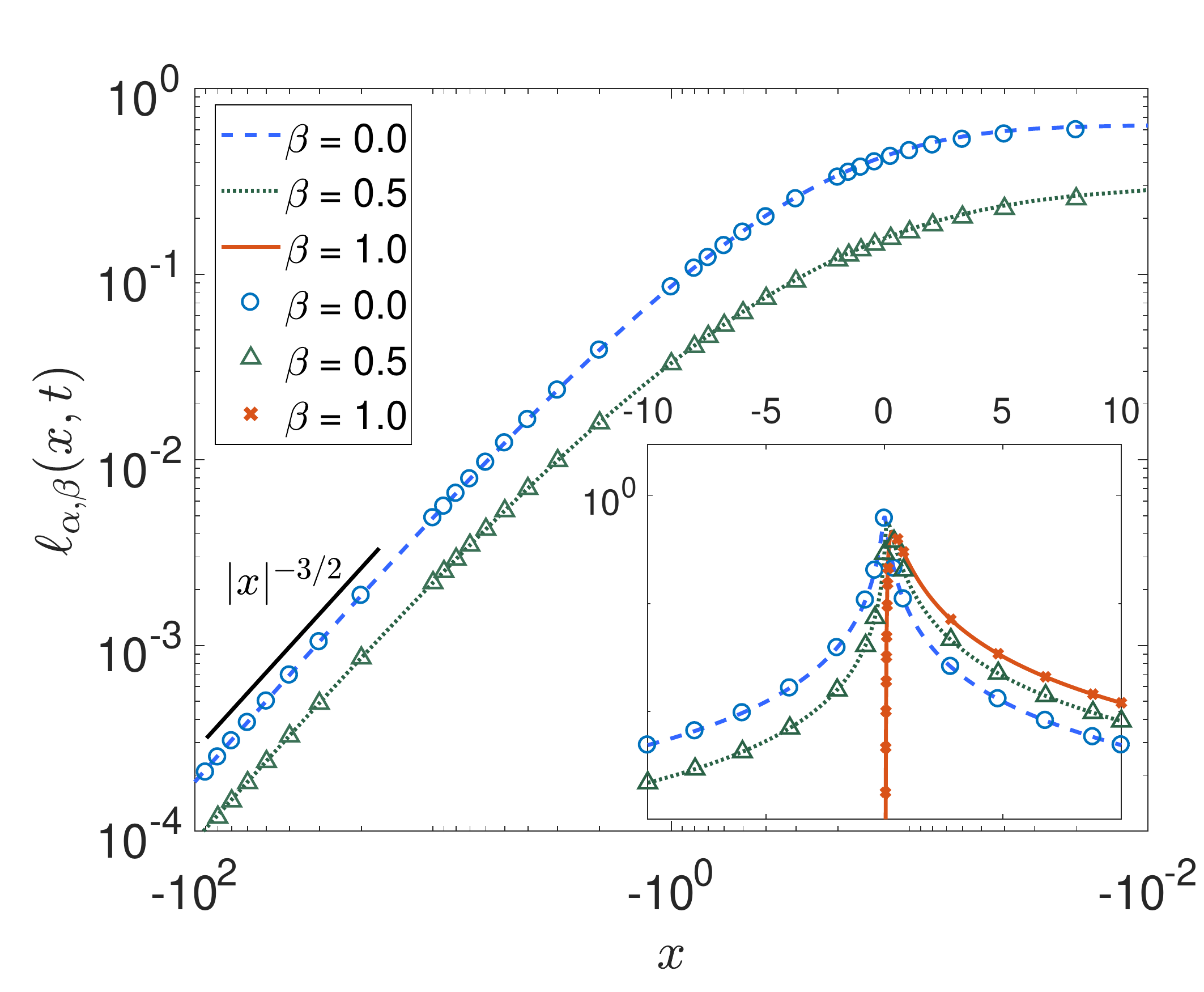}
\includegraphics[width=0.49\textwidth]{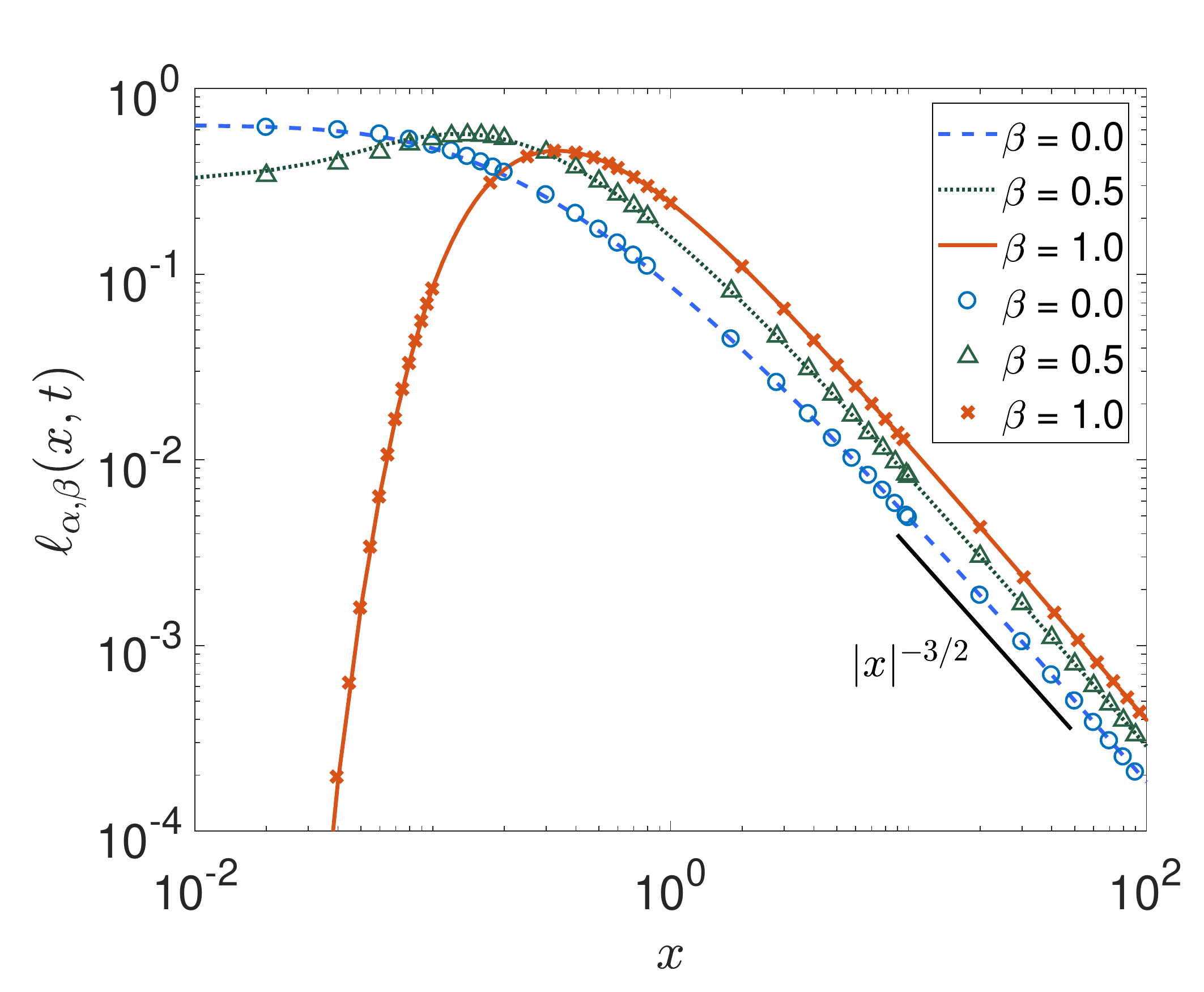}
\includegraphics[width=0.49\textwidth]{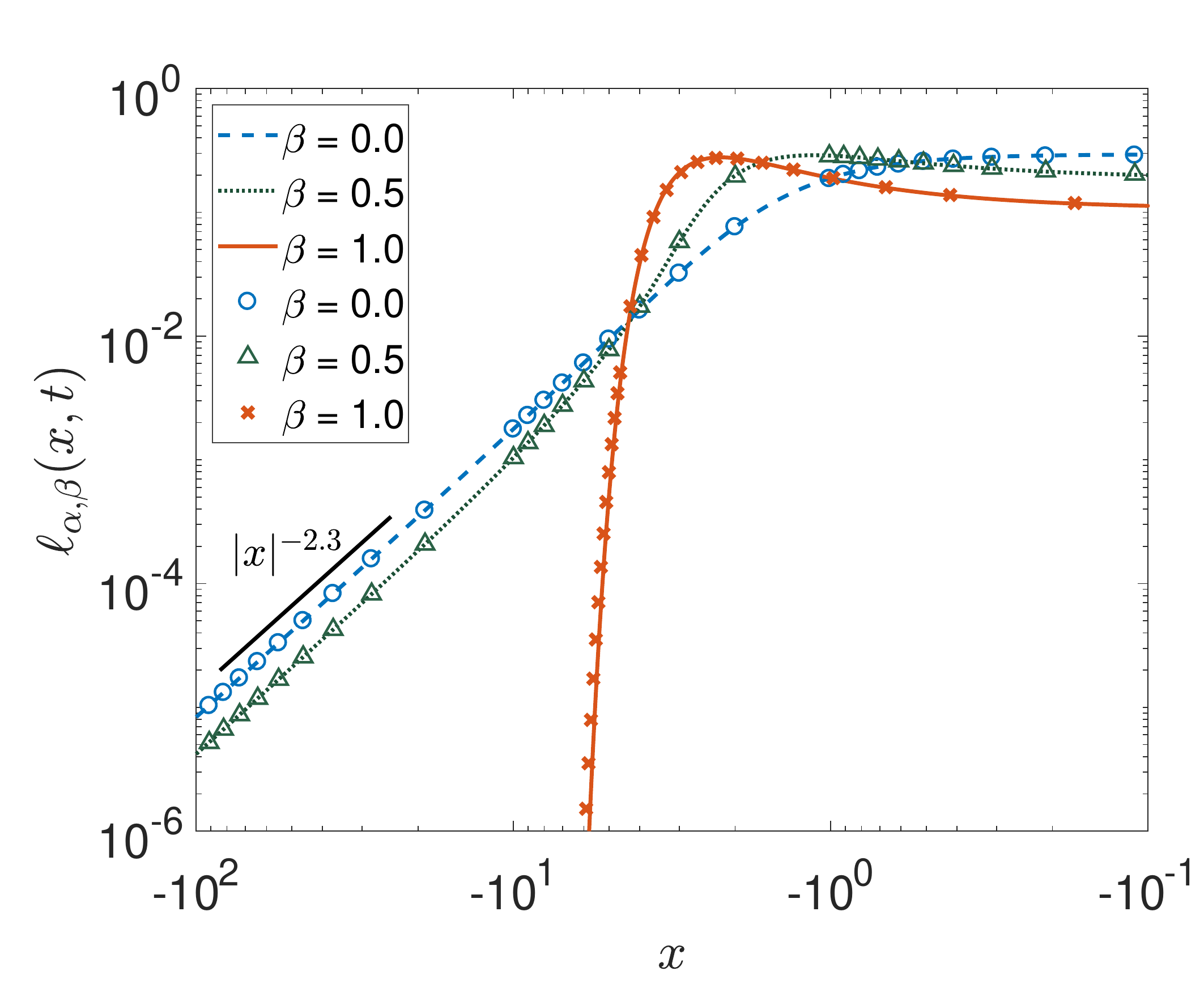}
\includegraphics[width=0.49\textwidth]{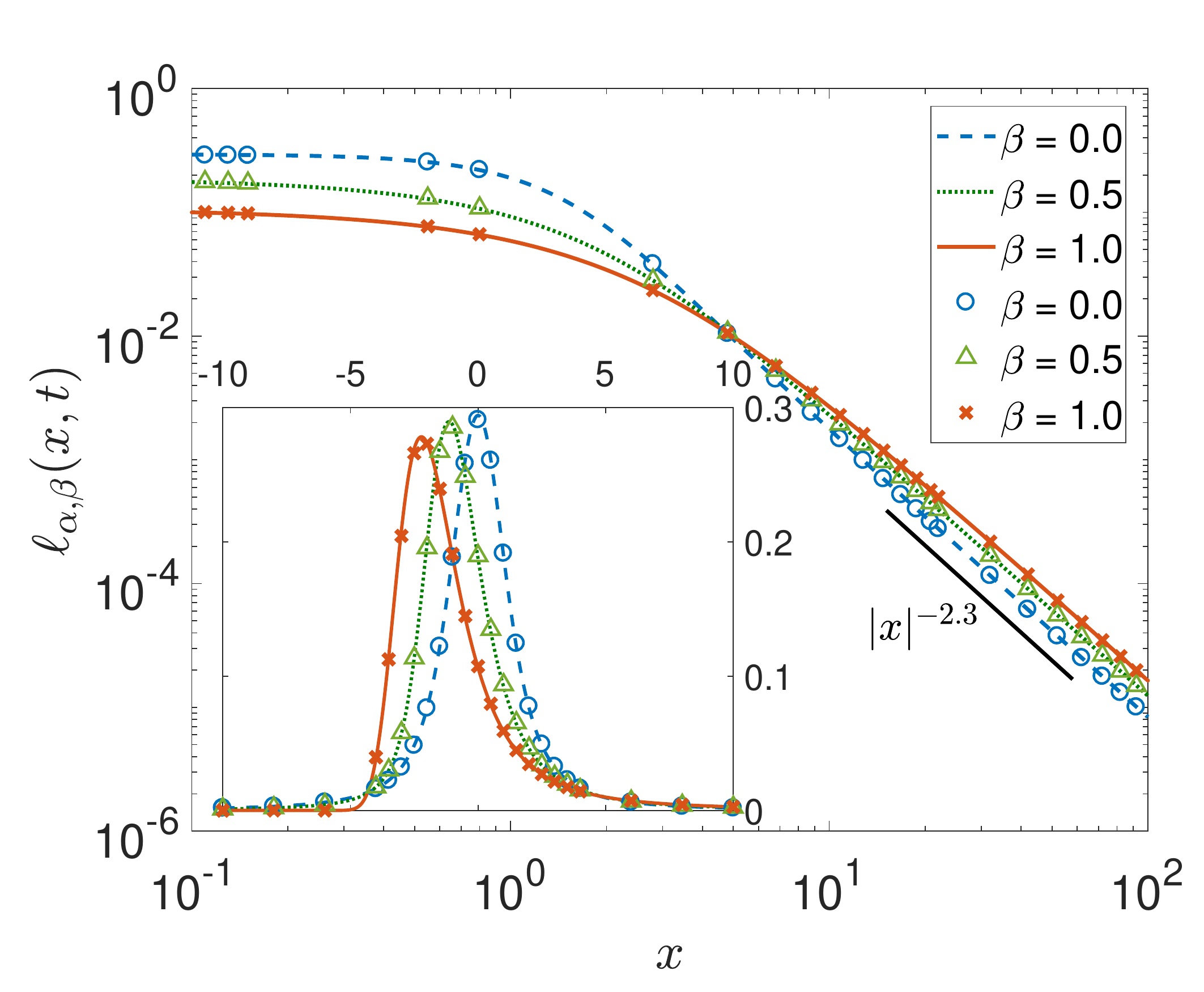}
\caption{Probability density function of $\alpha$-stable probability law for
different skewness parameters $\beta$ at time $t=1$. Top left: negative side
of the PDF for $\alpha=0.5$ (the inset focuses on the central part of the PDF
on the log-linear scale). Top right: positive side of the PDF for $\alpha=0.5$.
Bottom left: negative side of the PDF for $\alpha=1.3$. Bottom right: positive
side of the PDF for $\alpha=1.3$ (the inset shows the central part of the PDF
on log-linear scale). For each panel we use $L=100$, $\Delta t=0.001$, and $\Delta
x=0.01$. The symbols show the solution of the diffusion equation (\ref{eq:ffpe}) and
the lines show $\alpha$-stable distributions obtained from Fourier inversion.
The short black solid lines show the asymptotic behaviour of the PDF (main
panels).}
\label{fig:fig2}
\end{figure}

Asymmetric stable distributions with non-zero values of the skewness $\beta$
may occur in various situations, for instance, when in a random walk the left
and right diffusion coefficients are different. In figure \ref{fig:fig2} (top)
we plot the PDF with $\alpha=1/2$ at $t=1$ for different values of the skewness
parameter $\beta$. On the left side in the main panel the negative side of the
tails is shown, on the right side we display the positive side of the tails. Figure
\ref{fig:fig2} (bottom) analogously shows the PDF for $\alpha=1.3$ and different
$\beta$ at time $t=1$. Note that for $\beta=1$ \textcolor{black}{and $0<\alpha<1$}
the PDF is completely one-sided \textcolor{black}{on the positive axis} and does
not possess a left tail.

By comparison we see that again the numerical scheme for solving the space-fractional
diffusion equation produces solutions that are in very good agreement with the
numerical results for the $\alpha$-stable laws. In the following we study the
first-passage properties of random walk processes with $\alpha$-stable jump
length distribution obtained from two numerical methods: the space-fractional
diffusion equation and the Langevin approach. We also
compare the numerical method for solving the space-fractional diffusion
equation with results following from Skorokhod's theorem.

\section{Survival probability and first-passage properties}
\label{surv-first}

\begin{figure}
\centering
\includegraphics[width=0.60\textwidth]{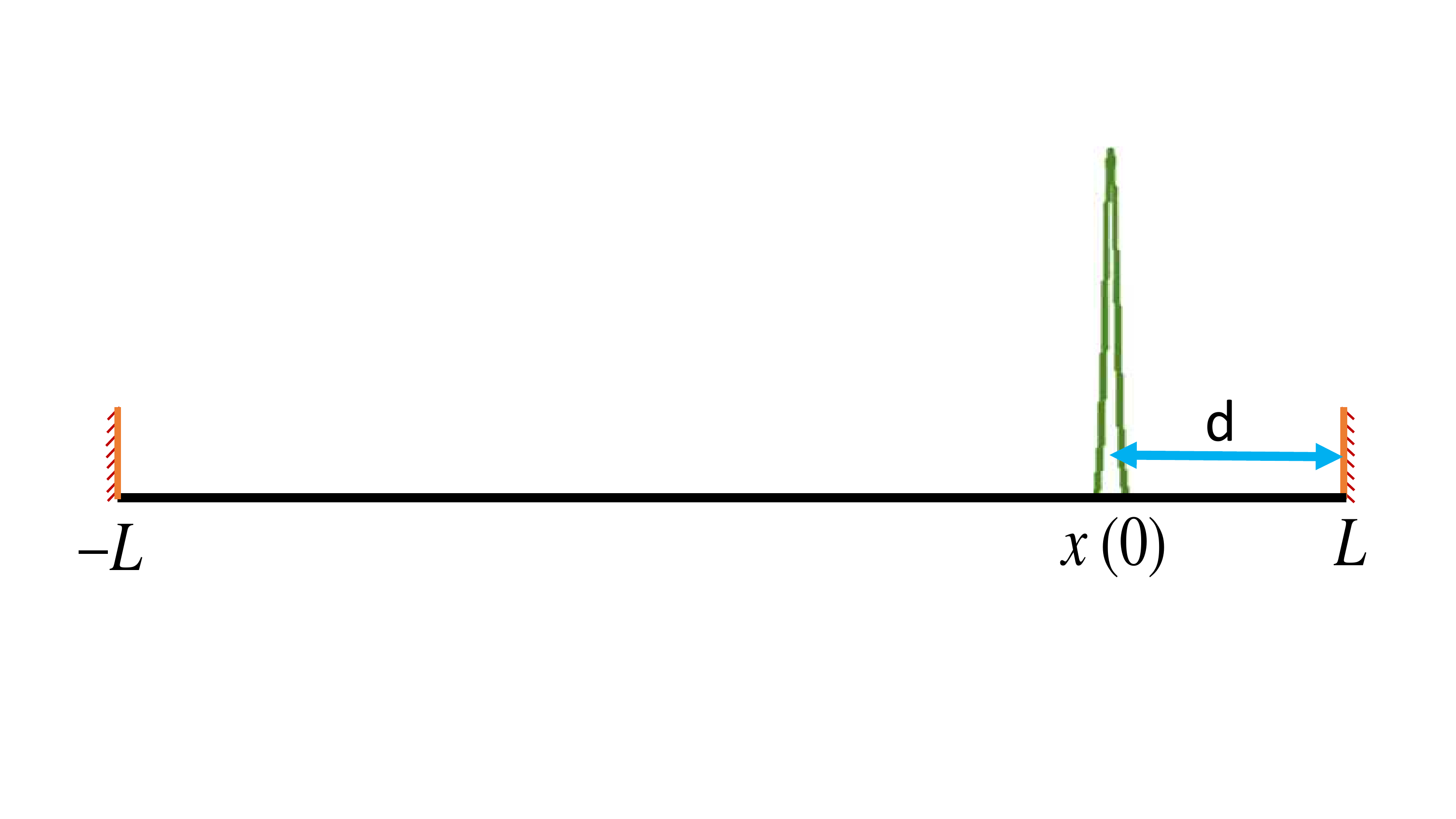}
\caption{Schematic of the setup used in our approach: in the interval of length $2L$
the initial probability density function is given by a $\delta$-distribution located
at $x(0)=L-d$, where $d$ is the distance from the right boundary. At both
interval boundaries we implement absorbing boundary conditions, that is, when the
particle hits the boundaries or attempts to move beyond them, it is absorbed.}
\label{fig:fig3}
\end{figure}

The survival probability and the first-passage time are observable statistical
quantities characterising the stochastic motion in bounded domains with absorbing
boundary conditions. In the following we investigate the properties of the survival
probability and the first-passage time density in a finite interval for symmetric
and asymmetric $\alpha$-stable laws underlying the space-fractional diffusion
equation. To this end we use the setup shown in figure \ref{fig:fig3}, in which
absorbing boundaries are located at $-L$ and $L$, and the initial point of the
initial $\delta$-distribution is located the distance $d$ away from the right
boundary.

The probability that at time $t$ the random walker is still present in the interval
$[-L, L]$ is defined as the survival probability \cite{SidneyRedner2014}
\begin{equation}
S(t)=\int\limits_{-L}^{L}\ell_{\alpha,\beta}(x,t)\mathrm{d}x,
\end{equation}
and the first-passage time PDF is given by the negative time
derivative,
\begin{equation}
\wp(t)=-\frac{\mathrm{d}S(t)}{\mathrm{d}t}.
\label{eq:fptandsurv}
\end{equation}
Therefore, in Laplace domain with initial condition $S(0)=1$ the relation between
the survival probability and the first-passage time reads
\begin{equation}
\wp(u)=1-uS(u).
\label{eq:fptandsurpLap}
\end{equation}
We now first consider the survival probability for symmetric $\alpha$-stable
laws in a semi-infinite and finite interval and demonstrate how the asymptotic
properties change with the length of the interval. Furthermore, we compare the
results obtained from the numerical difference scheme in section 3.1 with the
Langevin equation approach, before embarking for the study of asymmetric
$\alpha$-stable laws.

\subsection{First-passage processes for symmetric $\alpha$-stable laws}

\begin{figure}
\centering
\includegraphics[width=0.49\textwidth]{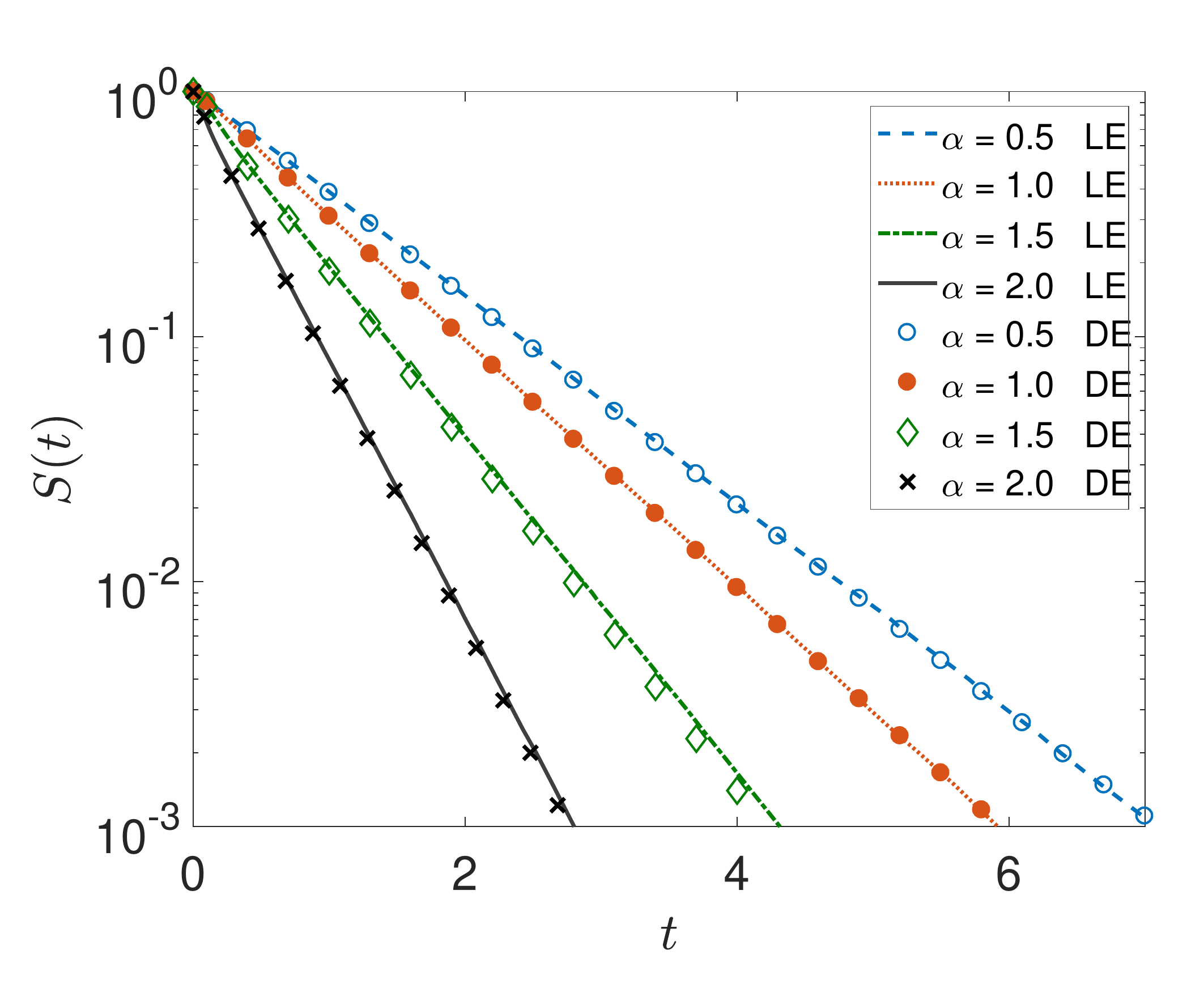}
\includegraphics[width=0.49\textwidth]{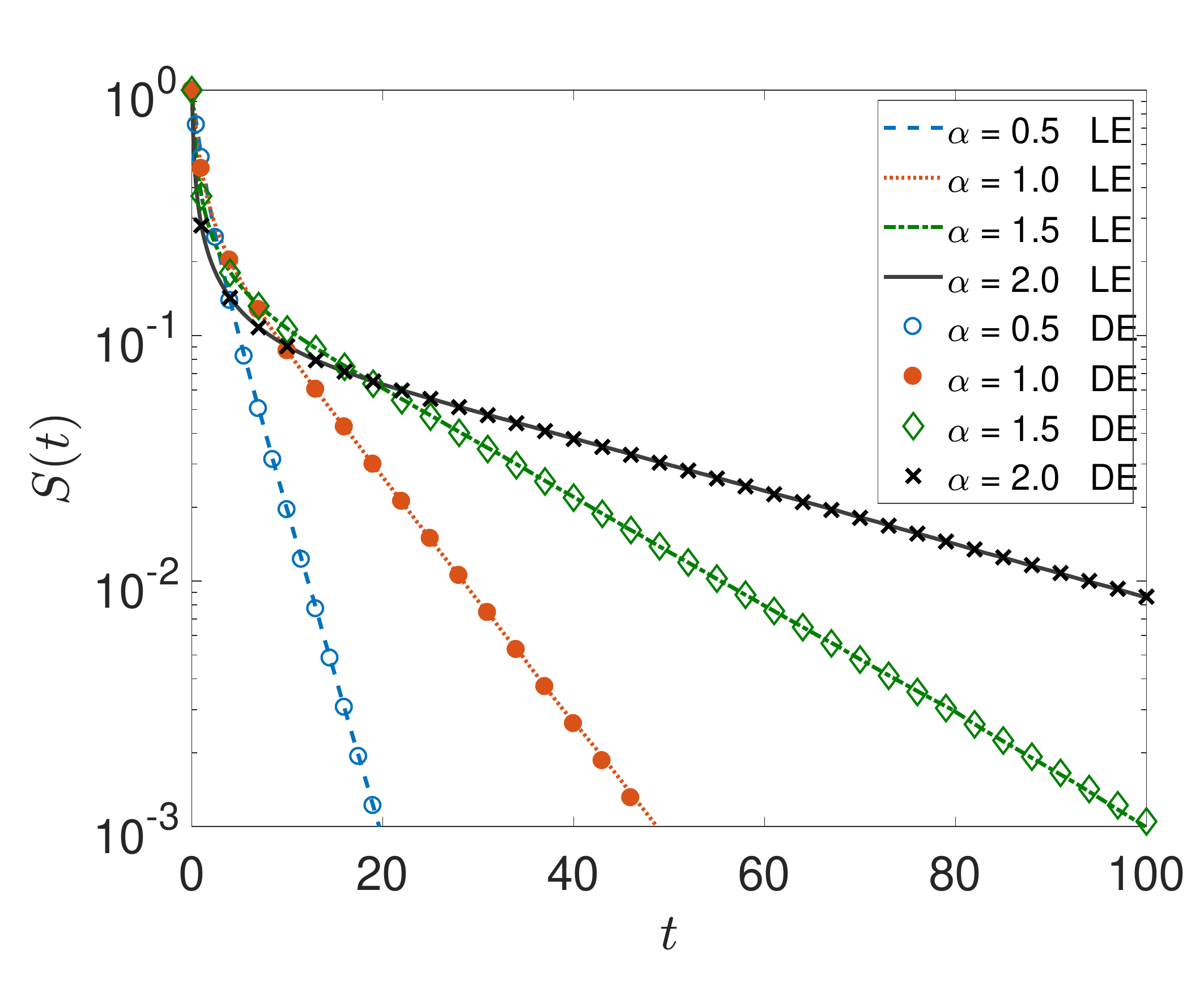}
\caption{Survival probability for symmetric $\alpha$-stable laws ($\beta=0$) in
log-linear scale with distance $d=0.5$ of the initial point from the right
boundary and interval half length $L=1$ (left) and $L=10$ (right), for
different indices $\alpha$. Symbols show results from numerical
solution of the space-fractional diffusion equation and lines correspond to the
Langevin equation simulation. Results constructed for the diffusion equation
use time step $\Delta t=0.0001$ and space increment $\Delta x=0.01$. In the
Langevin dynamics simulations the time step is $\Delta t=0.01$, data were
averaged over $N=10^6$ runs.}
\label{fig.fig4}
\end{figure}

Here we study the properties of $\alpha$-stable processes in domains
restricted by one or two boundaries. In figure \ref{fig.fig4} we show the
survival probability for different $\alpha$ and two different interval
lengths $L$ based on the difference scheme solution of the space-fractional
diffusion equation and simulations of the Langevin dynamics. The results
constructed with both methods are in very good agreement. The data in
figure \ref{fig.fig4} clearly show an exponential decay \textcolor{black}{(in
analogy to the escape dynamics of L{\'e}vy flights from a confining potential
\cite{levykramers,levykramers1}).} For the short interval the exponential
behaviour sets in almost immediately on the linear time axis in the plot,
while for the longer interval a crossover behaviour is visible, as we will
see below.

\begin{figure}
\centering
\includegraphics[height=6.4cm]{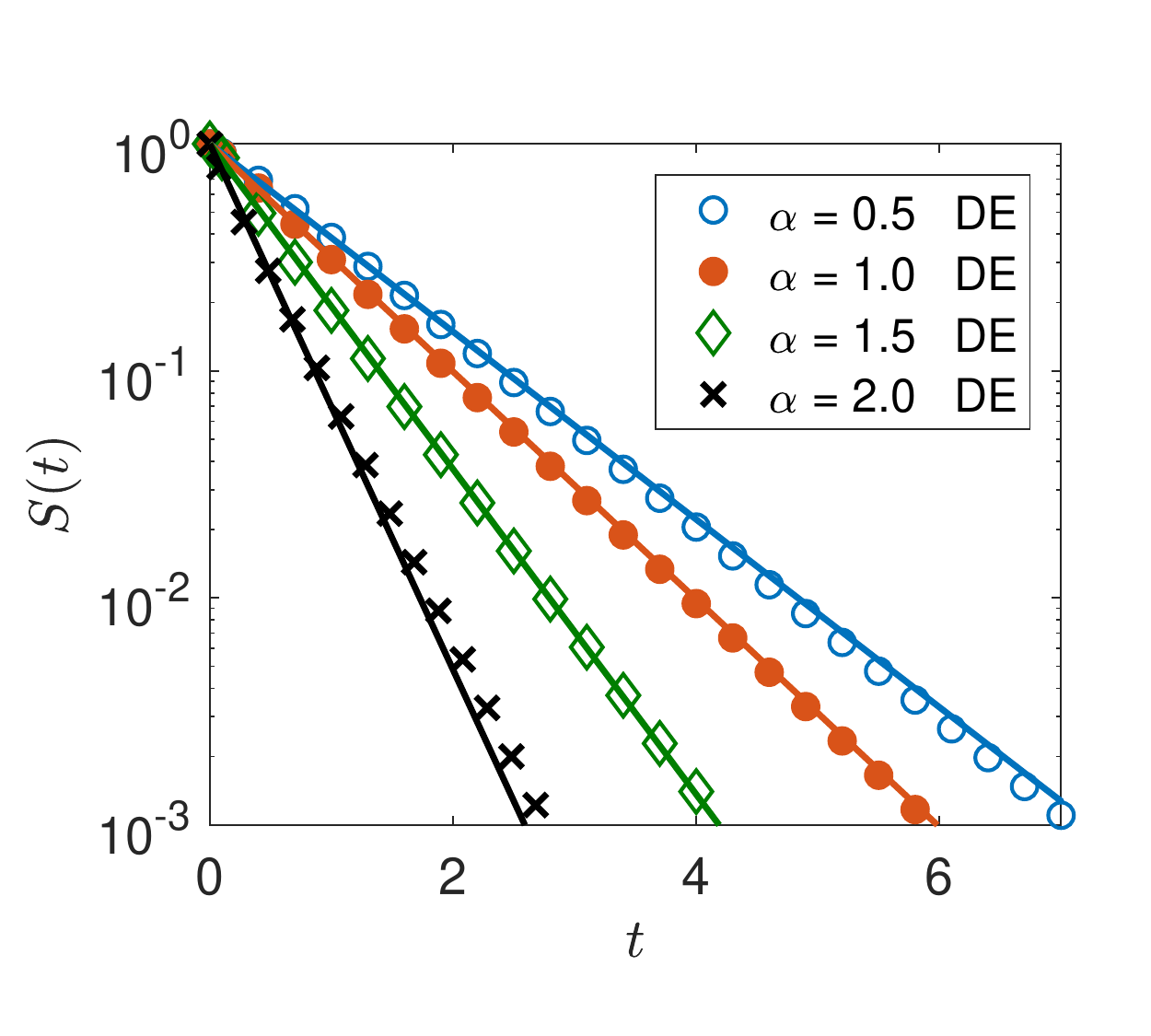}
\includegraphics[height=6.4cm]{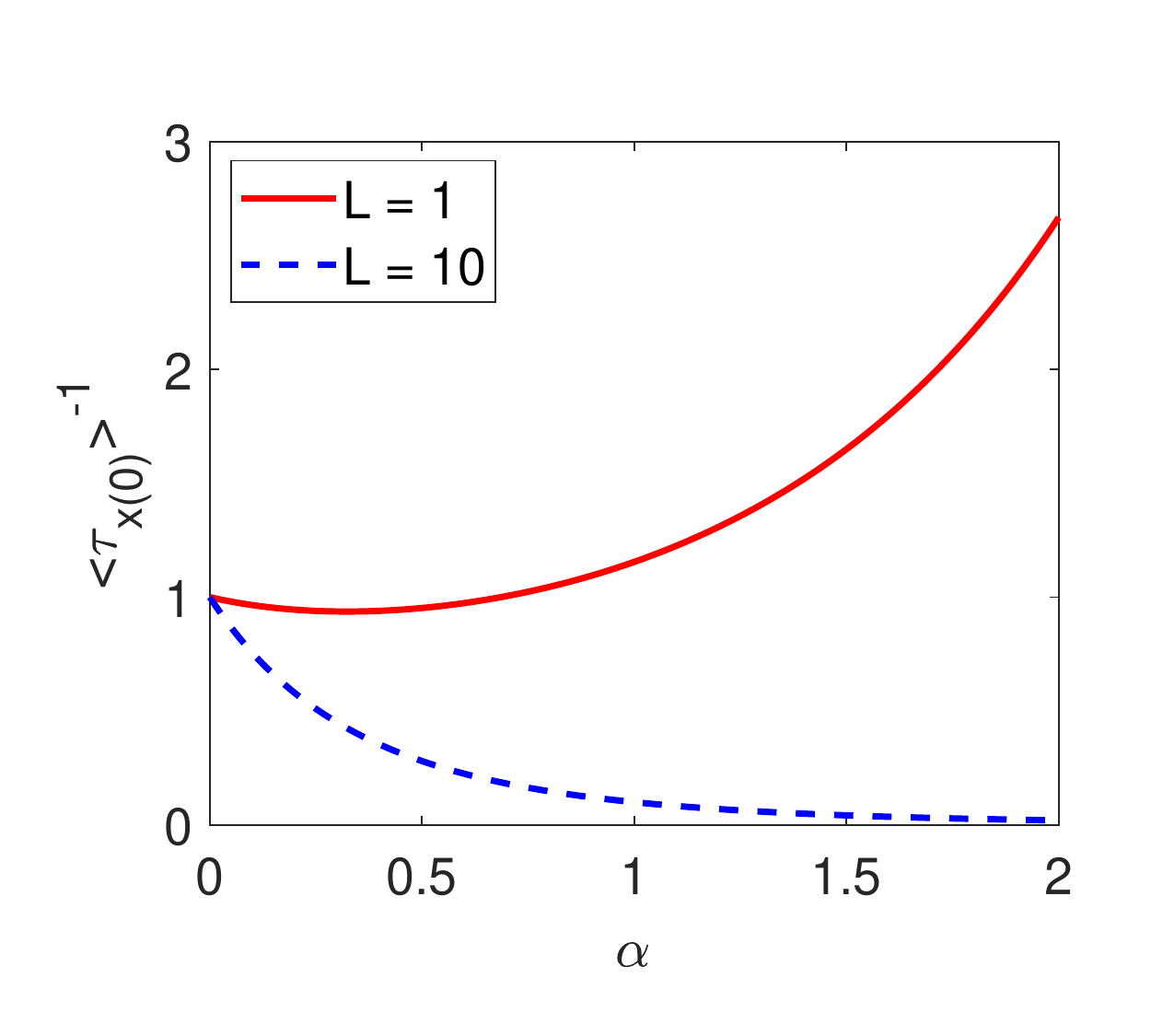}
\caption{\textcolor{black}{
Left: Survival probability for symmetric $\alpha$-stable processes ($\beta=0$)
in log-linear scale with $d=0.5$ and $L=1$, for different $\alpha$. Symbols
show results from the numerical solution of the space-fractional diffusion
equation and solid lines correspond to the exponential approximation
(\ref{eq:Surapprox}). Right: inverse mean first-passage time $\langle\tau_{x(0)}
\rangle^{-1}$ versus stable index $\alpha$ for two values of the interval half
length $L=1$ and $L=10$, for $d=0.5$.}}
\label{fig:fig4-Revise}
\end{figure}

\textcolor{black}{
Interestingly, we see in figure \ref{fig.fig4} that the trend of decay reverses
with respect to the stable index $\alpha$. To understand this behaviour of the
survival probability we use the following approximation of the survival probability
of symmetric L\'evy flights in finite intervals,
\begin{equation}
S(t|x(0))\approx\exp{\left(-\langle\tau_{x(0)}\rangle^{-1} t\right)},
\label{eq:Surapprox}
\end{equation}
where the mean first-passage time
\begin{equation}
\langle\tau_{x(0)}\rangle=\int_0^{\infty}S(t|x(0))\mathrm{d}t
\end{equation}
is given by \cite{Getoor1961,Buldyrev2001}
\begin{equation}
\label{eq:Surapprox1}
\langle\tau_{x(0)}\rangle=\frac{(L^2-|x(0)|^2)^{\alpha/2}}{\Gamma(1+\alpha)K_\alpha}.
\end{equation}
Figure \ref{fig:fig4-Revise} compares this approximation with the numerical solution
of the space-fractional diffusion equation. As we can see from the left panel,
relations (\ref{eq:Surapprox}) and (\ref{eq:Surapprox1}) agree very well with the 
numerical results for $L=1$: an increase of the interval length $L$ leads to a
decrease of $\langle\tau_{x(0)}\rangle^{-1}$ for larger $\alpha$ (right panel of
figure \ref{fig:fig4-Revise}).}

For a semi-infinite domain with absorbing boundary condition it is well known
that the first-passage time density for any symmetric jump length distribution
in a Markovian setting has the universal Sparre Andersen asymptotic $\wp(t)\simeq
t^{-3/2}$ (and thus $S(t)\simeq t^{-1/2}$ \cite{SidneyRedner2014,ESparre1953,
ESparre1954}. This is exactly our setting here for the symmetric case with
$\beta=0$, and the Sparre Andersen universality was consistently corroborated
for symmetric LFs in a number of works, inter alia, \cite{jpa,Dybiec2016,
Koren2007PhysicaA,Koren2007PRL}.

\begin{figure}
\centering
\includegraphics[width=0.49\textwidth]{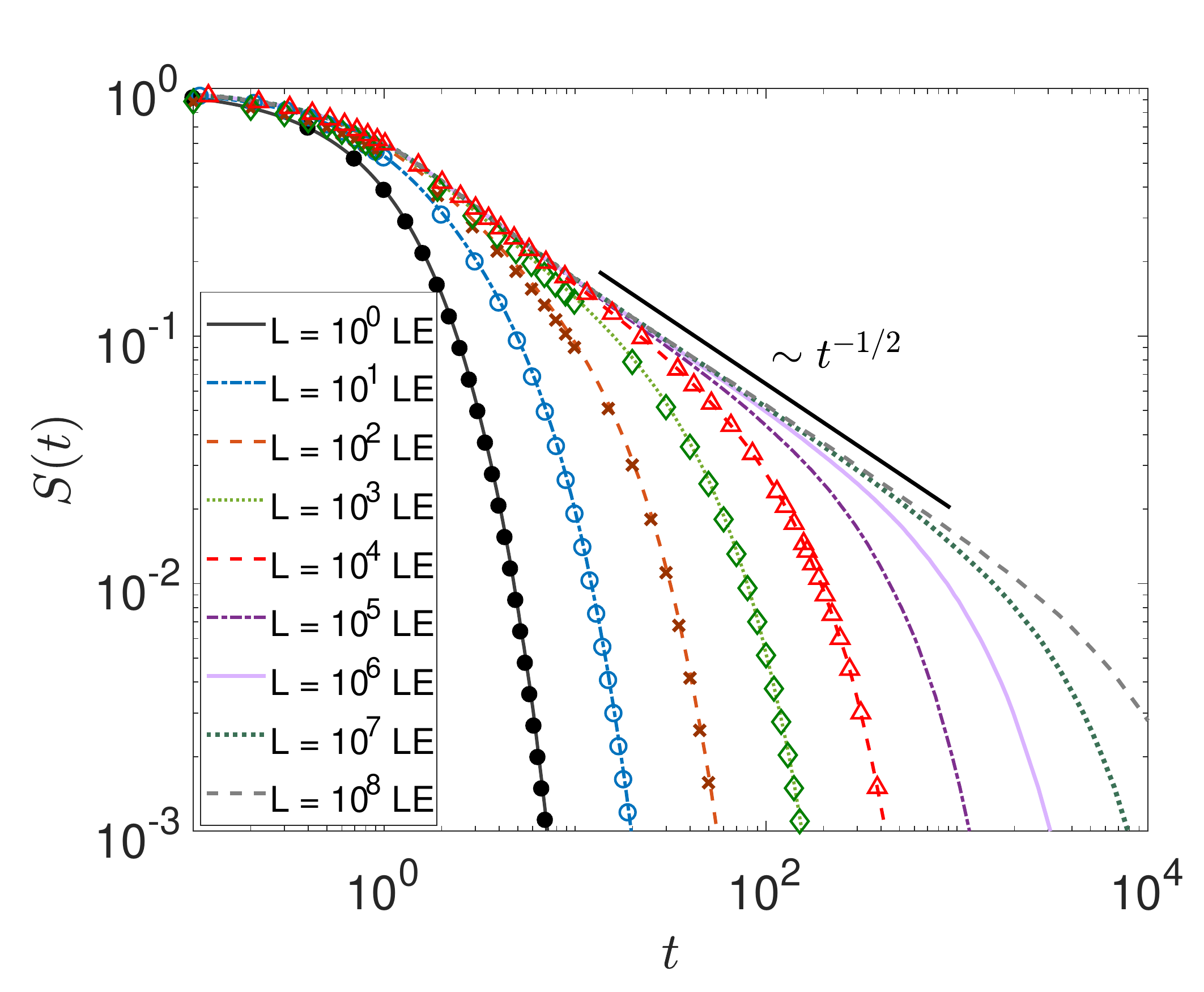}
\includegraphics[width=0.49\textwidth]{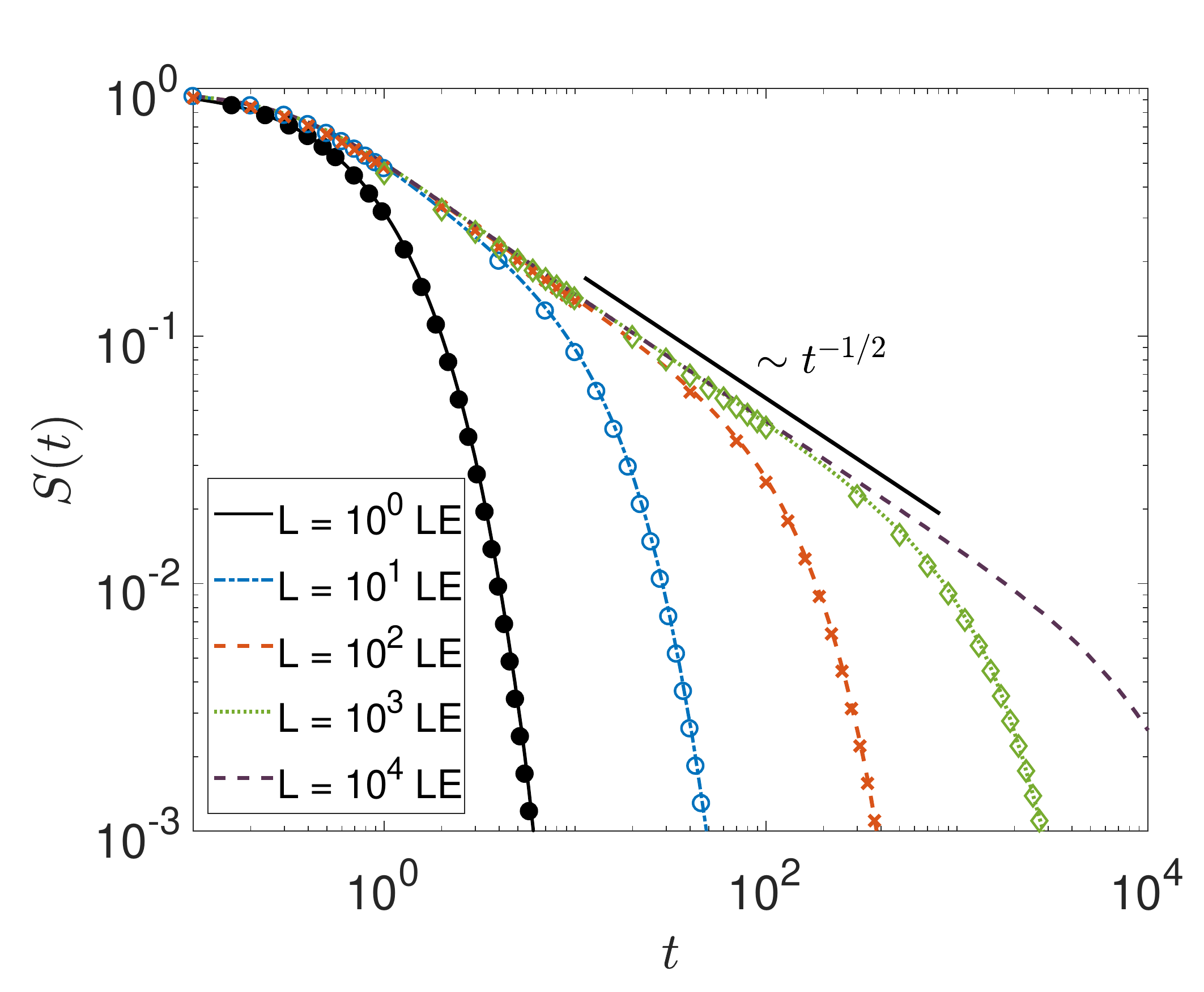}\\
\includegraphics[width=0.49\textwidth]{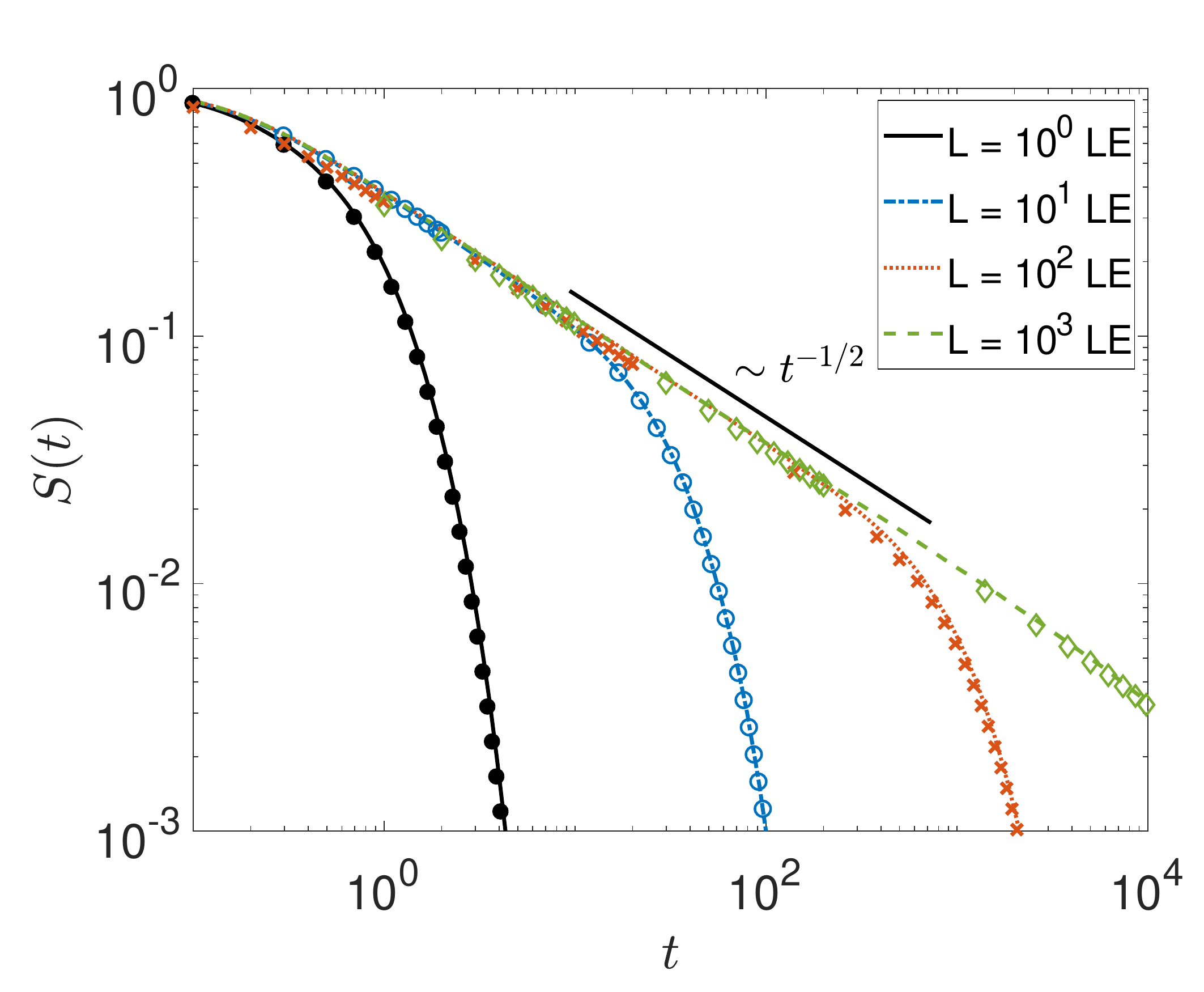}
\includegraphics[width=0.49\textwidth]{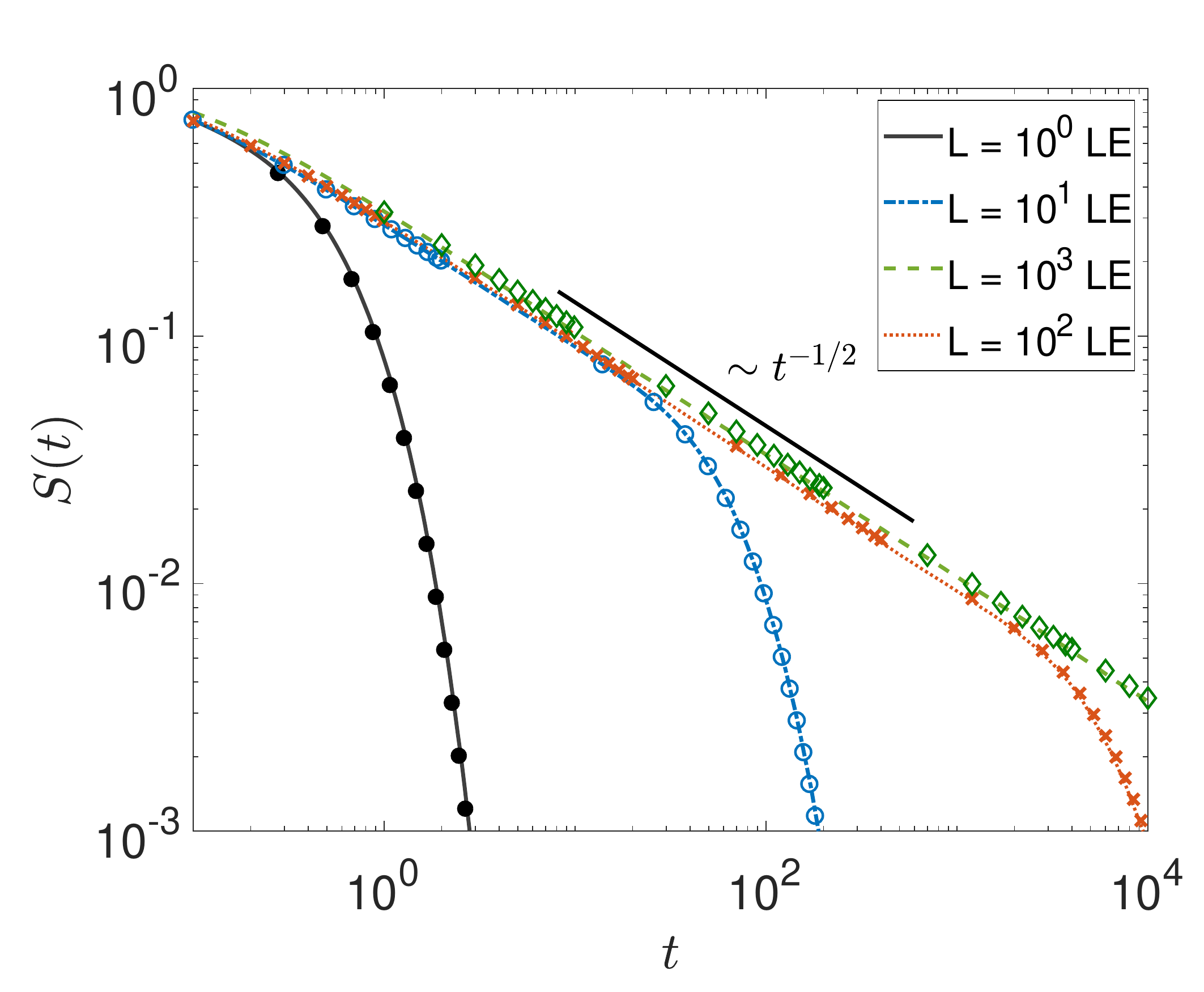}
\caption{Survival probability for different symmetric ($\beta=0$)
$\alpha$-stable densities with index $\alpha=0.5$ (top left), $\alpha=1.0$ (top
right), $\alpha=1.5$ (bottom left), and $\alpha=2.0$ (bottom right) for different
$L$ (see figure legends) with $d=0.5$. The lines represent simulation results of
the Langevin equation, while the symbols correspond to numerical results based on
the space-fractional diffusion equation. The black solid lines represent the
universal Sparre-Andersen scaling $S(t)\simeq t^{-1/2}$, which in the finite
interval is eventually cut off by an exponential shoulder.}
\label{fig:fig5}
\end{figure}

In figure \ref{fig:fig5} we study what happens at intermediate times in the case
of a finite interval, before the
terminal exponential shoulder cuts off the survival probability, as shown in
figure \ref{fig.fig4}. On the log-log scale of figure \ref{fig:fig5}, we indeed
recognise a transient power-law scaling with the universal Sparre Andersen
exponent $1/2$ for the survival probability. The onset of the hard exponential 
cutoff is shifted to longer times with increasing interval size $L$, in which,
on average, it takes the particles longer to explore the full extent of the
domain. This, of course, is fully consistent with results for normal diffusion
as well as continuous time random walk subdiffusion subordinated to regular
random walks \cite{aljaz,aljaz1,denis,SidneyRedner2014,bvp}, compare also the
discussion of the area filling dynamics of LFs \cite{johannesmahsa}.
Moreover, we see that the results from numerical solution of the fractional
diffusion equation and simulations of the Langevin equation almost perfectly
agree with each other. The lines without symbols in the top left of figure
\ref{fig:fig5} correspond to cases when the numerical approach based on the
fractional diffusion equation did not converge well. \textcolor{black}{We note
that one has to increase the value of $L$ with decreasing $\alpha$ in order
to meet the Sparre-Andersen scaling for a semi-infinite interval. This is
intuitively clear, as smaller $\alpha$ enhances the likelihood of longer
jumps and thus effects a higher probability of interaction with the interval
boundaries at fixed $L$.}

\begin{figure}
\centering
\includegraphics[width=0.49\textwidth]{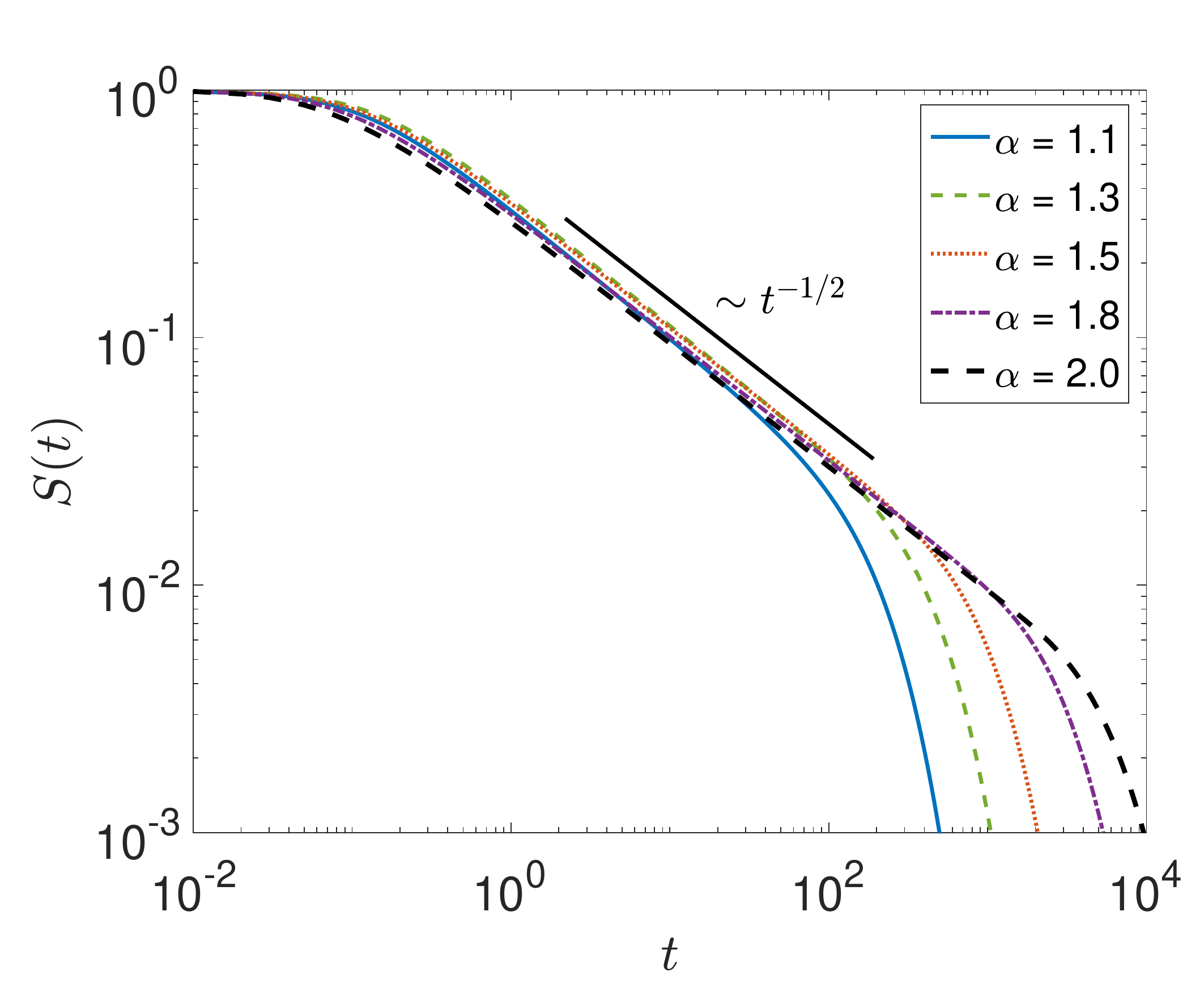}
\includegraphics[width=0.49\textwidth]{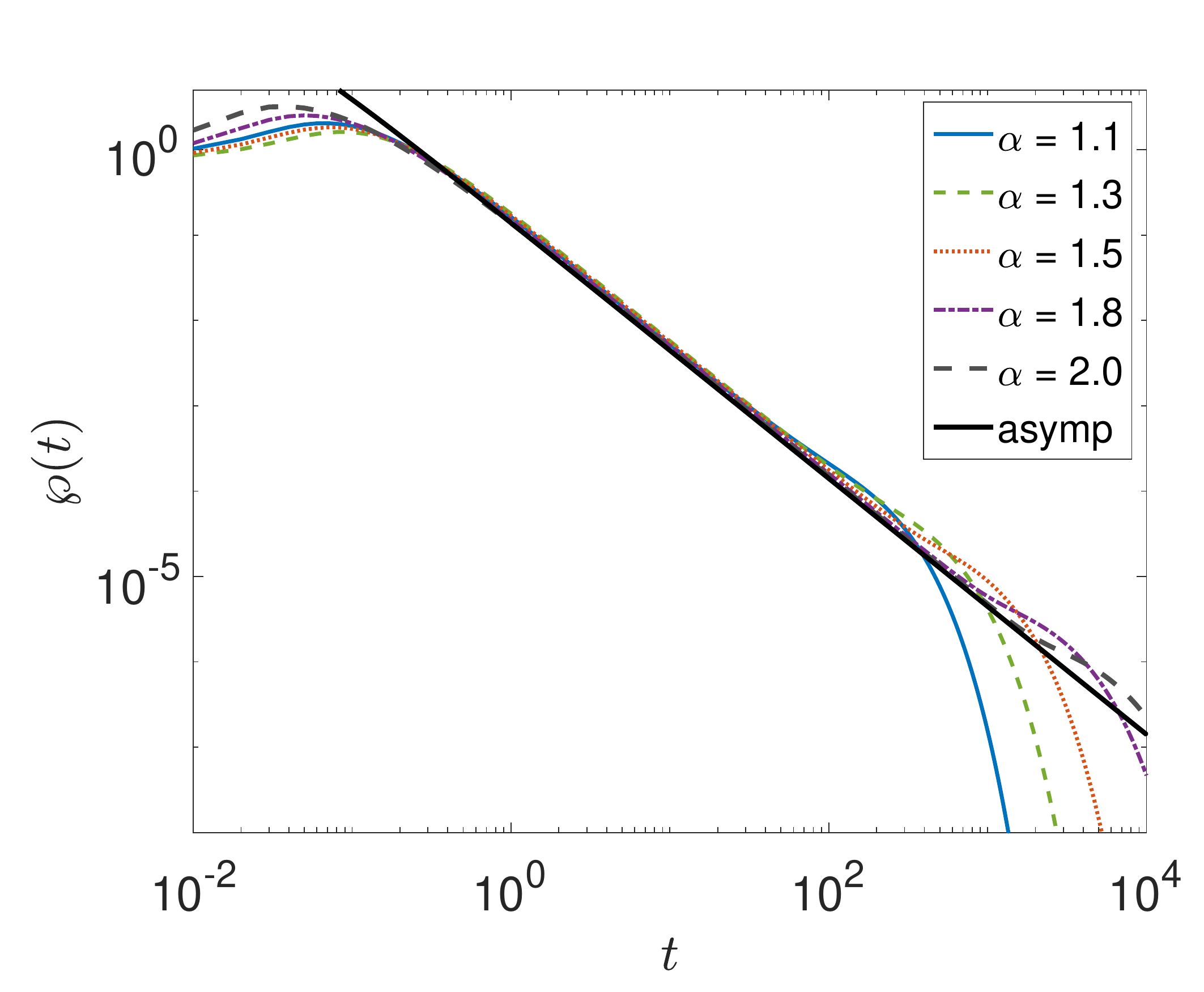}
\caption{Left: Survival probabilities for $d=0.5$, $L=100$ and the skewness
parameter $\beta=0$ for different sets of the index of stability $\alpha$
obtained from solving the space-fractional diffusion equation for $\Delta t=
0.01$ and $\Delta x=0.1$. The
solid short black line shows the Sparre Andersen scaling. Right: First passage
time probability density function. The
solid black line shows equation (\ref{eq:skorokhod16}) with the prefactor.}
\label{fig:fig6}
\end{figure}

In figure \ref{fig:fig6} for the interval size $L=100$ we show the survival
probability in the left panel along with the the first-passage time density
in the right panel, for different $\alpha$ and $\beta=0$. Consistently, the
transient
Sparre Andersen scaling is passed on from the power-law exponent $1/2$ for
the survival probability to the exponent $3/2$ of the first-passage density.

In the theory of a general class of L{\'e}vy processes, that is, homogeneous
random processes with independent increments, there
exists a theorem, that provides an analytical expression for the PDF of first
passage times in a semi-infinite interval, often referred to as the Skorokhod theorem
\cite{Gikhman-Skorokhod,Skorokhod1964}. Based on this theorem the asymptotic
expression for symmetric $\alpha$-stable laws, the first-passage time PDF is
(\ref{Appendix Skoro}) \cite{Koren2007PRL}
\begin{equation}
\wp(t)\sim\frac{d^{\alpha/2}}{\alpha \sqrt{\pi K_{\alpha} }\Gamma{(\alpha/2)}}
t^{-3/2},
\label{eq:skorokhod16}
\end{equation}
which specifies an exact expression for the prefactor of the Sparre Andersen
power-law. For Brownian motion ($\alpha=2$), the PDF for the first-passage
time has the well known L{\'e}vy-Smirnov form \cite{Feller1971}
\begin{equation}
\wp(t)=\frac{d}{\sqrt{4\pi K_2t^3}}\exp{\left(-\frac{d^2}{4K_2t}
\right)},
\label{eq:LevySmirnovFPT}
\end{equation}
that also emerges from the Skorokhod theorem in the limit $\alpha=2$. Equation
(\ref{eq:LevySmirnovFPT}) is exact for all times \cite{Feller1971,Redner}, and
apart from the Sparre Andersen law $\wp(t)\simeq t^{-3/2}$ it includes the hard
short time exponential cutoff reflecting the fact that it takes the particle a
typical time $\propto d^2/K_2$ to reach the absorbing boundary.

\subsection{First-passage processes for asymmetric $\alpha$-stable laws}
\label{DD_sec}

The case of asymmetric $\alpha$-stable laws is mathematically more involved and
also has been less well studied numerically. We now present results for the
survival probability and first-passage time PDF for different skewness parameters
$\beta$, addressing first the special cases of completely one-sided and extremal
two-sided laws.

\subsubsection{One-sided $\alpha$-stable laws.}

One-sided $\alpha$-stable laws with $\alpha\in(0,1)$, $\beta=1$ satisfy the
non-negativity of their
increments. Physically, such laws are appropriate for jump processes that always
move in the same direction, for instance, as a generalisation of shot noise.
Applying the Skorokhod theorem to this case one can show that the first-passage
time PDF in the permitted interval $\alpha\in(0,1)$
has the exact analytical solution (\ref{Appendix Skoro}) \cite{Koren2007PRL}
\begin{equation}
\wp(t)=\frac{\xi}{d^\alpha} M_\alpha\left(\frac{\xi t}{d^\alpha}\right),
\label{eq:fptoneside}
\end{equation}
with
\begin{equation}
\label{xidef}
\xi=\frac{K_{\alpha}}{|\cos(\alpha \pi/2)|},
\end{equation}
and where $M_\alpha$ is the Wright $M$-function (also called Mainardi function)
\cite{Podlubny1999,FMainardi2010} with series representation (\ref{eq:seriMfun})
and asymptotic exponential decay (\ref{eq:asymMfun}). At sufficiently long times
the first-passage time PDF reads
\begin{equation}
\wp(t)\sim A_1(\alpha)t^{(\alpha-1/2)/(1-\alpha)}\exp{(-B_1(\alpha)t^{1/(1-
\alpha)})},
\label{eq:fptasymp}
\end{equation}
where we used the coefficients
\begin{equation}
A_1(\alpha)=\frac{(\alpha\xi/d^\alpha)^{1/(2-2\alpha)}}{\alpha\sqrt{2\pi(1
-\alpha)}},\quad
B_1(\alpha)=\frac{1-\alpha}{\alpha}(\alpha\xi/d^\alpha)^{1/(1-\alpha)}.
\label{eq:asymoneprefa}
\end{equation}
\textcolor{black}{From equation (\ref{eq:fptasymp}) we see that for smaller $\alpha$
the first-passage time density decays faster which is intuitively clear since
L{\'e}vy flights with smaller $\alpha$ feature longer jumps in the positive
direction.} The value of $\wp(t)$ at $t=0$,
\begin{equation}
\wp(t=0)=\frac{\xi}{\Gamma{(1-\alpha)}d^\alpha},
\end{equation}
demonstrates that, in contrast to the Gaussian case, the probability density
does not vanish at $t=0$, indicating the possibility of immediate escape.

\begin{figure}
\centering
\includegraphics[height=6.8cm]{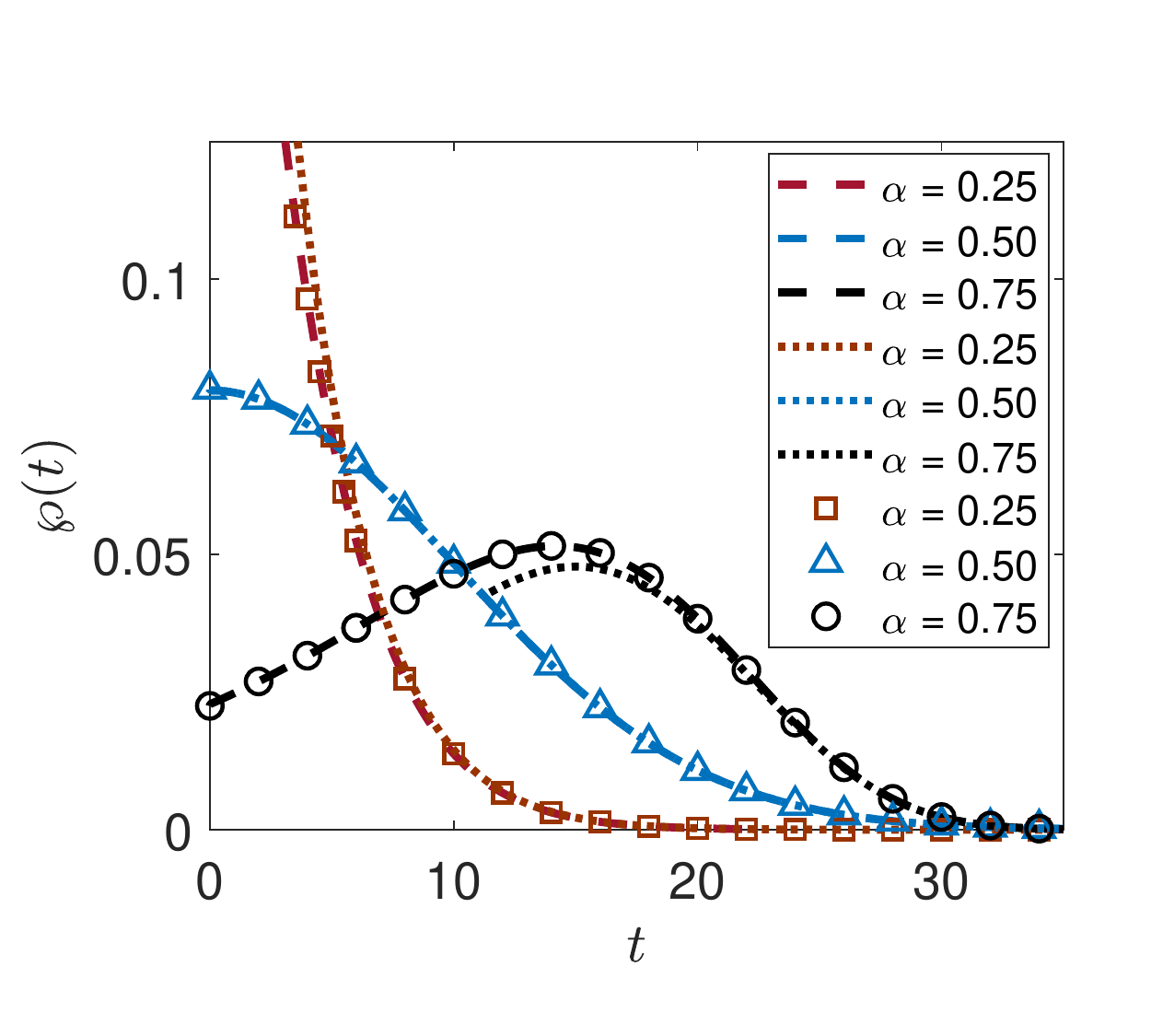}
\includegraphics[height=6.8cm]{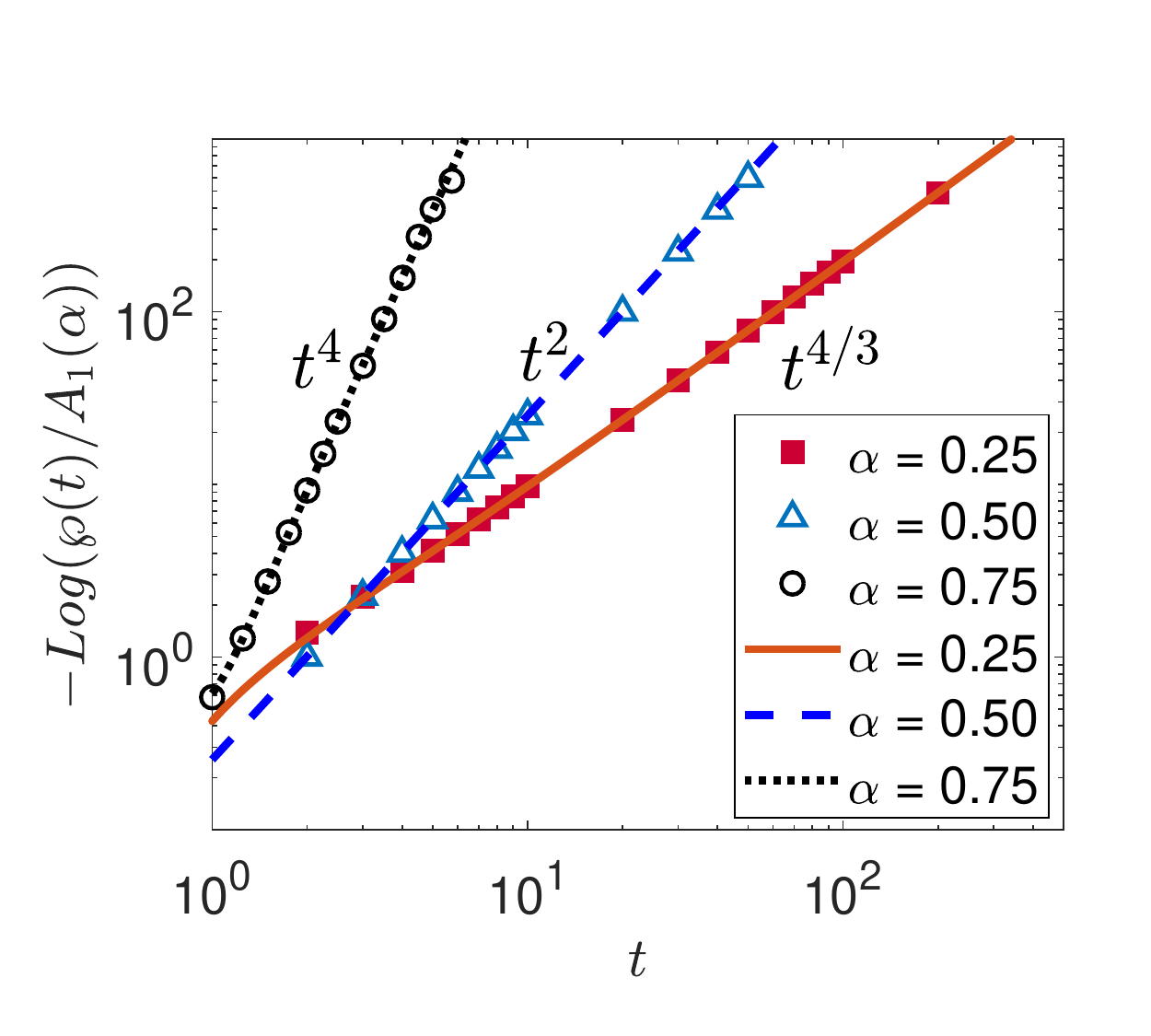}
\caption{Left: First passage time PDF for one-sided $\alpha$-stable laws with
$0<\alpha<1$ and $\beta=1$ with interval half length $L=100$ and $x(0)=0$. The
dashed lines represent numerical evaluations using the exact analytic expression
(\ref{eq:fptoneside}), the dotted lines represent the asymptotic behaviour
(\ref{eq:fptasymp}), and the symbols show simulation results based on the space
fractional diffusion equation. Right: Asymptotic first-passage time PDF for $L=1$
and $x(0)=0$. The lines represent (\ref{eq:asymoneprefa}) including the prefactors,
and the symbols show the simulation results based on the space-fractional diffusion
equation. Note the specific choice of the ordinate such that in this figure we see
a log-log plot of the power-law $t^{1/(1-\alpha)}$ in the exponential of equation
(\ref{eq:fptasymp}). We used the time step $\Delta t=0.001$ and space increment
$\Delta x=0.01$.}
\label{fig:fig7}
\end{figure}

Using equation (\ref{eq:fptandsurv}) the survival probability for $0<\alpha<1$,
$\beta=1$ can be expressed exactly in terms of the Wright function $W_{a,b}$
(see equation (\ref{eq:seriWfun})) or its series expansion,
\begin{equation}
S(t)=W_{-\alpha,1}(-\xi t/d^\alpha)=\displaystyle\sum_{n=0}^{\infty}\frac{(-\xi
t/d^\alpha)^n}{n!\Gamma(1-\alpha n)}.
\end{equation}
In the limit $\alpha=1/2$ this expression can be reduced to the simple form
\begin{equation}
S(t)=1-\mathrm{erf}\left(\frac{K_{1/2}t}{\sqrt{2d}}\right),
\end{equation}
and the corresponding first-passage time density is the half-sided Gaussian
\cite{Koren2007PhysicaA,iddo}
\begin{equation}
\wp(t)=K_{1/2}\sqrt{\frac{2}{\pi d}}\exp\left(-\frac{(K_{1/2}t)^{2}}{2d}\right).
\label{eq:rightoneside1}
\end{equation}
Figure \ref{fig:fig7} shows numerical and simulations results for $\wp(t)$,
lending excellent support for result (\ref{eq:fptasymp}).

\subsubsection{Extremal two-sided $\alpha$-stable probability laws.}
\label{Extremal-two-sided}

Stable probability laws with stability index $1<\alpha<2$ and skewness $\beta=1$
or $\beta=-1$ are called extremal two-sided skewed $\alpha$-stable laws.
When $\beta=1$ the PDF of an $\alpha$-stable random variable has a positive
power-law tail $x^{-1-\alpha}$, and a negative tail which is lighter than
Gaussian \cite{Skorokhod1954}, see figure 2. For $\beta=-1$ the properties of
the tails are exchanged. In \ref{Appendix Skoro} (see equation (\ref{eq:fptPDFtwobn}))
by applying the Skorokhod theorem it is shown that for $\beta=-1$ the PDF of the
first-passage time for extremal two-sided stable probability laws has the exact form
\begin{equation}
\wp(t)=\frac{t^{-1-1/\alpha}d}{\alpha\xi^{1/\alpha}}M_{1/\alpha}\left(\frac{d}{
(\xi t)^{1/\alpha}}\right)
\label{twosidedmainardi1}
\end{equation}
in terms of the Wright $M$-function $M_{1/\alpha}$. In the limit $\alpha=2$ we recover
the PDF (\ref{eq:LevySmirnovFPT}) for a Gaussian process. Moreover by using equation
(\ref{eq:fptandsurv}) the survival probability can be transformed to
\begin{equation}
S(t)=1-W_{-1/\alpha,1}\left(-\frac{d}{(\xi t)^{1/\alpha}}\right)=\displaystyle
\sum_{n=1}^{\infty}\frac{(-1)^{n-1}d^n(\xi t)^{-n/\alpha}}{n!\Gamma(1-n/\alpha)}.
\label{survbeta-1}
\end{equation}
Equation (\ref{twosidedmainardi1}) has the following limiting
behaviours: at short times $t\to 0$ corresponding to the asymptotic of large argument
in the Wright function, by using the asymptotic expression (\ref{eq:asymMfun}) we
find 
\begin{equation}
\wp(t)\sim A_2(\alpha)t^{-\frac{2\alpha-1}{2\alpha-2}}\exp{(-B_2(\alpha)t^{-
\frac{1}{\alpha-1}})},
\label{shorttime}
\end{equation}
where
\begin{equation}
A_2(\alpha)=\frac{(\alpha\xi/d^\alpha)^{-1/(2\alpha-2)}}{\sqrt{2\pi(\alpha-1)}},
\quad B_2(\alpha)=(\alpha-1)(\alpha^{\alpha}\xi/d^\alpha)^{-1/(\alpha-1)}.
\label{eq:asymtwoprefa}
\end{equation}
At long times, $t\to\infty$,
\begin{equation}
\wp(t)\sim\frac{\xi^{-1/\alpha}d}{\alpha\Gamma{(1-1/\alpha)}}t^{-1-1/\alpha},
\label{eq:longtime}
\end{equation}
consistent with the result in \cite{Koren2007PhysicaA}.

For the extremal two-sided $\alpha$-stable probability laws with index $1<\alpha<2$
and skewness $\beta=1$, by using the Skorokhod theorem the PDF of first-passage
times has the following series representation (see \ref{Appendix Skoro}, equation
(\ref{eq:fptPDFtwobp}))
\begin{equation}
\wp(t)=\frac{t^{-2+1/\alpha}d^{\alpha-1}}{\alpha\xi^{1-1/\alpha}}\displaystyle
\sum_{n=1}^{\infty}\frac{({\xi t/d^\alpha})^{-n+1}}{\Gamma{(\alpha n-1)}\Gamma{
(1+1/\alpha-n)}}.
\label{twosidedbeta1}
\end{equation}
For $\alpha=2$ we again consistently recover result (\ref{eq:LevySmirnovFPT}).
The asymptotic behaviour of equation (\ref{twosidedbeta1}) at long times is
\begin{equation}
\wp(t)\sim\frac{\xi^{-1+1/\alpha}d^{\alpha-1}}{\alpha\Gamma{(\alpha-1)}\Gamma{
(1/\alpha)}}t^{-2+1/\alpha},
\label{asymtwosidedbeta1lon}
\end{equation}
and in the limit $t\to0$, we find the finite value
\begin{equation}
\wp(0)=-\frac{\xi d^{-\alpha}}{\Gamma{(1-\alpha)}}.
\label{asymtwosidedbeta1shor}
\end{equation}

\begin{figure}
\centering
\includegraphics[height=6.8cm]{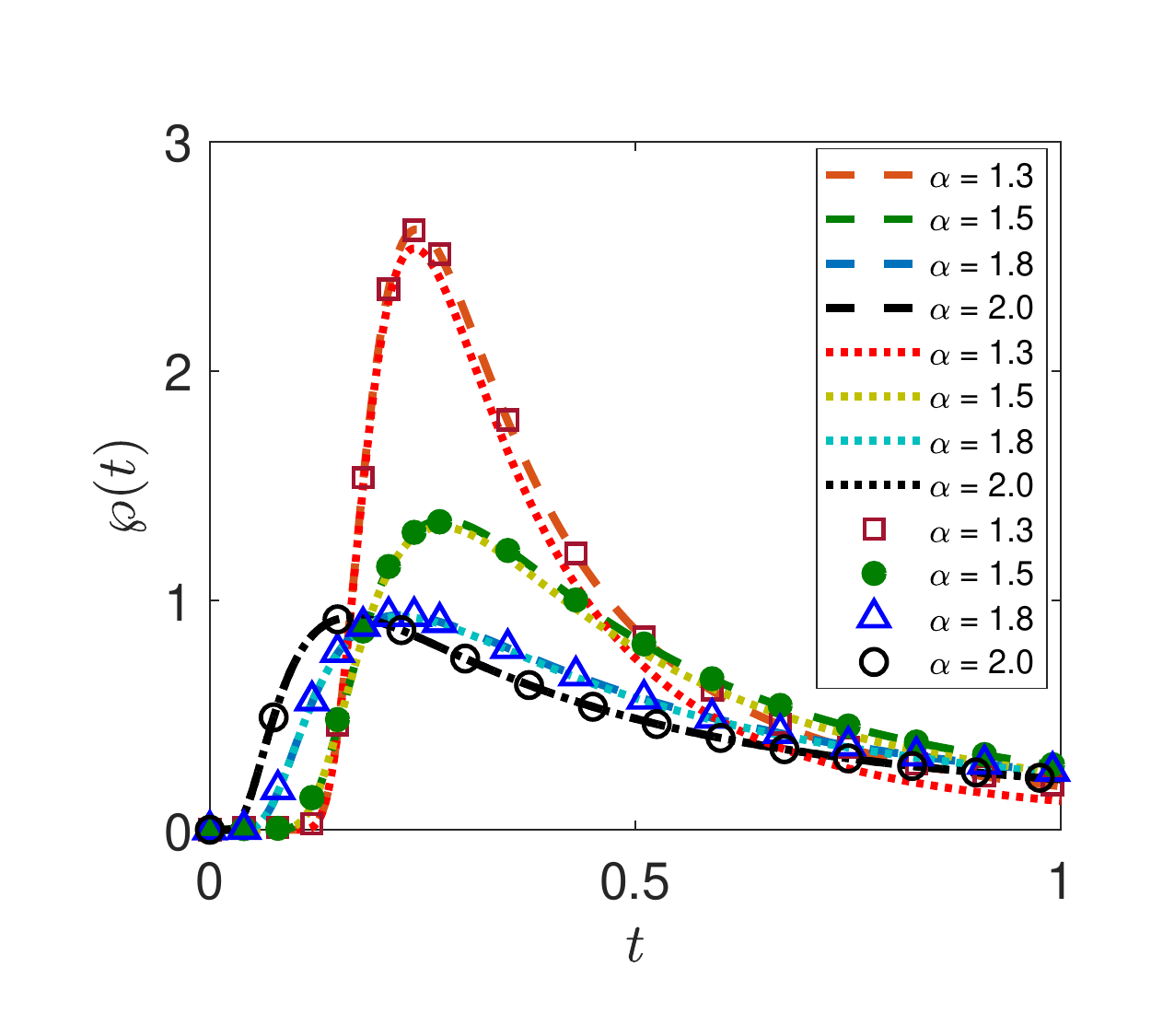}
\includegraphics[height=6.8cm]{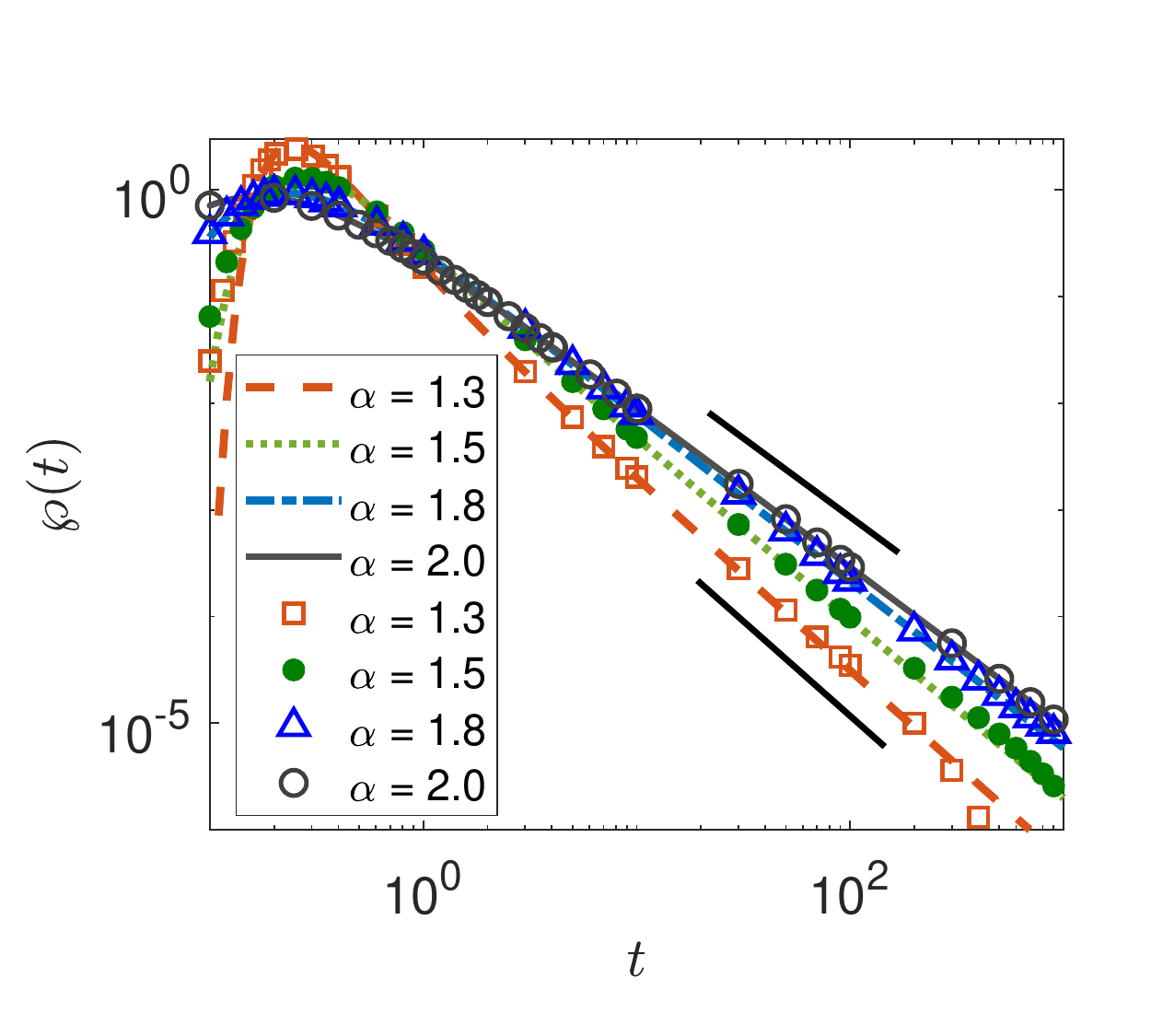}\\
\includegraphics[height=6.8cm]{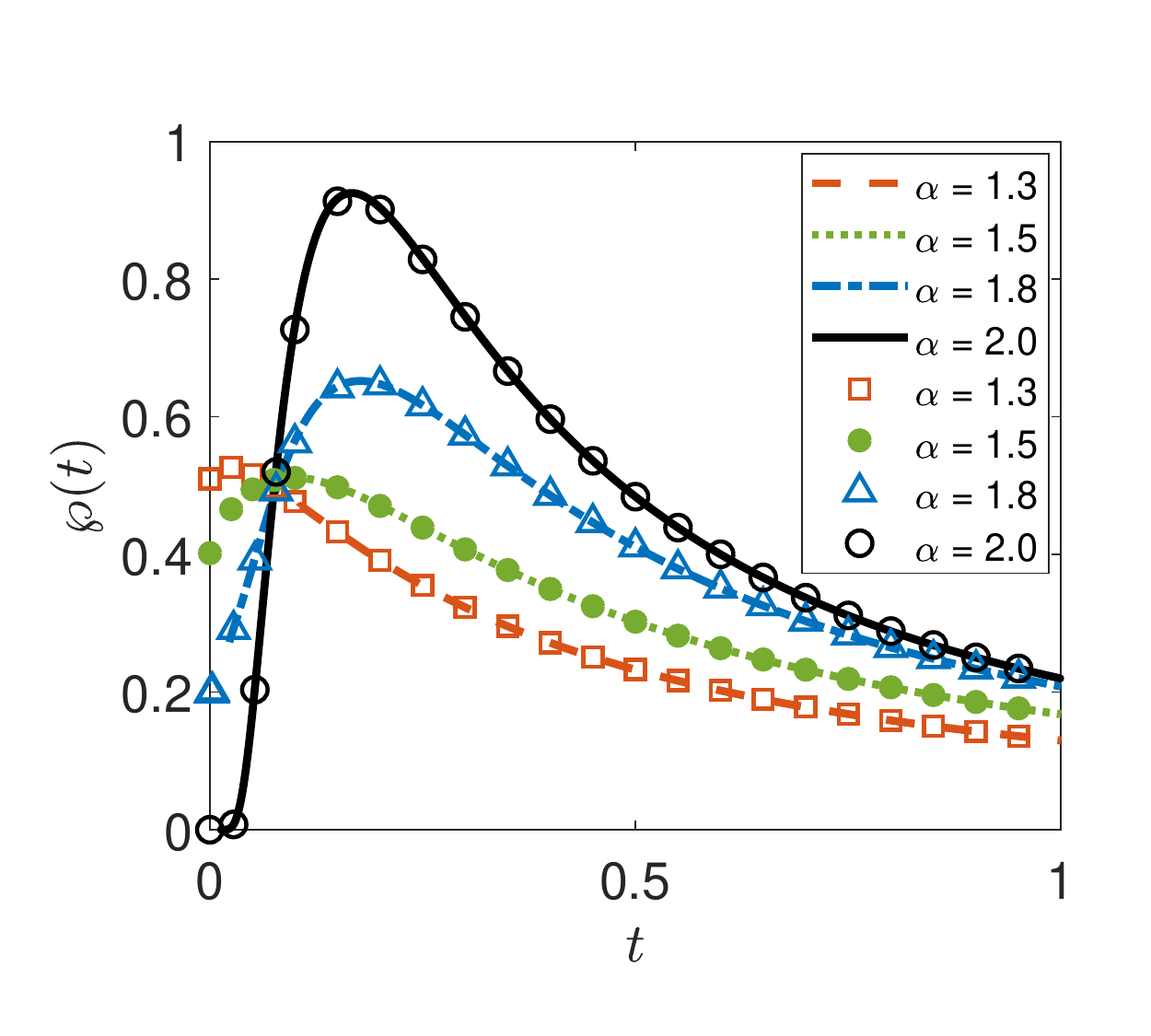}
\includegraphics[height=6.8cm]{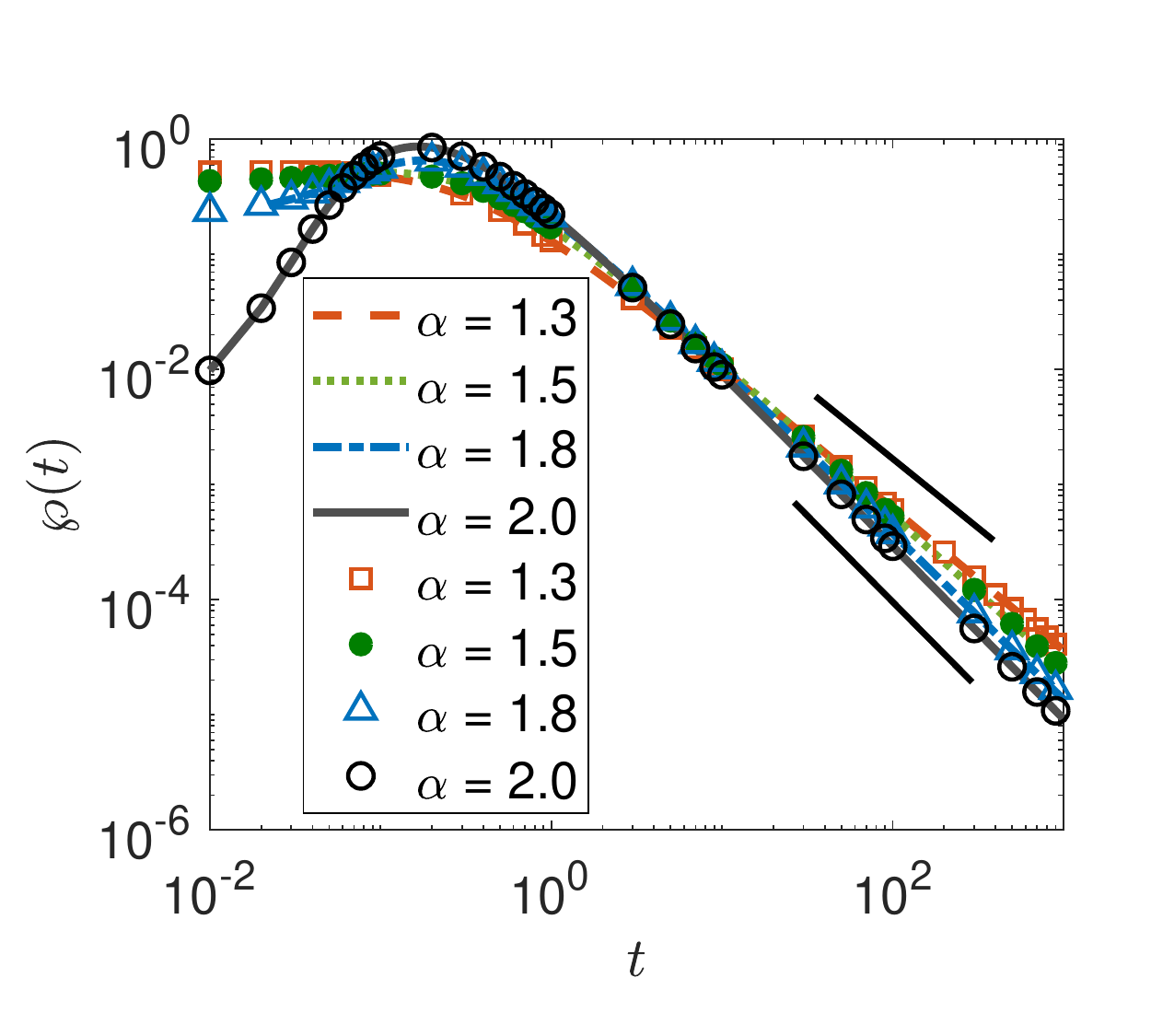}
\caption{First passage time PDF for two-sided $\alpha$-stable laws with $1<\alpha
\leq2$ and parameters $L=100$, $d=1$, $\Delta t=0.001$, and $\Delta x=0.02$. Top
left (skewness $\beta=-1$): The dashed lines represent numerical evaluations using
the exact analytic expression (\ref{twosidedmainardi1}), the dotted lines are for
the asymptotic behaviour (\ref{shorttime}), and the symbols show simulations
results based on the space-fractional diffusion equation. Top right ($\beta=-1$):
Asymptotic behaviour in log-log scale. The lines represent expression
(\ref{twosidedmainardi1}), symbols show the simulation results, and the black
lines show the power-law (\ref{eq:longtime}). Bottom left ($\beta=1$):
The lines represent numerical evaluations using the exact analytic expression
(\ref{twosidedbeta1}), and the symbols show simulation results. Bottom right
($\beta=1$): Asymptotic behaviour of the first-passage time on log-log scale.
Lines represent equation (\ref{twosidedbeta1}), symbols show simulations,
and the black line is the power-law (\ref{asymtwosidedbeta1lon}).}
\label{fig:fig8}
\end{figure}

By using the relation between the survival probability and the first-passage time
PDF in Laplace space (equation (\ref{eq:fptandsurpLap})) and applying the inverse
Laplace transform we obtain a series representation for the survival probability
for extremal two-sided $\alpha$-stable probability laws ($1<\alpha<2$, $\beta=1$)
in the form
\begin{equation}
S(t)=\displaystyle\sum_{n=1}^{\infty}\frac{(\xi t)^{1/\alpha-n}d^{n\alpha-1}}{
\Gamma{(\alpha n)}\Gamma(1-n+1/\alpha)}.
\label{survbeta+1}
\end{equation}

The first-passage time PDFs for extremal two-sided $\alpha$-stable probability laws
are displayed in figure {\ref{fig:fig8}}.

\subsubsection{$\alpha$-stable probability laws in general asymmetric form.}

\begin{figure}
\centering
\includegraphics[width=0.49\textwidth]{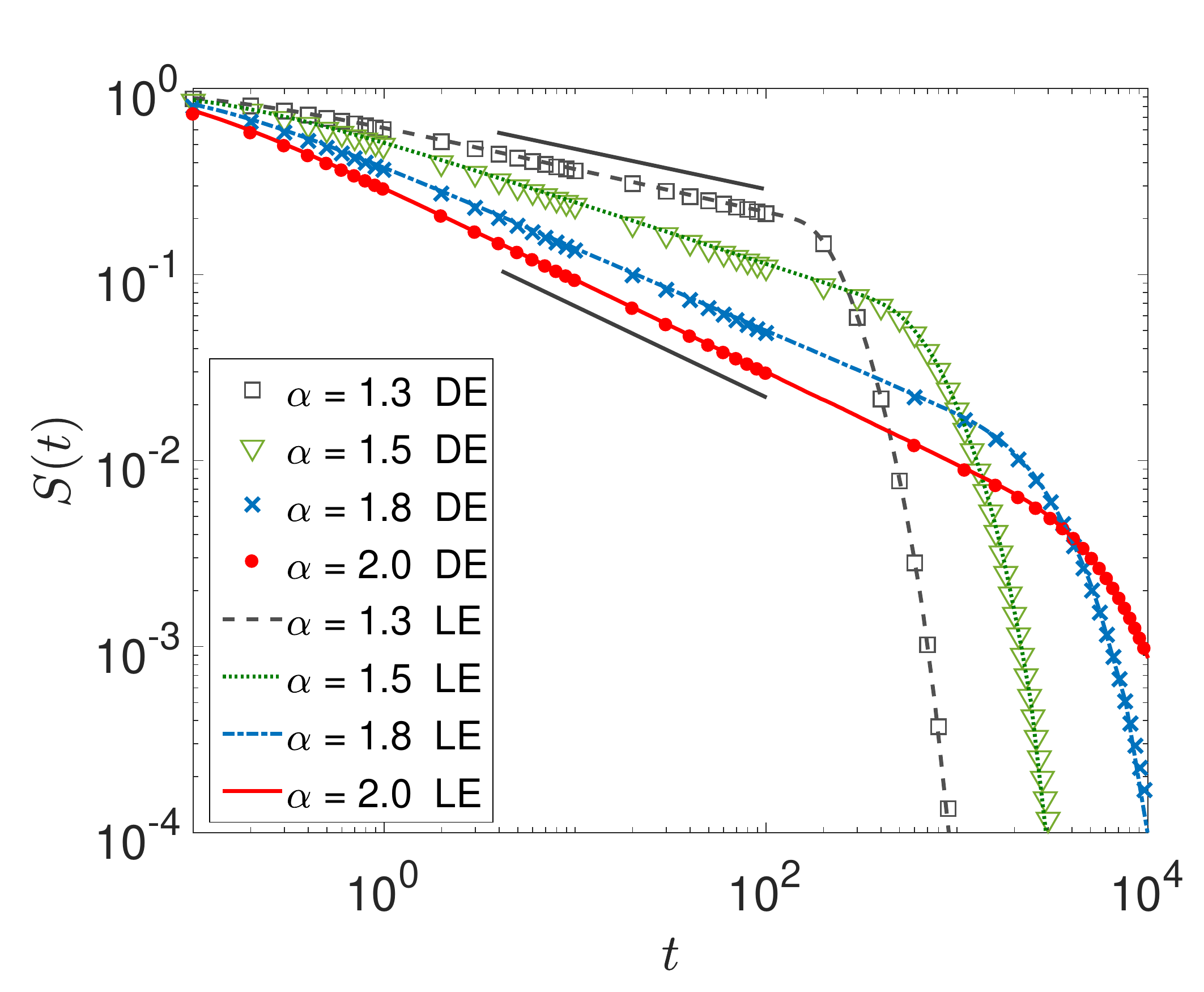}
\includegraphics[width=0.49\textwidth]{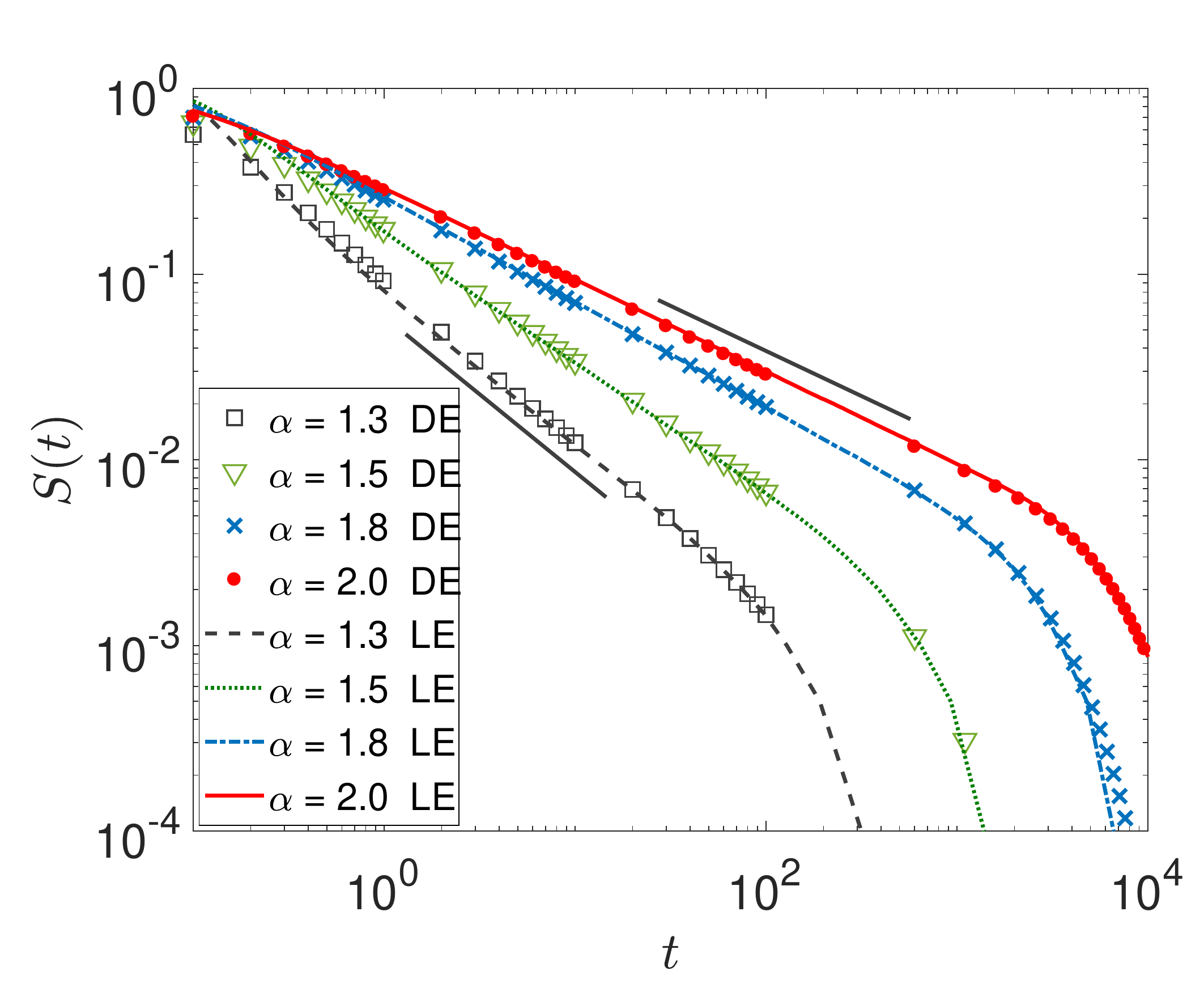}
\caption{Survival probability for two-sided $\alpha$-stable probability laws
with $d=0.5$, $L=100$ as well as $\beta=1$ (left) and $\beta=-1$ (right) for
different $\alpha$ with $1<\alpha\leq2$. Symbols represent simulation results
based on the space-fractional diffusion equation and lines show simulations
of the Langevin equation. The black lines depict the slope of the asymptotic
behaviour of the survival probability following from relation
(\ref{eq:generalasym}), concretely \textcolor{black}{$t^{-1+1/\alpha}$ (left)
and $t^{-1/\alpha}$ (right)}. Results are obtained with time step $\Delta t=0.01$ and
space increment $\Delta x=0.1$ for the solution of the space-fractional
diffusion equation. The time step $\Delta t=10^{-4}$ and averaging over $N=
10^5$ runs were chosen for the simulations of the Langevin equation.}
\label{fig:fig9}
\end{figure}

\begin{figure}
\centering
\includegraphics[width=0.49\textwidth]{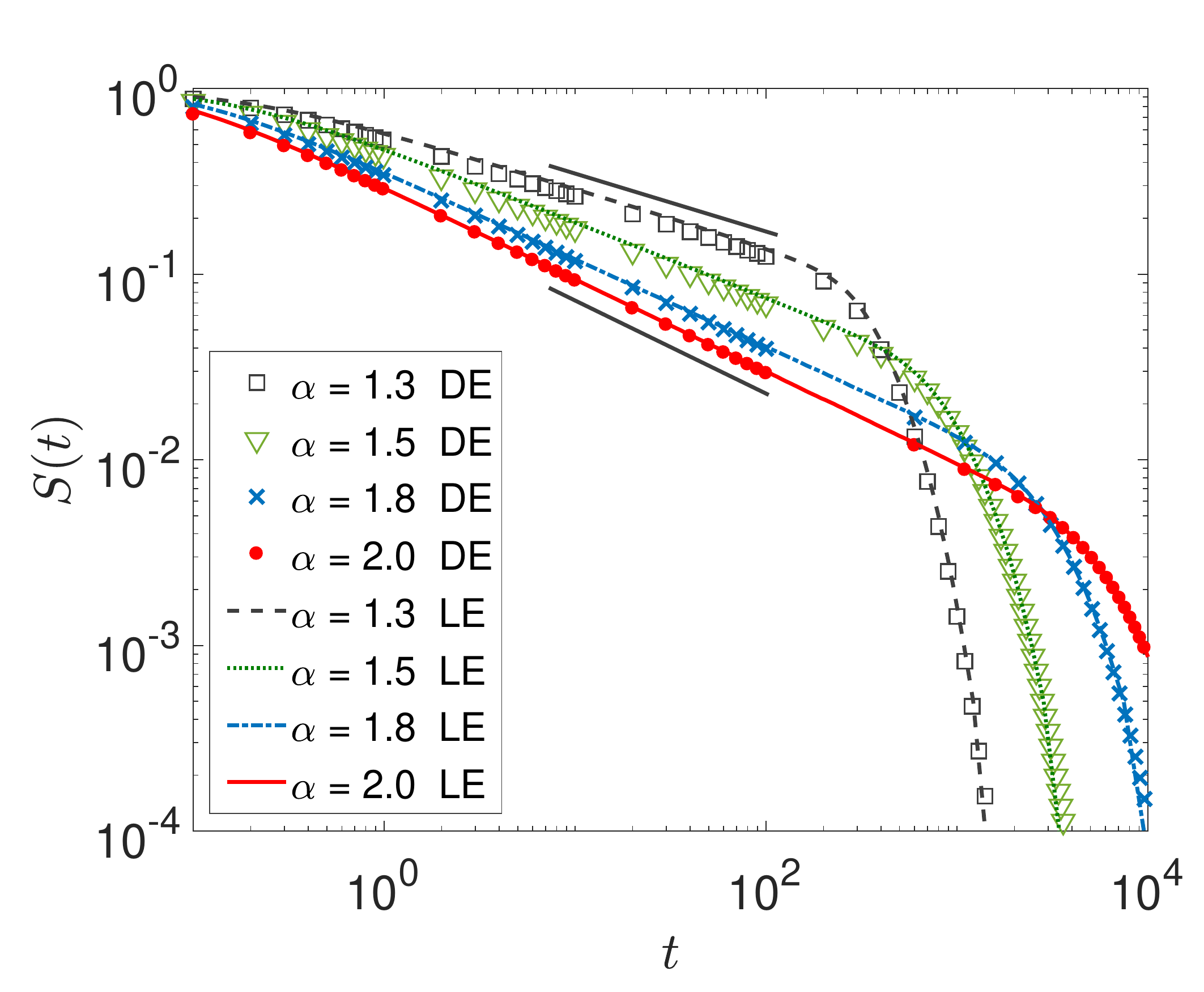}
\includegraphics[width=0.49\textwidth]{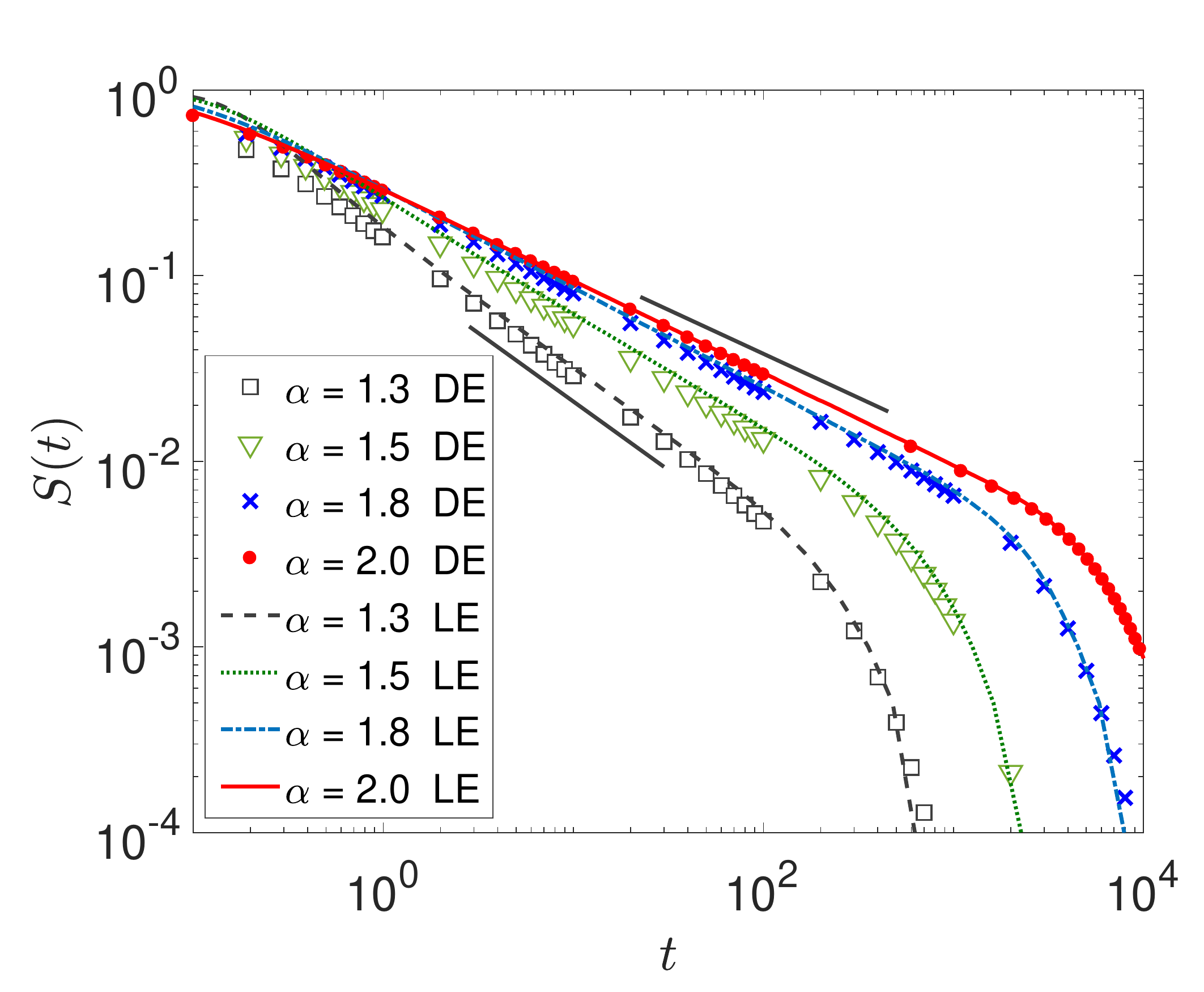}
\caption{Survival probability for asymmetric $\alpha$-stable probability laws
with $\beta=0.5$ (left) $\beta=-0.5$ (right). Symbols and parameters are analogous
to figure \ref{fig:fig9}. The black lines show the asymptotic $t^{-\rho}$, where
the exponent $\rho$ is defined in equation (\ref{sparreexponent}).}
\label{fig:fig10}
\end{figure}

We finally study the first-passage behaviour for asymmetric L{\'e}vy stable
laws of arbitrary skewness $\beta$, again based on the comparison between the
numerical solution of the space-fractional diffusion equation and simulations
of the Langevin approach for different stable index $\alpha$. The results are
shown in log-log scale in figures \ref{fig:fig9} and \ref{fig:fig10} in the
range $1<\alpha<2$. Figure \ref{fig:fig9} depicts the cases $\beta=1$ and
$\beta=-1$, while figure \ref{fig:fig10} shows the cases $\beta=0.5$ and $\beta
=-0.5$. As can be seen from both figures, a positive value of the skewness leads
\textcolor{black}{to a slower decay (shallower slope) than for the Gaussian case,
and opposite for negative values of $\beta$. Indeed, this behaviour is not
immediately intuitive, as a positive skewness means that the stable law has
a longer tail on the positive axis. However, what matters for the short and
intermediate first-passage time scales are values of the stable law around the
most likely value, and for positive skewness this is located on the negative
axis (compare bottom panels in figure \ref{fig:fig2}). Thus, L{\'e}vy flights
with a positive skewness experience an effective drift to the left, in our
setting away from the absorbing boundary.}

From the Skorokhod theorem for $\alpha\in(0,1)$ and $\beta\in(-1,1)$, $\alpha=1$
and $\beta=0$, as well as $\alpha\in(1,2]$ and $\beta\in[-1,1]$
we obtained the power-law decay (see \ref{skoroapp})
\begin{equation}
\wp(t)\sim\frac{\rho(K_\alpha(1+\beta^2\tan^2{(\alpha\pi/2)})^{1/2})^{-\rho}
d^{\alpha \rho}}{\Gamma(1-\rho)\Gamma(1+\alpha\rho)}t^{-\rho-1},
\label{eq:generalasym}
\end{equation}
where the positive parameter $\rho$ is defined as
\begin{equation}
\rho=\frac{1}{2}+\frac{1}{\pi\alpha}\arctan(\beta\tan(\pi\alpha/2)).
\label{sparreexponent}
\end{equation}
Following \cite{Bertoin1996} (page 218) we write this in a form with a general $\beta$.
This is the direct generalisation of the classical Sparre Andersen universality
for asymmetric stable jump length distributions. It is easy to check that this
result reduces to the known cases for vanishing skewness. We note that the
inapplicability of the Sparre Andersen law for asymmetric jump length distributions
was already pointed out by Spitzer (\cite{FSpitzer1976}, page 227). The analytical
predictions from relations (\ref{eq:generalasym}) and (\ref{sparreexponent}) are in
excellent agreement with the data shown in figures \ref{fig:fig9} and \ref{fig:fig10}.

\section{Discussion and unsolved problems}
\label{concl}

LFs are Markovian stochastic processes that are commonly used across disciplines
as models for jump processes that exhibit the distinct propensity for very long
jumps. The scale-free nature of the underlying, L{\'e}vy stable PDF of jump lengths
with its heavy power-law tail has been shown to effect a more efficient random
search strategy than the more conventional Brownian search processes. We here
combined numerical inversion methods of the solution of the space-fractional
diffusion equation and simulation of the Langevin equation fuelled by $\alpha$-stable
white noise to quantify the first-passage dynamics of LFs with a general asymmetric
jump length PDF. In particular, we demonstrated that in all cases both approaches
are in excellent agreement. As both approaches have advantages and disadvantages,
it is very useful to have available two equally potent methods. In addition, we
verified the crossover to an exponential behaviour of the first-passage time PDF
in a finite domain and the existence of a well-established
power-law decay at intermediate times,
before the random walker explores the full range of the finite domain and thus
behaves as if it were in a semi-infinite range. For symmetric $\alpha$-stable
laws this decay was shown to be fully consistent with the expected Sparre Andersen
universal law. For asymmetric cases, when the conditions of the Sparre Andersen
theorem are no longer fulfilled, we derived the analytical behaviour from the
Skorokhod theorem for specific values of the skewness. In the general case the
direct extension of this analytical law was shown to be fully consistent with
the numerical and simulations results. The results obtained here will be of use
in applications, as these typically are involved with search processes and thus
measure first-passage times. Concurrently, these results also further complete the
mathematical theory of L{\'e}vy stable processes.

\begin{savenotes}
\begin{table}
\centering
\renewcommand{\arraystretch}{2}
\begin{tabular}{|c|c|c|c|c|} \hline\hline
$\alpha$ & $\beta$ & Exact PDF & Long-time asymptotic & Prefactor\\
&& Equation & & Equation \\\hline\hline
2 & Irrelevant & \multicolumn{3}{c|}{$\frac{d}{\sqrt{4\pi K_2t^3}}
\exp\left(-\frac{d^2}{4K_2t}\right)$ (\ref{eq:LevySmirnovFPT}) \cite{Feller1971}}
\\\hline\hline
(0,2) & 0 & Unknown & $\simeq t^{-3/2}$ \cite{frisch,zukla95} & 
(\ref{eq:skorokhod16}) \cite{Koren2007PRL} \\\hline\hline
(0,1) & \multirow{2}{*}{1} & (\ref{eq:fptoneside}) \cite{Koren2007PRL} & $\sim
A_1(\alpha)t^{(\alpha-1/2)/(1-\alpha)}\exp[-B_1(\alpha)
t^{1/(1-\alpha)}]$ & (\ref{eq:asymoneprefa}) \cite{Koren2007PRL}\\\cline{1-1}
\cline{3-5}
$1/2$ & & \multicolumn{3}{c|}{$K_{\alpha}\sqrt{2/\pi d}\exp[-(K_{\alpha}t)^2/2d]$
(\ref{eq:rightoneside1}) \cite{Koren2007PhysicaA,Koren2007PRL,iddo}}\\\hline\hline
\multirow{2}{*}{(1,2)} & -1 & (\ref{twosidedmainardi1}) & $\simeq t^{-1-1/\alpha}$
\cite{Koren2007PhysicaA} & (\ref{eq:longtime}) \cite{peskir}\footnotemark
\\\cline{2-5}
& 1 & (\ref{twosidedbeta1}) & $\simeq t^{-2+1/\alpha}$ & (\ref{asymtwosidedbeta1lon})
\\\hline\hline
(0,1) & {\multirow{2}{*}{(-1,1)}} & Unknown & \multirow{2}{*}{$\simeq t^{-3/2
-(\pi\alpha)^{-1}\tan^{-1}[\beta \tan (\pi \alpha/2)]}$ \cite{Bertoin1996}} &
\multirow{2}{*}{(\ref{eq:generalasym}) \cite{bingham1,bingham2}}\\\cline{1-1}\cline{3-3}
(1,2) &  & Unknown & & \\\hline\hline
\end{tabular}
\caption{First-passage time PDF for different stable indices $\alpha$ and skewness
parameters $\beta$. The fifth column refers to the equation number for the full
prefactor of the asymptotic law in column four.
\footnotetext{Note that the result
(\ref{eq:longtime}) differs from that of \cite{peskir} by a factor which appears
due to the use of two different forms of the characteristic function for the
$\alpha$-stable process.}}
\label{tab:table1}
\end{table}
\end{savenotes}

The first-passage time properties of general $\alpha$-stable probability
laws are summarised in table \ref{tab:table1}. For the known cases we include
the references to the relevant equations of the exact PDF as well as the
asymptotic prefactor. Some special cases are included, as well. Those cases
with previously known results refer to the relevant references.

\textcolor{black}{
It is possible to extend the studied setup to higher dimensions \cite{refa,refb,
refc,refd}. In this case, the scalar noise term $\zeta(t)$ in the Langevin equation
(\ref{eq:langevin}) for $x(t)$ is replaced by multivariate L{\'e}vy noise in the
higher-dimensional Langevin equation for $\mathbf{x}(t)$. Here, multivariate
$\alpha$-stable variables are characterised by a spectral measure defined on the
unit circle \cite{Samorodnitsky-Taqqu}. For the numerical scheme of the
multi-dimensional space-fractional diffusion equation we refer to \cite{Duo-Zhang2018,
Meerschaert2006,Tadjeran2007,Chen-Liu2008,Zhuang-Liu2007,Liu-Liu2008,Changpin-Fanhai2015,Castillo-Negrete2013}.
We note that to the best knowledge of the authors no multi-dimensional generalisation
of Skorokhod's theorem exists. Thus, the extension of the analytical and numerical
approaches presented here to higher dimensions represents a challenging problem
requiring further studies.
}

Generally the formulation of non-local and/or correlated stochastic processes is
not always an easy task and, in some cases, still not fully understood. Apart
from LFs, we may allude to the debate on the formulation and solution of boundary
value problems for fractional Brownian motion, a process fuelled with Gaussian yet
long-range correlated noise \cite{molchan,vojta,tobias}. For LFs, in addition to
the results obtained here it will be interesting to generalise the results obtained
for symmetric $\alpha$-stable laws in the presence of an external drift in 
\cite{vladimir}. Similarly, it will be of interest to investigate the PDF of
first \emph{arrival\/} times, related to the probability of hitting a small
target in an otherwise unbounded environment, for the general case of asymmetric
L{\'e}vy stable laws.

\ack

AP acknowledges funding from the Ministry of Science, Research and Technology
of Iran and University of Potsdam. This research was supported in part by PLGrid
Infrastructure. The computer simulations were performed
at Potsdam University and the Academic Computer Center Cyfronet, Akademia
G\'orniczo-Hutnicza (Krak\'ow, Poland) under CPU grant DynStoch. AC and RM
acknowledge support from DFG project ME 1535/7-1. RM also thanks the
Foundation for Polish Science for support within an Alexander von Humboldt
Polish Honorary Research Scholarship.

\appendix

\section{Parametrisation of characteristic function for $\alpha$-stable processes}
\label{Appendix Char}

There are several forms for the parametrisation of $\alpha$-stable laws appearing
in literature, basically because of historical reasons. Each form might be useful
in a particular situation. For example, one of them is preferable for analytical
calculations, whereas the other ones can be more convenient for numerical purposes
or for fitting of data. Following the exposition of the various forms of stable
laws in \cite{Zolotarev1986,Uchaikin}, we here present four parameterisation forms
for the characteristic functions.

In the main text we use the A-form of the characteristic function,
\begin{equation}
\hat{\ell}_{\alpha,\beta_A}^A(k,t)=\exp\Big(-t K_{\alpha}^A|k|^{\alpha}[1-
\mathrm{i}\beta_A\mathrm{sgn}(k)\omega_A(k,\alpha)]+\mathrm{i}\mu_Akt\Big),
\end{equation}
where
\begin{equation}
\omega_A(k,\alpha)=\left\{\begin{array}{ll}\tan(\frac{\pi\alpha}{2}),&\alpha\neq1,\\
-\frac{2}{\pi}\ln|k|,&\alpha=1\end{array}\right..
\end{equation}
In this paper we exclude the case $\alpha=1$ and $\beta\neq0$. The B-form is helpful
from an analytical point of view, it is given by
\begin{equation}
\hat{\ell}_{\alpha,\beta_B}^B(k,t)=\exp\Big(-t K_{\alpha}^B|k|^{\alpha}\omega_B
(k,\alpha,\beta_B)+\mathrm{i}\mu_Bkt\Big),
\label{eq:eqBpara}
\end{equation}
where (for $\alpha\neq1$)
\begin{equation}
\omega_B(k,\alpha,\beta_B)=\exp\Big(-\mathrm{i}\frac{\pi}{2}\beta_B K(\alpha)
\mathrm{sgn}(k)\Big)
\end{equation}
and $K(\alpha)=\alpha-1+\mathrm{sgn}(1-\alpha)$. The parameters have the same
domains of variation as in the A-form,
\begin{equation}
\fl\beta_A=\cot\left(\frac{\alpha\pi}{2}\right)\tan\left(\beta_BK(\alpha)\frac{\pi}{2}
\right),\quad\mu_A=\mu_B,
K_{\alpha}^A=\cos\left(\beta_BK(\alpha)\frac{\pi}{2}\right)
K_{\alpha}^B.
\label{eq:relacharAandB}
\end{equation}
The M-form is used in numerical simulations and reads
\begin{equation}
\hat{\ell}_{\alpha,\beta_M}^M(k,t)=\exp\Big(-tK_{\alpha}^M|k|^{\alpha}[1+\mathrm{i}
\beta_M\mathrm{sgn}(k)\omega_M(k,\alpha,t)]+\mathrm{i}\mu_M(t)kt\Big),
\end{equation}
where ($\alpha\neq1$)
\begin{equation}
\omega_M(k,\alpha,t)=\tan\left(\frac{\pi\alpha}{2}\right)\left[(K_{\alpha}^Mt)^{
1/\alpha-1}|k|^{1-\alpha}-1\right].
\end{equation}
The domain of variation of the parameters in the A- and M-forms is the same and
connected by the following relations
\begin{equation}
\beta_M=\beta_A,\quad\mu_M(t)=\mu_A+\frac{(K_{\alpha}^At)^{1/\alpha}}{t}\beta_A\tan
\left(\frac{\pi\alpha}{2}\right),\quad K_{\alpha}^M=K_{\alpha}^A. 
\label{eq:relparAandM}
\end{equation}
Finally, the Z-form is represented by
\begin{equation}
\hat{\ell}_{\alpha,\rho}^Z(k,t)=\exp\Big(-tK_{\alpha}^Z(\mathrm{i}k)^{\alpha}\exp
\left[-\mathrm{i}\pi\alpha\rho\mathrm{sgn}(k)\right]\Big), 
\label{eq:eqCmpara}
\end{equation}
where the parameters $\alpha$ and $\rho$ vary within the bounds
\begin{equation}
0<\alpha\leq2,1-\min(1,1/\alpha)\leq\rho\leq\min(1,1/\alpha),\quad t>0,
\end{equation}
and the relation between the parameters in the A- and Z-forms is as follows,
\begin{equation}
\fl\rho=\frac{1}{2}+\frac{1}{\alpha \pi}\arctan\left(\beta_A\tan\left(\frac{\alpha
\pi}{2}\right)\right),\quad K_{\alpha}^Z=K_{\alpha}^A\left(1+\left[\beta_A\tan\left(
\frac{\alpha\pi}{2}\right)\right]^2\right)^{1/2}.
\label{eq:Prcm1}
\end{equation}
In the A-, B-, and M-forms $\beta_A=1$ corresponds to $\beta_M=1$ and $\beta_B=1$,
while in the case of the Z-form the value $\beta_A=1$ corresponds to the value
$\rho=1$ if $\alpha<1$ and to the value $\rho=1-1/\alpha$ if $\alpha>1$.

\section{Some details of the numerical scheme}
\label{Appendix Num}

With the use of equations (\ref{eq:fraccoef}) and (\ref{eq:disctime}) we can write
equation (\ref{eq:ffpe}) on a discrete space-time grid as
\begin{equation}
\frac{f_i^{j+1}-f_i^j}{\Delta t}=K_{\alpha}\left[L_{\alpha,\beta}\,_aD_x^{\alpha}
D{f(x_i,t_j)}+R_{\alpha,\beta}\,_xD_b^{\alpha}D{f(x_i,t_j)}\right].
\label{eq:discffpe}
\end{equation}
Here we consider discretisation schemes for the three cases $0<\alpha<1$, $1<\alpha
\leq2$, and $\alpha=1$ separately.

\subsection{$0<\alpha<1$}

By substitution of equations (\ref{eq:disctime}) to (\ref{eq:discright1}) into
equation (\ref{eq:discffpe}) and using the relation
\begin{equation}
\int\limits_a^b\frac{1}{[\pm(x-y)]^\gamma}\mathrm{d}y=\frac{1}{\pm(\gamma-1)}
\left[(\pm(x-b))^{1-\gamma}-(\pm(x-a))^{1-\gamma}\right]
\end{equation}
where the sign $+$ is taken for $x>b>a$ and $-$ for $x<a<b$, we obtain
\begin{eqnarray}
\fl\frac{f_i^{j+1}-f_i^j}{\Delta t}&=&K_{\alpha}\left[\frac{L_{\alpha,\beta}}{
\Gamma(2-\alpha)}\displaystyle\sum_{k=1}^i\frac{f_k^j-f_{k-1}^{j}}{\Delta x}
\left((x_i-x_{k-1})^{1-\alpha}-(x_i-x_k)^{1-\alpha}\right)\right.\\
&&\left.+\frac{R_{\alpha,\beta}}{\Gamma(2-\alpha)}\displaystyle\sum_{k=i}^{I-1}
\frac{f_k^j-f_{k+1}^j}{\Delta x}\left((x_{k+1}-x_i)^{1-\alpha}-(x_k-x_i)^{1-
\alpha}\right)\right].
\label{eq:descretdiffeq1}
\end{eqnarray}
Defining 
\begin{equation}
\fl\lambda_n=n^{1-\alpha}-(n-1)^{1-\alpha},\quad\Omega_L=\frac{L_{\alpha,\beta}
K_{\alpha}\Delta t}{ \Gamma(2-\alpha)(\Delta x)^{\alpha}},\quad\Omega_R=\frac{
R_{\alpha,\beta}K_{\alpha}\Delta t}{\Gamma(2-\alpha)(\Delta x)^{\alpha}},
\end{equation}
we rewrite equation (\ref{eq:descretdiffeq1}) as
\begin{equation}
f_i^{j+1}-f_i^j=\Omega_L\displaystyle\sum_{k=1}^i\left(f_k^j-f_{k-1}^j\right)
\lambda_{i-k+1}+\Omega_R\displaystyle\sum_{k=i}^{I-1}\left(f_k^j-f_{k+1}^j\right)
\lambda_{k-i+1}.
\end{equation}
Changing $f_i\to\theta f_i^{j+1}+(1-\theta)f_i^j$, $0\leq\theta\leq1$, on the right
hand side,
\begin{eqnarray}
\nonumber
\fl-\theta\Omega_L\displaystyle\sum_{k=1}^i\left(f_k^{j+1}-f_{k-1}^{j+1}\right)
\lambda_{i-k+1}+f_i^{j+1}-\theta\Omega_R\displaystyle\sum_{k=i}^{I-1}\left(f_k^
{j+1}-f_{k+1}^{j+1}\right)\lambda_{k-i+1}\\
\fl=(1-\theta)\Omega_L\displaystyle\sum_{k=1}^i\left(f_k^j-f_{k-1}^{j}\right)
\lambda_{i-k+1}+f_i^j+(1-\theta)\Omega_R\displaystyle\sum_{k=i}^{I-1}\left(
f_k^j-f_{k+1}^j\right)\lambda_{k-i+1}.
 \end{eqnarray}
Then the matrices $\mathbf{A}$ and $\mathbf{B}$ in equation (\ref{eq:matrixab}) are
\begin{equation}
\fl \mathbf{A}=\left(\begin{array}{cccc}A_c&A_{1,R}&\cdots& A_{I,R}\\
A_{1,L}&\ddots&\ddots&\vdots\\\vdots&\ddots&\ddots&A_{1,R}\\
A_{I,L}&\cdots&A_{1,L}& A_{c}\end{array}\right),\qquad
\mathbf{B}=\left(\begin{array}{cccc}B_c&B_{1,R}&\cdots&B_{I,R}\\
B_{1,L}&\ddots&\ddots&\vdots\\\vdots&\ddots&\ddots&B_{1,R}\\
B_{I,L}&\cdots&B_{1,L}&B_{c}\end{array}\right),
\end{equation}
where
\begin{equation}
\begin{array}{ll}
A_c=1-\theta(\Omega_L+\Omega_R)\lambda_1 \\
A_{i,L}=\theta\Omega_L(\lambda_i-\lambda_{i+1}),\quad i=1,2,\ldots,I\\
A_{i,R}=\theta\Omega_R(\lambda_i-\lambda_{i+1}),\quad i=1,2,\ldots,I
\end{array}
\end{equation}
and
\begin{equation}
\begin{array}{ll}
B_c=1+(1-\theta)(\Omega_L+\Omega_R)\lambda_1 \\
B_{i,L}=-(1-\theta)\Omega_L(\lambda_i-\lambda_{i+1}),\quad i=1,2,\ldots,I\\
B_{i,R}=-(1-\theta)\Omega_R(\lambda_i-\lambda_{i+1}),\quad i=1,2,\ldots,I.
\end{array}
\end{equation}

\subsection{$1<\alpha<2$}

Substituting equations (\ref{eq:disctime}), (\ref{eq:discleft2}), and
(\ref{eq:discright2}) into equation (\ref{eq:discffpe}) we get
\begin{eqnarray}
\nonumber
\fl\frac{f_i^{j+1}-f_i^{j}}{\Delta t}&=&K_{\alpha}\left[\frac{L_{\alpha,\beta}}{
\Gamma(3-\alpha)}\displaystyle\sum_{k=1}^i\frac{f_{k+1}^j-2f_k^j+f_{k-1}^j}{
(\Delta x)^2}\left((x_i-x_{k-1})^{2-\alpha}-(x_i-x_k)^{2-\alpha}\right)\right.\\
&&\hspace*{-1.2cm}
+\left.\frac{R_{\alpha,\beta}}{\Gamma(3-\alpha)}\displaystyle\sum_{k=i}^{I-1}\frac{
f_{k+1}^j-2f_k^j+f_{k-1}^j}{(\Delta x)^2}\left((x_{k+1}-x_i)^{2-\alpha}-(x_k-x_i)^{
2-\alpha}\right)\right]
\end{eqnarray}
with the definition
\begin{equation}
\fl\lambda_n=n^{2-\alpha}-(n-1)^{2-\alpha},\quad\Omega_L=\frac{L_{\alpha,\beta}
K_{\alpha}\Delta t}{\Gamma(3-\alpha)(\Delta x)^{\alpha}},\quad\Omega_R=\frac{R_{
\alpha,\beta} K_{\alpha}\Delta t}{\Gamma(3-\alpha)(\Delta x)^{\alpha}},
\end{equation}
and changing the notation as above, we obtain
\begin{eqnarray}
\nonumber
\fl&&-\theta\Omega_L\displaystyle\sum_{k=1}^i\left(f_{k+1}^{j+1}-2f_k^{j+1}+f_{k-1}^{
j+1}\right)\lambda_{i-k+1}+f_i^{j+1}\\
\nonumber
\fl&&-\theta\Omega_R\displaystyle\sum_{k=i}^{I-1}
\left(f_{k+1}^{j+1}-2f_k^{j+1}+f_{k-1}^{j+1}\right)\lambda_{k-i+1}\\
\nonumber
\fl&=&(1-\theta)\Omega_L\displaystyle\sum_{k=1}^i\left(f_{k+1}^j-2f_k^j+f_{k-1}^j
\right)\lambda_{i-k+1}+f_i^j\\
&&+(1-\theta)\Omega_R\displaystyle\sum_{k=i}^{I-1}
\left(f_{k+1}^j-2f_k^j+f_{k-1}^{j}\right)\lambda_{k-i+1}.
\end{eqnarray}
Then the elements of the matrix A and B in equation (\ref{eq:matrixab}) are
\begin{equation}
\begin{array}{ll}
A_c=1-\theta(\Omega_L+\Omega_R)(\lambda_2-2\lambda_1) \\
A_{1,L}=-\theta\Omega_L(\lambda_3-2\lambda_2+\lambda_1)-\theta\Omega_R\lambda_1\\
A_{1,R}=-\theta\Omega_R(\lambda_3-2\lambda_2+\lambda_1)-\theta\Omega_L\lambda_1\\
A_{i,L}=-\theta\Omega_L(\lambda_{i+2}-2\lambda_{i+1}+\lambda_{i}),\quad i=2,3,
\ldots,I\\
A_{i,R}=-\theta\Omega_R(\lambda_{i+2}-2\lambda_{i+1}+\lambda_{i}),\quad i=2,3,
\ldots,I
\end{array}
\end{equation}
and
\begin{equation}
\begin{array}{ll}
B_c=1+(1-\theta)(\Omega_L+\Omega_R)(\lambda_2-2\lambda_1) \\
B_{1,L}=(1-\theta)\Omega_L(\lambda_3-2\lambda_2+\lambda_1)+(1-\theta)\Omega_R\lambda_1\\
B_{1,R}=(1-\theta)\Omega_R(\lambda_3-2\lambda_2+\lambda_1)+(1-\theta)\Omega_L\lambda_1\\
B_{i,L}=(1-\theta)\Omega_L(\lambda_{i+2}-2\lambda_{i+1}+\lambda_{i}),\quad i=2,3,
\ldots,I\\
B_{i,R}=(1-\theta)\Omega_R(\lambda_{i+2}-2\lambda_{i+1}+\lambda_{i}),\quad i=2,3,
\ldots,I
\end{array}.
\end{equation}

\subsection{$\alpha=1$, $\beta=0$}

By substituting equations (\ref{eq:Hermitd1}) and (\ref{eq:disctime}) into
equation (\ref{eq:hilbertdiffeq}) we obtain
\begin{eqnarray}
\fl\frac{f_i^{j+1}-f_i^j}{\Delta t}=-\frac{2K_{\alpha}}{\pi}\left[\displaystyle
\sum_{k=1}^i\frac{f^j_k-f^j_{k-1}}{\Delta x}\frac{1}{2(i-k)+1}+\displaystyle
\sum_{k=i}^{I-1}\frac{f^j_k-f^j_{k+1}}{\Delta x}\frac{1}{2(k-i)+1}\right].
\end{eqnarray}
Defining 
\begin{equation}
\lambda_n=\frac{1}{2n+1},\quad\Omega_L=\frac{2L_{\alpha,\beta} K_{\alpha}\Delta t}{
\Delta x},\quad\Omega_R=\frac{2R_{\alpha,\beta}K_{\alpha}\Delta t}{\Delta x},
\end{equation}
changing notation as above, we obtain
\begin{eqnarray}
\nonumber
\fl\theta\Omega_L\displaystyle\sum_{k=1}^i\left(f_k^{j+1}-f_{k-1}^{j+1}\right)
\lambda_{i-k}+f_i^{j+1}+\theta\Omega_R\displaystyle\sum_{k=i}^{I-1}\left(f_k^{
j+1}-f_{k+1}^{j+1}\right)\lambda_{k-i}\\
\fl=-(1-\theta)\Omega_L\displaystyle\sum_{k=1}^i\left(f_k^j-f_{k-1}^j\right)\lambda_{
i-k}+f_i^j-(1-\theta)\Omega_R\displaystyle\sum_{k=i}^{I-1} \left(f_k^j-f_{k+1}^j
\right)\lambda_{k-i}.
\end{eqnarray}
Then the elements of the matrices A and B in equation (\ref{eq:matrixab}) are
\begin{equation}
\begin{array}{ll}
A_c=1+\theta(\Omega_L+\Omega_R)\lambda_0 \\
A_{i,L}=\theta\Omega_L(\lambda_{i}-\lambda_{i-1}),\quad i=1,2,\ldots,I\\
A_{i,R}=\theta\Omega_R(\lambda_{i}-\lambda_{i-1}),\quad i=1,2,\ldots,I
\end{array}
\end{equation}
and
\begin{equation}
\begin{array}{ll}
B_c=1-(1-\theta)(\Omega_L+\Omega_R)\lambda_0 \\
B_{i,L}=-(1-\theta)\Omega_L(\lambda_{i}-\lambda_{i-1}),\quad i=1,2,\ldots,I\\
B_{i,R}=-(1-\theta)\Omega_R(\lambda_{i}-\lambda_{i-1}),\quad i=1,2,\ldots,I.
\end{array}
\end{equation}

\section{Error estimation of the difference scheme}
\label{error}

\begin{figure}
\centering
\includegraphics[width=0.49\textwidth]{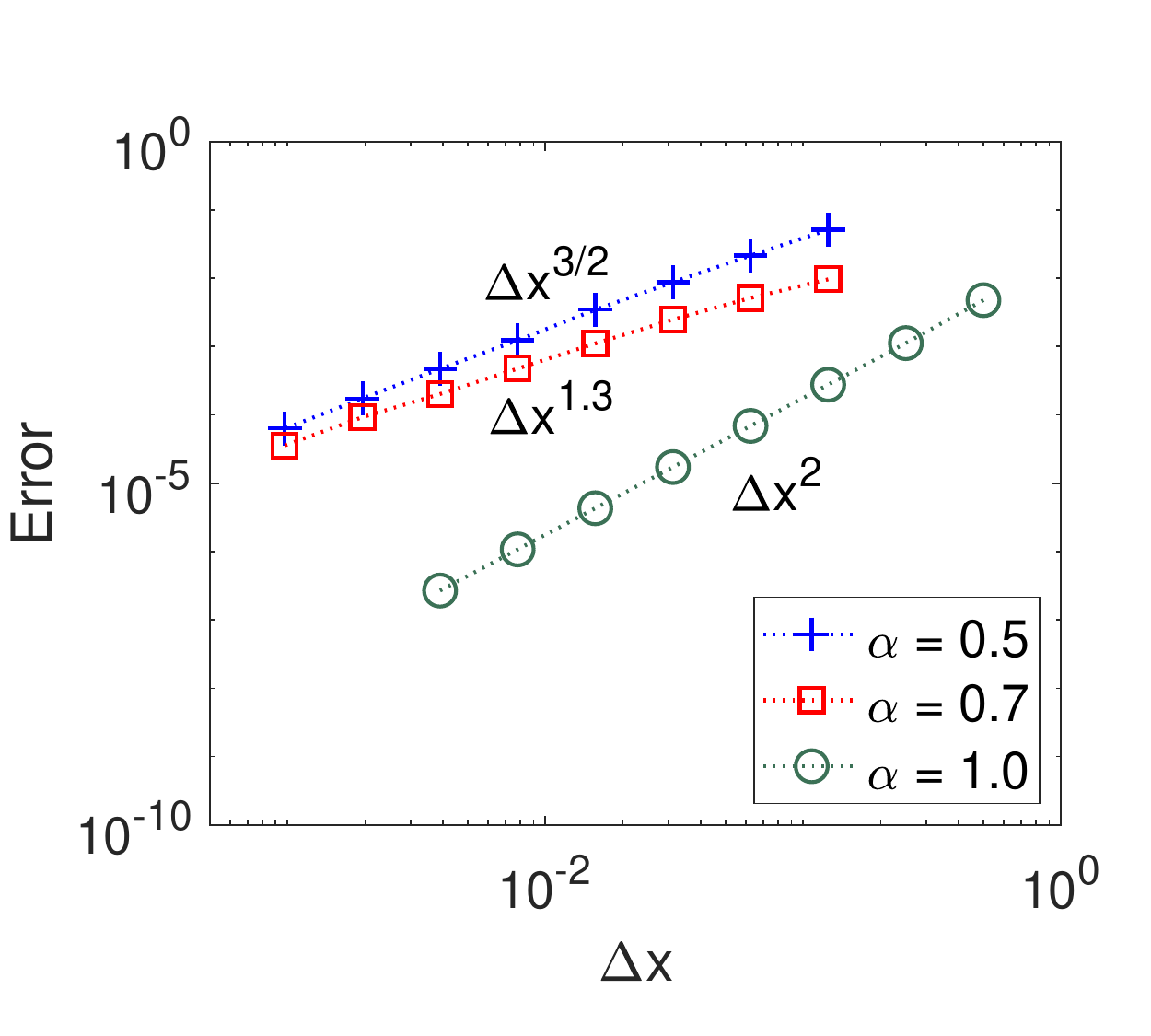}
\includegraphics[width=0.49\textwidth]{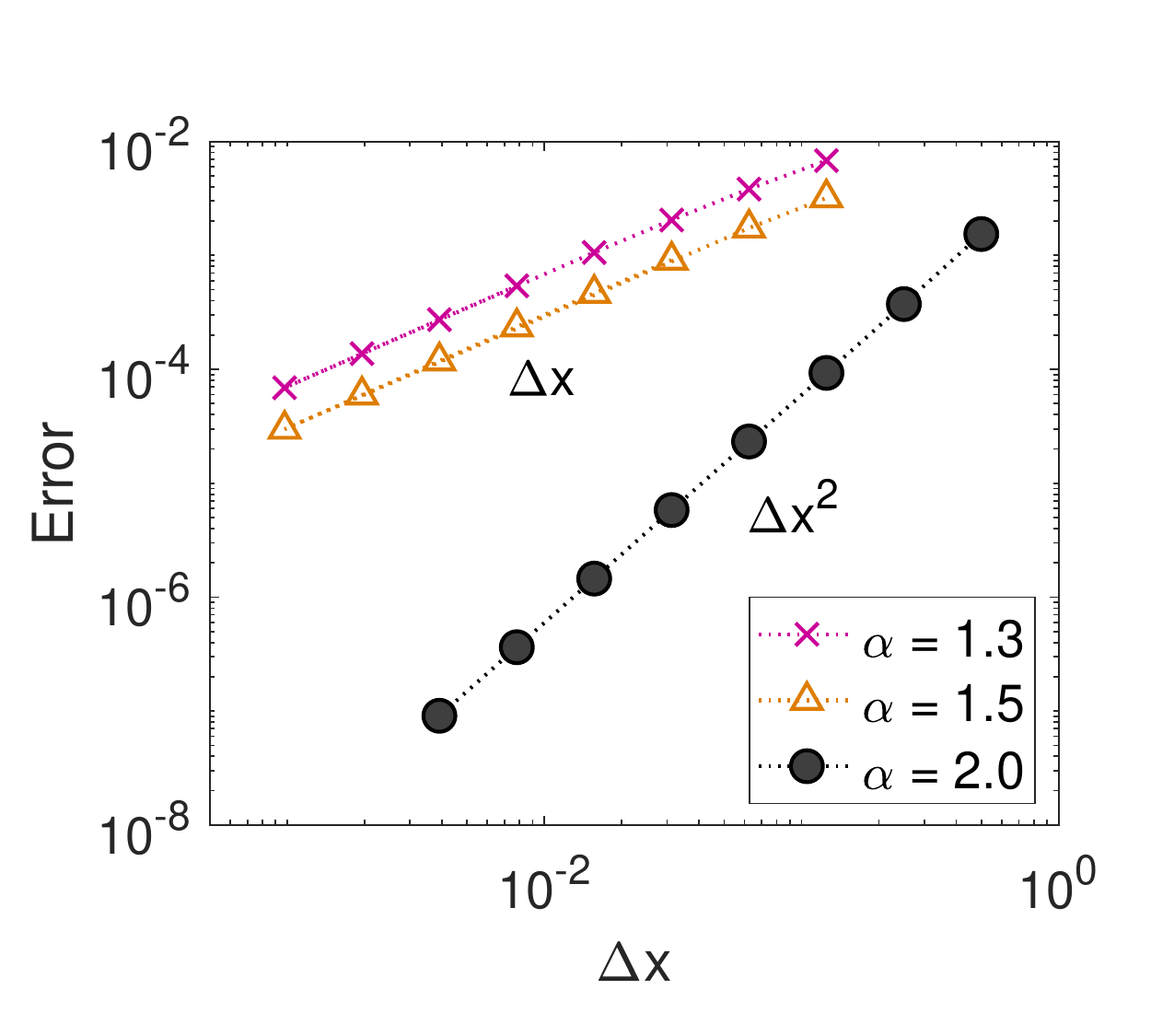}
\includegraphics[width=0.49\textwidth]{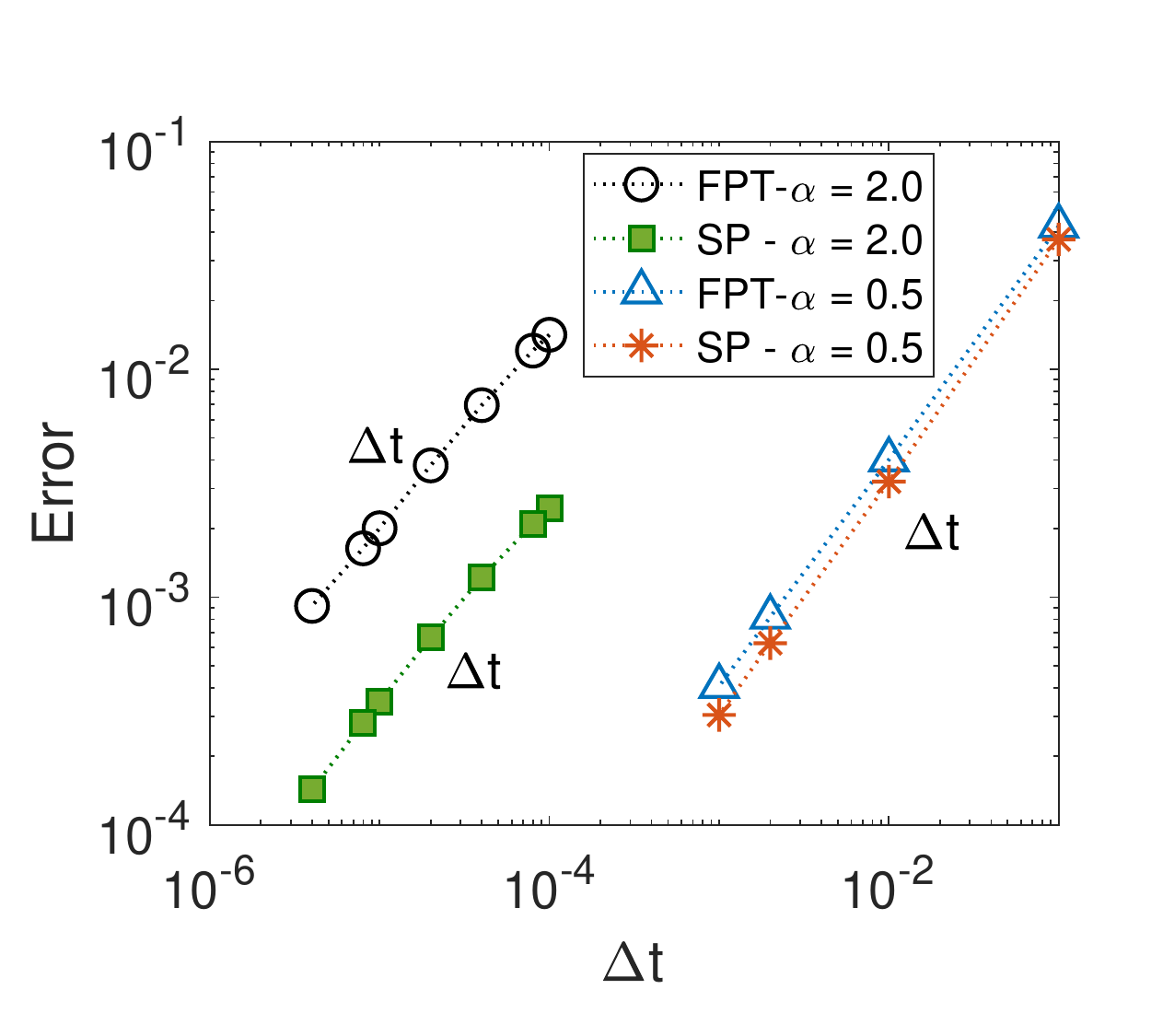}
\includegraphics[width=0.49\textwidth]{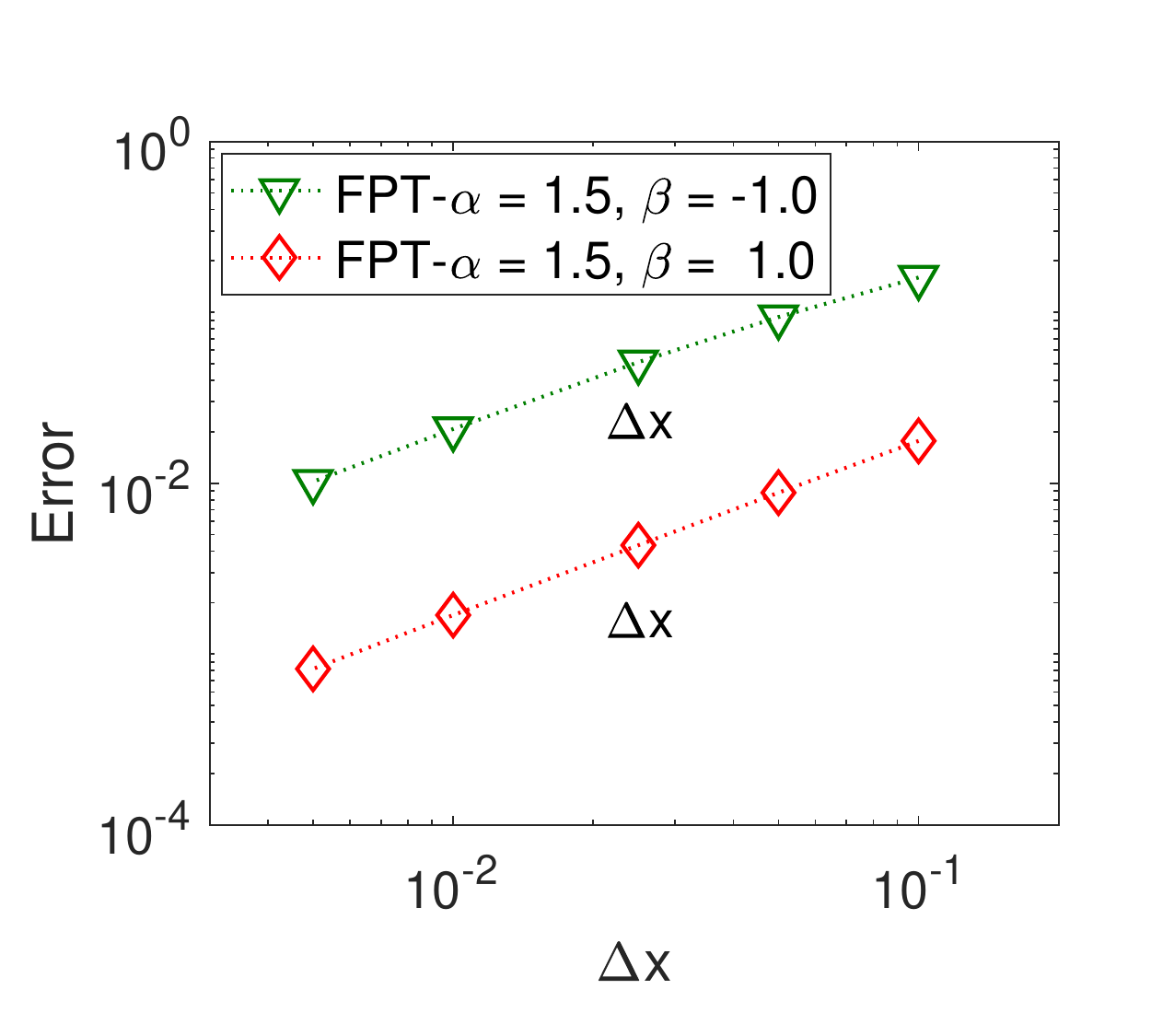}
\caption{\textcolor{black}{
Error analysis of the numerical schemes in section 2. Top left:
Truncation error for $\ell_{\alpha,\beta}(x,t)$ in the L1 scheme ($\alpha=0.5,
0.7$) (equations (13)) and Hilbert discretising scheme for $\alpha=1$,
$\beta=0$ (equation (15)). For this panel we use $\Delta t=10^{-4}$, $t=1$,
$L=16$ in the case $\alpha=0.5, 0.7$ and $\Delta t=10^{-5}$, $t=1$, $L=8$ in
the case $\alpha=1$. Top right: Truncation error for $\ell_{\alpha,\beta}(x,t)$
in the L2 scheme ($\alpha=1.3, 1.5$) (equations (14)) and central difference
discretising scheme for $\alpha=2$. For this panel we use $\Delta t=10^{-5}$,
$t=1$, $L=8$ in the case $\alpha=1.3, 1.5$ and $\Delta t=10^{-6}$,
$t=1$, $L=4$ in the case $\alpha=2$. Bottom left: Truncation error for the
first-passage time PDF and survival probability versus time step for Brownian
motion (equation (25)) and one-sided $\alpha$-stable probability law with
$\alpha=0.5$ and $\beta=1$ (equation (33)). For this panel we use $\Delta
x=5\times10^{-3}$, $t=10$ and $L=20$. Bottom right: Truncation error for
the first-passage time PDF of extremal two-sided $\alpha$-stable probability
laws with stable index $\alpha=1.5$ and skewness parameter $\beta=-1, 1$
(equations (34) and (39)) versus space increment $\Delta x$. For this panel
we use $\Delta t=10^{-5}$, $t=5$ and $L=10$.}}
\label{fig:fig1-Error}
\end{figure}

\textcolor{black}{We here provide some details on the error estimate of our
difference scheme. For the case $0<\alpha<1$ (equation (13))
\begin{equation}
\int\limits_{-L}^{x_{i}}\frac{f^{(1)}(\zeta,t_{j})}{(x_{i}-\zeta)^{\alpha}}
\mathrm{d}\zeta\approx \displaystyle\sum_{k=1}^{i}\frac{f^{j}_{k}-f^{j}_{k-1}}{
\Delta x}\int\limits_{x_{k-1}}^{x_{k}}\frac{1}{(x_i-\zeta)^{\alpha}}\mathrm{d}
\zeta+\mathcal{O}(\Delta x^{2-\alpha}),
\label{eq:discleft1a}
\end{equation}
for the left side derivative, and
\begin{equation}
\int\limits_{x_i}^{L}\frac{f^{(1)}(\zeta,t_j)}{(\zeta-x_i)^{\alpha}}\mathrm{d}
\zeta\approx \displaystyle\sum_{k=i}^{I-1}\frac{f^j_{k+1}-f^j_k}{ \Delta x}\int
\limits_{x_k}^{x_{k+1}}\frac{1}{(\zeta-x_i)^{\alpha}}\mathrm{d}\zeta+\mathcal{O}
(\Delta x^{2-\alpha}),
\label{eq:discright1a}
\end{equation}
for the right side derivative. This efficient way to approximate the
Caputo derivative of order $\alpha$ ($0<\alpha<1$) is called L1 scheme
\cite{Oldham-Spanier1974,Langlands-Henry2005,Changpin-Fanhai2012} and its
error estimate is $\mathcal{O}(\Delta x^{2-\alpha})$ (see figure 2 top
left panel) \cite{Langlands-Henry2005,Sun-Wu2006,Lin-Xu2007}. For the case
$1<\alpha<2$ the suitable method to discretise the Caputo derivative is the L2
scheme \cite{Oldham-Spanier1974,Changpin-Fanhai2012,Lynch2003}
\begin{equation}
\fl\int\limits_{-L}^{x_i}\frac{f^{(2)}(\zeta,t_{j})}{(x_{i}-\zeta)^{\alpha-1}}
\mathrm{d}\zeta\approx\displaystyle\sum_{k=1}^{i}\frac{f^{j}_{k+1}-2f^{j}_{k}
+f^{j}_{k-1}}{(\Delta x)^2}\int\limits_{x_{k-1}}^{x_{k}}\frac{1}{(x_{i}-\zeta)
^{\alpha-1}}\,\mathrm{d}\zeta+\mathcal{O}(\Delta x) ,
\label{eq:discleft2a}
\end{equation}
for the left side, and
\begin{equation}
\fl\int\limits_{x_{i}}^{L}\frac{f^{(2)}(\zeta,t_{j})}{(\zeta-x_{i})^{\alpha-1}}
\mathrm{d}\zeta\approx \displaystyle\sum_{k=i}^{I-1}\frac{f^{j}_{k+1}-2f^{j}_k
+f^{j}_{k-1}}{(\Delta x)^2}\int\limits_{x_{k}}^{x_{k+1}}\frac{1}{(\zeta-x_{i})^{
\alpha-1}}\,\mathrm{d}\zeta+\mathcal{O}(\Delta x),
\label{eq:discright2a}
\end{equation}
for the right side. The truncation error for this scheme is $\mathcal{O}(\Delta
x)$ (see figure 2 top right panel) \cite{Lynch2003,Sousa2010}.  For the special
case $\alpha=2$ the central difference scheme has truncation error $\mathcal{O}
(\Delta x^{2})$ (see figure \ref{fig:fig1-Error} top right panel). For comparison,
in \cite{Sun-Wu2006} an error estimate of order $\mathcal{O}(\Delta x^{3-\alpha})$
is presented for $1<\alpha<2$. In \cite{Odibat2009} a computational algorithm for
approximating the Caputo derivative was developed, and the convergence order is
$\mathcal{O}(\Delta x^2)$ for $0<\alpha\leq2$. Another difference method of order
two was derived in \cite{Sousa2010} for $1<\alpha\leq2$.}

\textcolor{black}{
For the special case $\alpha=1$, $\beta=0$,
\begin{eqnarray}
\nonumber
-\frac{d}{d x}(H f(x,t))&\approx&-\frac{2}{\pi}\displaystyle\sum_{k=1}^{i} 
\frac{f^{j}_{k}-f^{j}_{k-1}}{\Delta x}\frac{1}{2(i-k)+1}\\
&&-\frac{2}{\pi}
\displaystyle\sum_{k=i}^{I-1}\frac{f^{j}_{k}-f^{j}_{k+1}}{\Delta x}\frac{1}{
2(k-i)+1}+\mathcal{O}(\Delta x^{2}),
\label{eq:Hilbert}
\end{eqnarray}
the truncation error is $\mathcal{O}(\Delta x^{2})$ (see figure \ref{fig:fig1-Error}
top left panel). To evaluate the truncation error we used the relation
\begin{equation}
\|e(x)\|_2=\|f(x_i,t_j)-f_{i}^{j}\|_2=\sqrt{\frac{1}{I}\displaystyle\sum_{i=1}^{I}(
f(x_i,t_j)-f_{i}^{j})^2} ,
\end{equation}
where $f(x_i,t_j)$ is the exact solution and $f_{i}^{j}$ is the approximated
solution of function $f(x,t)$. For $\|e(t)\|_2$ this is similar and we use
\begin{equation}
\|e(t)\|_2=\sqrt{\frac{1}{J}\displaystyle\sum_{j=1}^{J}(f(x_i,t_j)-f_{i}^{j})^2} .
\end{equation}
The results of the error analysis for $\ell_{\alpha,\beta}(x,t)$, survival
probability and the first-passage time PDF are shown in figure \ref{fig:fig1-Error}.}

\section{A short review of the Skorokhod theorem}
\label{Appendix Skoro}

The Skorokhod theorem establishes a general formula for the Laplace transform
$\wp(\lambda)$ of the first-passage time PDF in the semi-infinite domain for a
broad class of homogeneous
random processes with independent increments and thus has a pivotal role in the
theory of first-passage processes \cite{Gikhman-Skorokhod,Skorokhod1964}. For
the process starting at $x=0$ with a boundary at $x=d$,
\begin{equation}
\wp(\lambda)=\langle e^{-\lambda t}\rangle=\int\limits_0^{\infty}e^{-\lambda t}
\wp(t)\mathrm{d}t=1-p_+(\lambda,d).
\label{eq:LaplFPTandp}
\end{equation}
Here the auxiliary measure $p_+(\lambda, x)$ is defined via its Fourier transform as 
\begin{equation}
\fl q_+(\lambda, k)=\int\limits_{-\infty}^{\infty}e^{\mathrm{i}kx}\frac{\partial p_+(\lambda,
x)}{\partial x}\mathrm{d}x=\exp\left\{\int\limits_0^{\infty}\frac{e^{-\lambda t}}{t}
\int\limits_0^{\infty}(e^{\mathrm{i}kx}-1)f(x,t)\mathrm{d}x\mathrm{d}t\right\},
\label{eq:skorokhod8}
\end{equation}
where the function $f(x,t)$ is the PDF of the process, that is $\ell_{\alpha,\beta}(
x,t)$ in our case. The boundary condition reads $p_+(\lambda,x)=0$ at $x=0$.
Below, for didactic purposes we first calculate $\wp(t)$ for Brownian motion and
then proceed to symmetric ($0<\alpha\leq2$ and $\beta=0)$, one-sided ($0<\alpha<1$
and $\beta=\pm1$), extremal two-sided ($1<\alpha<2$ and $\beta=\pm1$), and finally
to the general case ($0<\alpha<2$, $\alpha\neq1$, and $-1<\beta<1$).

\subsection{First passage time PDF for Brownian motion}

For Brownian motion the PDF reads
\begin{equation}
f(x,t)=\frac{1}{\sqrt{4\pi K_2 t}}\exp\left(-\frac{x^2}{4K_2t}\right).
\label{eq:skorokhod9}
\end{equation}
Substitution into equation (\ref{eq:skorokhod8}) yields
\begin{eqnarray}
\nonumber
q_{+}(\lambda, k)&=&\exp\left\{ \int\limits_0^{\infty}\frac{e^{-\lambda t}}{t}
\int\limits_0^{\infty}(e^{\mathrm{i}kx}-1)\frac{e^{-x^2/(4K_2 t)}}{\sqrt{4\pi K_2 t}}
\mathrm{d}x\mathrm{d}t \right\}\\
\nonumber
&=&\exp\left\{\int\limits_0^{\infty}\frac{e^{-\lambda t}}{t}\frac{1}{2}\left\{
e^{-t K_2 k^2}\mathrm{erfc}\left(-\mathrm{i}k\sqrt{t K_2}\right)-1\right\}\right\}\\
&=&\exp\left\{\frac{1}{2}\left[\ln{\left(\frac{\lambda}{\lambda+K_2 k^2}\right)}
+\ln{\left(\frac{1+\mathrm{i}k\sqrt{\frac{K_2}{\lambda}}}{1-\mathrm{i}k\sqrt{\frac{
K_2}{\lambda}}}\right)}\right]\right\},
\end{eqnarray}
therefore
\begin{equation}
q_+(\lambda,k)=\left[\frac{\lambda}{\lambda+K_2k^2}\left(\frac{1+\mathrm{i}k\sqrt{\frac{K_2}{
\lambda}}}{1-\mathrm{i}k\sqrt{\frac{K_2}{\lambda}}}\right)\right]^{1/2} .
\end{equation}
Now, we use the relation
\begin{equation}
\frac{1}{1-\mathrm{i}k\sqrt{K_2/\lambda}}=\frac{1+\mathrm{i}k\sqrt{K_2/\lambda}}{1+K_{2}k^2/\lambda} 
\end{equation}
to get to
\begin{equation}
q_+(\lambda,k)=\frac{\lambda}{\lambda+K_2k^2}+\frac{\mathrm{i}k\sqrt{K_2\lambda}}{
\lambda+K_2k^2}.
\label{eq:skorokhod10}
\end{equation}
After an inverse Fourier transform according to equation (\ref{eq:skorokhod8}) we
arrive at
\begin{equation}
\fl\frac{d}{d x}p_+(\lambda,x)=\frac{1}{2\pi}\int\limits_{-\infty}^{\infty}e^{-\mathrm{i}kx}
q_+(\lambda,k)\mathrm{d}k=\frac{1}{2\pi}\int\limits_{-\infty}^{\infty}e^{-\mathrm{i}kx}
\left(\frac{\lambda}{\lambda+K_2k^2}+\frac{\mathrm{i}k\sqrt{K_2\lambda}}{\lambda+K_2k^2}
\right)\mathrm{d}k.
\label{eq:SkorokhodBrow6}
\end{equation}
For the first integral we have
\begin{equation}
\fl\frac{1}{2\pi}\int\limits_{-\infty}^{\infty}e^{-\mathrm{i}kx}\left(\frac{\lambda}{\lambda
+K_2 k^2}\right)\mathrm{d}k=\frac{\lambda}{2\pi K_2}\int\limits_{-\infty}^{\infty}
\frac{e^{-\mathrm{i}kx}}{\left(k-\mathrm{i}\sqrt{\frac{\lambda}{ K_2}}\right)\left(k
+\mathrm{i}\sqrt{\frac{\lambda}{K_2}}\right)}
\mathrm{d}k.
\end{equation}
With the residue theorem,
\begin{equation}
\frac{\lambda}{2\pi K_2}\int\limits_{-\infty}^{\infty}\frac{e^{-\mathrm{i}kx}}{\left(
k-\mathrm{i}\sqrt{\frac{\lambda}{K_2}}\right)\left(k+\mathrm{i}\sqrt{\frac{\lambda}{
K_2}}\right)}\mathrm{d}k=
\frac{1}{2}\sqrt{\frac{\lambda}{K_2}}\exp{\left(-\sqrt{\frac{\lambda}{K_2}}x\right)}, 
\label{eq:SkorokhodBrow8}
\end{equation}
For the second integral, we write
\begin{equation}
\fl\frac{1}{2\pi}\int\limits_{-\infty}^{\infty}e^{-\mathrm{i}kx}\left(\frac{\mathrm{
i}k\sqrt{K_2\lambda}}{\lambda+K_2k^2}\right)\mathrm{d}k=\frac{\mathrm{i}}{2\pi}\sqrt{
\frac{\lambda}{K_2}}\int\limits_{-\infty}^{\infty}\frac{k\e^{-\mathrm{i}kx}}{\left(k-
\mathrm{i}\sqrt{\frac{\lambda}{K_2}}\right)\left(k+\mathrm{i}\sqrt{\frac{\lambda}{K_
2}}\right)}\mathrm{d}k.
\end{equation}
Again, with the residue theorem,
\begin{equation}
\fl\frac{\mathrm{i}}{2\pi }\sqrt{\frac{\lambda}{K_2}}\int\limits_{-\infty}^{\infty}
\frac{k\e^{-\mathrm{i}kx}}{(k-\mathrm{i}\sqrt{\frac{\lambda}{ K_2}})(k+\mathrm{i}
\sqrt{\frac{\lambda}{ K_2}})}\mathrm{d}k=\frac{1}{2}\sqrt{\frac{\lambda}{K_2}}\exp{
\left(-\sqrt{\frac{\lambda}{K_2}}x\right)}.
\label{eq:SkorokhodBrow11}
\end{equation}
Therefore, by substitution of equations (\ref{eq:SkorokhodBrow8}) and
(\ref{eq:SkorokhodBrow11}) into equation (\ref{eq:SkorokhodBrow6}) we obtain
($x>0$)
\begin{equation}
\frac{d}{d x}p_+(\lambda,x)=\frac{1}{2\pi}\int\limits_{-\infty}^{\infty}e^{-\mathrm{i}kx}
\hat{q}_+(\lambda,k)\mathrm{d}k=\sqrt{\frac{\lambda}{K_2}}\exp{\left(-\sqrt{\frac{
\lambda}{K_2}}x\right)}.
\end{equation}
Using the boundary condition we get
\begin{equation}
p_{+}(\lambda,x)=\sqrt{\frac{\lambda}{K_2}}\int\limits_0^x\exp{\left(-\sqrt{\frac{
\lambda}{K_2}}x\right)}\mathrm{d}x=1-\exp{\left(-\sqrt{\frac{\lambda}{K_2}}x\right)},
\end{equation}
and thus with the help of equation (\ref{eq:LaplFPTandp}),
\begin{equation}
\wp(\lambda)=\exp\left(-\sqrt{\frac{\lambda}{K_2}}d\right) ,
\label{eq:skorokhod11}
\end{equation}
Finally, by inverse Laplace transform we get
\begin{equation}
\wp(t)=\frac{d}{\sqrt{4\pi K_2 t^3}}\exp\left(-\frac{d^2}{4K_2t}\right).
\label{eq:skorokhod12}
\end{equation}
This is the famous L{\'e}vy-Smirnov law representing a well-known result of
Brownian motion \cite{Feller1971}. This derivation is, of course, overly
complicated for the Gaussian case, but the same procedure can be applied
to the general case of an asymmetric L{\'e}vy stable law, as we now show.

\subsection{First passage time PDF for symmetric $\alpha$-stable processes}

Due to the symmetry of the PDF the function $\ln q_+(k,\lambda)$, equation
(\ref{eq:skorokhod8}), can be
written as
\begin{equation}
\ln {q_+(k,\lambda)}=A(\lambda,k)+\mathrm{i}B(\lambda,k),
\label{eq:skorokhod13}
\end{equation}
where 
\begin{equation}
\fl A(\lambda,k)=\int\limits_0^{\infty}\frac{e^{-\lambda t}}{t}\int\limits_0^{
\infty}(\cos{(k x)}-1)\ell_{\alpha,0}(x,t)\,\mathrm{d}x\mathrm{d}t=\frac{1}{2}
\ln{\frac{\lambda}{\lambda+K_{\alpha}|k|^{\alpha}}},
\label{eq:skorokhod14}
\end{equation}
and
\begin{equation}
B(\lambda,k)=\int\limits_0^{\infty}\frac{e^{-\lambda t}}{t}\int\limits_0^{\infty}
\sin{(k x)}\ell_{\alpha,0}(x,t)\mathrm{d}x\mathrm{d}t.
\label{eq:Blambda}
\end{equation}
To find $B(\lambda,k)$ at small $\lambda$ the self-similar property of the
$\alpha$-stable process comes in useful,
\begin{equation}
\ell_{\alpha,0}(x,t)=\frac{1}{(K_\alpha t)^{1/\alpha}}\ell_{\alpha,0}\left(\frac{x}{
(K_\alpha t)^{1/\alpha}},1\right),
\end{equation}
where
\begin{equation}
\ell_{\alpha,0}(y,1)=\ell_{\alpha,0}(-y,1)=\frac{1}{\pi}\int\limits_0^{\infty}
\cos{(ky)}e^{-|k|^\alpha}\mathrm{d}k.
\end{equation}
When $\lambda$ tends to zero, from equation (\ref{eq:Blambda}) we get
\begin{eqnarray}
\nonumber
\fl B(\lambda,k)&=&\int\limits_0^{\infty}\frac{e^{-\lambda t}}{t}\int\limits_0^{
\infty}\sin{(k x)}\frac{1}{(K_\alpha t)^{1/\alpha}}\ell_{\alpha,0}\left(x/(K_\alpha
t)^{1/\alpha},1\right)\,\mathrm{d}x\mathrm{d}t\\
\nonumber
&\approx&\int\limits_0^{\infty}\frac{1}{t}\int\limits_0^{\infty}\sin{(ky(K_\alpha
t)^{1/\alpha})}\ell_{\alpha,0}(y,1)\mathrm{d}y\mathrm{d}t\\
&=&\alpha\int\limits_0^{\infty}\ell_{\alpha,0}(y,1)\int\limits_0^{\infty}\frac{
\sin{(kys)}}{s}\mathrm{d}s\mathrm{d}y=\frac{\alpha\pi}{4}\mathrm{sgn}(k).
\label{eq:skorokhod15}
\end{eqnarray}
By substitution of equations (\ref{eq:skorokhod14}) and (\ref{eq:skorokhod15})
into equation (\ref{eq:skorokhod13}) we get
\begin{equation}
q_+(\lambda,k)\approx\frac{\sqrt{\lambda}}{\sqrt{K_\alpha}|k|^{\alpha/2}}e^{\mathrm{i}
\mathrm{sgn}(k)\alpha\pi/4},\quad\lambda\to0.
\end{equation}
Then the inverse Fourier transform of the above equation renders
\begin{equation}
\frac{d}{dx}p_+(\lambda,x)=\frac{1}{2\pi}\int\limits_{-\infty}^{\infty}e^{-\mathrm{i}kx}
\hat{q}_+(\lambda,k)\mathrm{d}k=\frac{\sqrt{\lambda}}{\sqrt{K_\alpha}\Gamma(\alpha
/2)}x^{-1+\alpha/2},
\end{equation} 
and, after applying the boundary condition,
\begin{equation}
p_+(\lambda,x)=\frac{2\sqrt{\lambda}}{\alpha\sqrt{K_\alpha}\Gamma(\alpha/2)}x^{
\alpha/2}.
\end{equation}
Recalling equation (\ref{eq:LaplFPTandp}),
\begin{equation}
\wp(\lambda)\approx 1-\frac{2d^{\alpha/2}}{\alpha\sqrt{K_\alpha}\Gamma(\alpha/2)}
\lambda^{1/2}.
\end{equation}
Now, with the help of the Tauberian theorem \cite{Feller1971} (Chapter XIII,
section 5) we find that the
small-$\lambda$ asymptotic of the Laplace transform
\begin{equation}
\psi(\lambda)\approx1-b_2\lambda^{\mu},\quad b_2=b_1\Gamma(1-\mu)/\mu,\quad
\lambda\to0
\label{eq:Tauberian2}
\end{equation}
corresponds to the long-time asymptotic of the PDF (\cite{Sokla}, chapter 3)
\begin{equation}
\psi(t)\sim b_1t^{-1-\mu},\quad0<\mu<1,\quad b_1>0.
\label{eq:Tauberian1}
\end{equation}
Therefore, the long-time asymptotic of the first-passage time PDF for symmetric
$\alpha$-stable process has the form \cite{Koren2007PRL}
\begin{equation}
\wp(t)\approx\frac{d^{\alpha/2}}{\alpha\sqrt{\pi K_{\alpha}}\Gamma{(\alpha/2)}}
t^{-3/2}.
\label{eq:skorokhod16Apendix}
\end{equation}

\subsection{First passage time PDF for one-sided $\alpha$-stable processes,
$0<\alpha<1$ and $\beta=1$}

Due to the monotonic growth of the process in this case there exists a simple
relation between the cumulative probabilities of the first-passage time and the
$\alpha$-stable process itself, see, e.g., \cite{iddo}. However, for a didactic
purpose in this Appendix we obtain the result by the use of Skorokhod's method.
Since for one-sided $\alpha$-stable processes with $0<\alpha<1$ and $\beta=1$
the PDF $\ell_{\alpha,1}(x,t)$ vanishes for $x<0$, we get
\begin{eqnarray}
\nonumber
\fl\int\limits_0^{\infty}\left(e^{\mathrm{i}kx}-1\right)\ell_{\alpha,1}(x,t)\mathrm{d}x
&=&\int\limits_{-\infty}^{\infty}\left(e^{\mathrm{i}kx}-1\right)\ell_{\alpha,1}(x,t)
\mathrm{d}x\\
&=&\exp{\left[-tK_\alpha|k|^\alpha\left(1-\mathrm{i}\mathrm{sgn}(k)\tan(\alpha\pi/2)
\right)\right]}-1.
\label{eq:skorokhod17}
\end{eqnarray}
After plugging this into equation (\ref{eq:skorokhod8}),
\begin{equation}
\fl q_{+}(\lambda,k)=\exp\left\{\int\limits_0^{\infty}\frac{e^{-\lambda t}}{t}\left(
\exp{\left[-tK_\alpha|k|^\alpha\left(1-\mathrm{i}\mathrm{sgn}(k)\tan(\alpha\pi/2)
\right)\right]}-1\right)\mathrm{d}t\right\}.
\end{equation}
Then
\begin{equation}
q_{+}(\lambda,k)=\exp{\left\{\ln{\left(\frac{\lambda}{\lambda+\zeta}\right)}
\right\}}=\frac{\lambda}{\lambda+\zeta},
\label{eq:skorokhod18}
\end{equation}
where
\begin{equation}
\zeta=K_{\alpha}|k|^{\alpha}\left(1-\mathrm{i}\mathrm{sgn}(k)\tan{\frac{\pi\alpha}{
2}}\right)=\frac{K_{\alpha}}{\cos{(\alpha\pi/2)}}(-\mathrm{i}k)^{\alpha}.
\label{eq:skorokhod19}
\end{equation}
Therefore, $\partial p_+(\lambda,x)/\partial x$ follows from result
(\ref{eq:skorokhod18}) by inverse Fourier transform,
\begin{equation}
\fl\frac{d}{dx}p_+(\lambda,x)=\frac{1}{2\pi}\int\limits_{-\infty}^{\infty}e^{-
\mathrm{i}kx}q_+(\lambda,k)\mathrm{d}k=\frac{1}{2\pi}\int\limits_{-\infty}^{
\infty}e^{-\mathrm{i}kx}\left(\frac{\lambda}{(-\mathrm{i}k)^{\alpha}\frac{K_\alpha}{
\cos{(\alpha\pi/2)}}+\lambda}\right)\mathrm{d}k.
\end{equation}
Defining $s=-\mathrm{i}k$, we have
\begin{equation}
\frac{\mathrm{d}}{\mathrm{d} x}p_{+}(\lambda,x)=\frac{1}{2\pi\mathrm{i}}\int\limits_{
-\mathrm{i}\infty}^{\mathrm{i}\infty}e^{sx}\frac{\lambda}{s^{\alpha}K_\alpha/[\cos(
\alpha\pi/2)]+\lambda}\mathrm{d}s.
\end{equation}
Recalling relation (\ref{eq:LaplderivMittag}) we obtain
\begin{equation}
\frac{\mathrm{d}}{\mathrm{d}x}p_+(\lambda,x)=-\frac{\mathrm{d}}{\mathrm{d}x}E_
\alpha\left(-\frac{\lambda\cos{(\alpha\pi/2)}}{K_\alpha}x^\alpha\right),
\end{equation}
where $E_{\alpha}(z)=\sum_{n=0}^{\infty}z^n/\Gamma(1+\alpha n)$ is the
Mittag-Leffler function, see \cite{Podlubny1999,FMainardi2010} and
\ref{Appendix Spec-func}. With the boundary
condition $p_+(\lambda,x\le0)=0$ we get
\begin{equation}
p_+(\lambda,x)=1-E_\alpha\left(-\frac{\lambda\cos{(\alpha\pi/2)}}{K_\alpha}
x^\alpha\right) .
\end{equation}
Thus, for the Laplace transform of the first-passage time PDF $\wp(\lambda)=1-p_+
(\lambda,d)$ we have
\begin{equation}
\wp(\lambda)=E_{\alpha}(-\lambda\cos(\alpha\pi/2)d^{\alpha}/K_{\alpha}).
\label{eq:skorokhod21}
\end{equation}
The Laplace inversion is then immediately accomplished in terms of the Wright
function (see equation (\ref{eq:LaplMfun})) for $0<\alpha<1$ \cite{Koren2007PRL},
\begin{equation}
\wp(t)=\frac{K_{\alpha}}{\cos(\alpha\pi/2)d^{\alpha}}M_{\alpha}\left(\frac{K_{
\alpha}t}{\cos(\alpha\pi/2)d^{\alpha}}\right).
\label{eq:skorokhod23}
\end{equation}
By using relation (\ref{eq:M-normalization}) we make sure that $\wp(t)$ is normalised,
\begin{equation}
\int\limits_0^{\infty}\wp(t)\mathrm{d}t=\int\limits_0^{\infty}\frac{K_{\alpha}}{
\cos(\alpha\pi/2)d^{\alpha}}M_{\alpha}\left(\frac{K_{\alpha}t}{\cos(\alpha\pi/2)
d^{\alpha}}\right)\mathrm{d}t=1.
\end{equation}
This can be also shown by taking the integral form (\ref{eq:intMfun}) of the
$M$-function. By changing the order of integration we get
\begin{eqnarray}
\nonumber
\fl\int\limits_0^{\infty}\wp(t)\mathrm{d}t&=&\int\limits_0^{\infty}\frac{K_{\alpha}}{
\cos(\alpha\pi/2)d^{\alpha}}\frac{1}{2\pi\mathrm{i}}\int\limits_{\mathrm{Ha}}\exp{
\left(\sigma-\frac{K_{\alpha}t}{\cos(\alpha\pi/2)d^{\alpha}}\sigma^{\alpha}
\right)}\frac{\mathrm{d}\sigma}{\sigma^{1-\alpha}}\mathrm{d}t\\
\nonumber
\fl&=&\frac{1}{2\pi\mathrm{i}}\int\limits_{\mathrm{Ha}}\sigma^{\alpha-1}\e^{\sigma}
\frac{
K_{\alpha}}{\cos(\alpha\pi/2)d^{\alpha}}\int\limits_0^{\infty}\exp\left(-\frac{K_{
\alpha}\sigma^{\alpha}}{\cos(\alpha\pi/2)d^{\alpha}}t\right)\mathrm{d}t\mathrm{d}
\sigma\\
\fl&=&\frac{1}{2\pi\mathrm{i}}\int\limits_{\mathrm{Ha}}\sigma^{-1}\e^{\sigma}\mathrm{d}
\sigma=1,
\end{eqnarray}
where Ha denotes the Hankel path, and in the last step we made use of equation
(\ref{eq:intGamma}).

\subsection{First passage time PDF for extremal two-sided $\alpha$-stable process,
$1<\alpha<2$ and $\beta=-1$}

To apply the Skorokhod theorem we need to calculate the following integral
\begin{equation}
\int\limits_0^{\infty}\left(e^{\mathrm{i}kx}-1\right)\ell_{\alpha,-1}(x,t)\mathrm{d}x.
\label{eq:skorokhodtwosidebn1}
\end{equation}
To this end we use the Laplace transform of $\alpha$-stable processes with $1<
\alpha\leq2$ and $\beta_B=-1$, which is derived in \cite{Zolotarev1986} (page 169,
equation (2.10.9)) in dimensionless B-form with $K_\alpha^B=1$ and $t=1$,
\begin{equation}
\frac{1}{\alpha}E_{1/\alpha}\left(-s\right)=\int\limits_0^{\infty} e^{-sx}\ell_{
\alpha,-1}^B(x,1)\mathrm{d}x.
\end{equation}
In dimensional variables this equation reads
\begin{equation}
\frac{1}{\alpha}E_{1/\alpha}\left(-s\left(K_\alpha^{B} t \right)^{1/\alpha}\right)
=\int\limits_0^{\infty}e^{-sx}\ell_{\alpha,-1}^B(x,t)\mathrm{d}x.
\end{equation}
With the help of equation (\ref{eq:relacharAandB}) we have in the A-form
\begin{equation}
\int\limits_0^{\infty}e^{-sx}\ell_{\alpha,-1}^{A}(x,t)\mathrm{d}x=\frac{1}{\alpha}
E_{1/\alpha}\left(-s\left(\frac{K_\alpha^At}{|\cos{(\alpha\pi/2)}|}\right)^{1/
\alpha}\right).
\label{eq:laplacetwosidezolota}
\end{equation}
Now we go back to equation (\ref{eq:skorokhodtwosidebn1}) which can be written as
(we again omit the index $A$ in what follows)
\begin{equation}
 \int\limits_0^{\infty}e^{-sx}\ell_{\alpha,-1}(x,t)|_{s=-\mathrm{i}k}\mathrm{d}x-\int
\limits_0^{\infty}e^{-sx}\ell_{\alpha,-1}(x,t)|_{s=0}\mathrm{d}x.   
\end{equation}
Using equation (\ref{eq:laplacetwosidezolota}) we find
\begin{equation}
\fl\int\limits_0^{\infty}\left(e^{\mathrm{i}kx}-1\right)\ell_{\alpha,-1}(x,t)\mathrm{
d}x=\frac{1}{\alpha}\left[E_{1/\alpha}\left(\mathrm{i}k\left(\frac{K_{\alpha}t}{|\cos{
(\alpha\pi/2)}|}\right)^{1/\alpha}\right)-1\right],
\end{equation}
and after plugging this expression into equation (\ref{eq:skorokhod8}),
\begin{equation}
\fl q_+(\lambda,k)=\exp\left\{\frac{1}{\alpha}\int\limits_0^{\infty}\frac{e^{-\lambda
t}}{t}\left[E_{1/\alpha}\left(\mathrm{i}k\left(\frac{K_{\alpha}t}{|\cos{(\alpha
\pi/2)}|}\right)^{1/\alpha}\right)-1\right]\mathrm{d}t\right\}.
\label{quplusone}
\end{equation}
To calculate expression (\ref{quplusone}) we first find
\begin{eqnarray}
\nonumber
\fl\frac{\partial}{\partial\lambda}\ln{q_{+}(\lambda,k)}&=&-\frac{1}{\alpha}\int
\limits_0^{\infty}\e^{-\lambda t}\left[E_{1/\alpha}\left(\mathrm{i}k\left(\frac{K_{
\alpha} t}{|\cos{(\alpha\pi/2)}|}\right)^{1/\alpha}\right)-1\right]\mathrm{d}t\\
\fl&=&\frac{1}{\alpha\lambda}\left[1-\frac{\lambda^{1/\alpha}}{\lambda^{1/\alpha}
-\mathrm{i}k\left(\frac{K_\alpha}{|\cos(\alpha\pi/2)|}\right)^{1/\alpha}}\right],
\end{eqnarray}
where we employ the Laplace transform (\ref{eq:LaplMittag}) of the Mittag-Leffler
function. By taking the indefinite integral over $\lambda$ we obtain
\begin{equation}
q_+(\lambda,k)=\frac{\lambda^{1/\alpha}}{\lambda^{1/\alpha}-\mathrm{i}k\left(\frac{
K_\alpha}{|\cos(\alpha\pi/2)|}\right)^{1/\alpha}},
\label{quplus}
\end{equation}
and then from equation (\ref{quplus}), by inverse Fourier transform,
\begin{eqnarray}
\nonumber
\fl\frac{d}{dx}p_+(\lambda,x)&=&\frac{1}{2\pi}\int\limits_{-\infty}^{\infty}
e^{-\mathrm{i}kx}q_+(\lambda,k)\mathrm{d}k\\
\fl&=&\left(\frac{|\cos{(\alpha\pi/2)}|}{K_\alpha}\right)^{1/\alpha}\lambda^{1/\alpha}
\exp{\left[-\left(\frac{|\cos{(\alpha\pi/2)}|}{K_\alpha}\right)^{1/\alpha}x\lambda^{
1/\alpha}\right]}.
\end{eqnarray}
With the boundary condition $p_+(\lambda,x=0)=0$ we get
\begin{equation}
p_+(\lambda,x)=1-\exp{\left[-\left(\frac{|\cos{(\alpha\pi/2)}|}{K_\alpha}\right)^{
1/\alpha}x\lambda^{1/\alpha}\right]}.
\end{equation}
Thus, for the Laplace transform $\wp(\lambda)$ we obtain
\begin{equation}
\wp(\lambda)=1-p_+(\lambda,d)=\exp{\left[-d\left(\frac{|\cos{(\alpha\pi/2)}|}{
K_\alpha}\right)^{1/\alpha}\lambda^{1/\alpha}\right]},
\end{equation}
which is of a stretched exponential form. Recalling now the Laplace transform pair
(\ref{eq:LaplPair1}) for the $M$-function, we finally arrive at the first-passage
time PDF for the extremal $\alpha$-stable process with $1<\alpha\leq2$ and $\beta
=-1$,
\begin{equation}
\wp(t)=\frac{t^{-1-1/\alpha}d}{\alpha\left(\frac{K_\alpha}{|\cos{(\alpha\pi/2)}|}
\right)^{1/\alpha}}M_{1/\alpha}\left(d\left(\frac{K_\alpha t}{|\cos{(\alpha\pi/2)}|}
\right)^{-1/\alpha}\right).
\label{eq:fptPDFtwobn}
\end{equation}
Let us show the normalisation of this function. By using the integral form
(\ref{eq:intMfun}) of the $M$-function and changing the order of integration we have
\begin{eqnarray}
\nonumber
\fl\int\limits_0^{\infty}\wp(t)\mathrm{d}t&=&\frac{d}{\alpha\left(\frac{K_\alpha}{
|\cos(\alpha\pi/2)|}\right)^{1/\alpha}}\int\limits_{0}^{\infty} t^{-1-1/\alpha}
M_{1/\alpha}\left(d\left(\frac{K_\alpha t}{|\cos{(\alpha\pi/2)}|}\right)^{-1/
\alpha}\right)\mathrm{d}t\\
\nonumber
\fl&=&\int\limits_0^{\infty}\frac{t^{-1}}{\alpha}\left[\frac{K_{\alpha}t}{d^{\alpha}
|\cos{(\alpha\pi/2)}|}\right]^{-1/\alpha}\\
\nonumber
&&\times\frac{1}{2\pi i}\int\limits_{\mathrm{Ha}}
\exp{\left(\sigma-\left[\frac{K_{\alpha}t}{d^{\alpha}|\cos{(\alpha\pi/2)}|}\right]^{
-1/\alpha}\sigma^{1/\alpha}\right)}\frac{\mathrm{d}\sigma}{\sigma^{1-1/\alpha}}
\mathrm{d}t\\
\nonumber
\fl&=&\frac{1}{2\pi i}\int\limits_{\mathrm{Ha}}\e^{\sigma}\sigma^{1/\alpha-1}\int
\limits_0^{\infty}\frac{t^{-1}}{\alpha}\left[\frac{K_{\alpha}t}{d^{\alpha}|\cos{
(\alpha\pi/2)}|}\right]^{-1/\alpha}\\
&&\times\exp{\left(-\left[\frac{K_{\alpha}}{d^{\alpha}
|\cos{(\alpha\pi/2)}|\sigma}\right]^{-1/\alpha}t^{-1/\alpha}\right)}\mathrm{d}t
\mathrm{d}\sigma.
\end{eqnarray}
By change of variable $u=\left[{K_{\alpha}/d^{\alpha}|\cos{(\alpha\pi/2)}|\sigma}
\right]^{-1/\alpha}t^{-1/\alpha}$, performing the inner integral and using the
Hankel formula (\ref{eq:intGamma}) for the Gamma function we obtain the necessary
normalisation condition. Now, if we employ the series expansion (\ref{eq:seriMfun})
for the $M$-function, from equation (\ref{eq:fptPDFtwobn}) we arrive at a series
which corresponds to that in equation (2.25) of \cite{peskir}. Note that in our
case the additional factor $K_{\alpha}/|\cos(\alpha\pi/2)|$ appears due to a
different starting form for the characteristic function of the $\alpha$-stable
process.

\subsection{First passage time PDF for extremal two-sided $\alpha$-stable processes,
$1<\alpha<2$, $\beta=1$}

Similar to above, at first we obtain the Laplace transform of the $\alpha$-stable
PDF with $1<\alpha\leq2$ and $\beta=1$. We write 
\begin{equation}
\int\limits_0^{\infty}f(u)\mathrm{d}u=\int\limits_{-\infty}^{\infty}f(u)\mathrm{d}
u-\int\limits_{-\infty}^0f(u)\mathrm{d}u,
\end{equation}
and then use the property $\ell_{\alpha,\beta}(x,t)=\ell_{\alpha,-\beta}(-x,t)$ to get
\begin{equation}
\int\limits_0^{\infty}e^{\mathrm{i}kx}\ell_{\alpha,1}(x,t)\mathrm{d}x=\int\limits_{
-\infty}^{\infty}e^{\mathrm{i}kx}\ell_{\alpha,1}(x,t)\mathrm{d}x-\int\limits_{-
\infty}^0e^{\mathrm{i}kx}\ell_{\alpha,-1}(-x,t)\mathrm{d}x. 
\label{eq:skorokhodtwosidebp1}
\end{equation}
The second integral on the right side can be written as
\begin{equation}
\int\limits_{-\infty}^0e^{\mathrm{i}kx}\ell_{\alpha,-1}(-x,t)\mathrm{d}x=\int
\limits_0^{\infty}e^{-\mathrm{i}kx}\ell_{\alpha,-1}(x,t)\mathrm{d}x.
\label{simple}
\end{equation}
To take the first integral on the right hand side of equation
(\ref{eq:skorokhodtwosidebp1}) we use the characteristic function in the A-form.
To take the second integral (\ref{simple}) we employ the Laplace transform of the PDF
with $1<\alpha<2$ and $\beta=-1$ given by equation (\ref{eq:laplacetwosidezolota})
with $s=\mathrm{i}k$. Thus, relation (\ref{eq:skorokhodtwosidebp1}) in the A-form
has the shape
\begin{equation}
\fl\int\limits_0^{\infty}e^{\mathrm{i}kx}\ell_{\alpha,1}^A(x,t)\mathrm{d}x=\exp{
\left(\frac{(-\mathrm{i}k)^{\alpha}K_{\alpha}^At}{|\cos{(\alpha\pi/2)}|}\right)}
-\frac{1}{\alpha}E_{1/\alpha}\left(\left[\frac{(-\mathrm{i}k)^{\alpha}K_{\alpha}
^At}{|\cos{(\alpha\pi/2)}|}\right]^{1/\alpha}\right).
\label{eq:laplacetwosidebposi1}
\end{equation}
With the help of equation (\ref{eq:laplacetwosidebposi1}) we write (again the index
$A$ is omitted in what follows)
\begin{eqnarray}
\nonumber
\fl\int\limits_0^{\infty}\left(e^{\mathrm{i}kx}-1\right)\ell_{\alpha,1}(x,t)
\mathrm{d}x&=&\int\limits_0^{\infty}e^{-sx}\ell_{\alpha,1}(x,t)|_{s=-\mathrm{i}k}
\mathrm{d}x-\int\limits_0^{\infty}e^{-sx}\ell_{\alpha,1}(x,t)|_{s=0}\mathrm{d}x\\
\fl&&\hspace*{-2.8cm}
=\left[\exp{\left(\frac{(-\mathrm{i}k)^{\alpha}K_\alpha t}{|\cos{(\alpha\pi/2)}|}
\right)}-\frac{1}{\alpha}E_{1/\alpha}\left(\left[\frac{(-\mathrm{i}k)^{\alpha}K_\alpha
t}{|\cos{(\alpha\pi/2)}|}\right]^{1/\alpha}\right)-1+\frac{1}{\alpha}\right].
\end{eqnarray}
By substituting this expression into equation (\ref{eq:skorokhod8}) we get
\begin{eqnarray}
\nonumber
\ln{q_+(\lambda,k)}&=&\int\limits_0^{\infty}\frac{e^{-\lambda t}}{t}\left[\exp{
\left(\frac{(-\mathrm{i}k)^{\alpha}K_\alpha t}{|\cos{(\alpha\pi/2)}|}\right)}\right.\\
&&\left.-\frac{1}{\alpha}E_{1/\alpha}\left(\left[\frac{(-\mathrm{i}k)^{\alpha}K_\alpha
t}{|\cos{(\alpha\pi/2)}|}\right]^{1/\alpha}\right)-1+\frac{1}{\alpha}\right]\mathrm{d}
t.
\end{eqnarray}
The derivative with respect to $\lambda$ reads 
\begin{eqnarray}
\nonumber
\fl\frac{\partial}{\partial\lambda}\ln{q_+(\lambda,k)}&=&-\int\limits_0^{\infty}\e^{
-\lambda t} \left[\exp{\left(\frac{(-\mathrm{i}k)^{\alpha}K_\alpha t}{|\cos{(\alpha
\pi/2)}|}\right)}\right.\\
\nonumber
\fl&&\left.-\frac{1}{\alpha}E_{1/\alpha}\left(\left[\frac{(-\mathrm{i}k)^{\alpha}
K_\alpha t}{|\cos{(\alpha\pi/2)}|}\right]^{1/\alpha}\right)-1+\frac{1}{\alpha}
\right]\mathrm{d}t\\
\fl&=&-\frac{1}{\lambda-\frac{(-\mathrm{i}k)^{\alpha}K_\alpha}{|\cos{(\alpha\pi/2)}|
}}+\frac{1}{
\alpha}\frac{\lambda^{1/\alpha-1}}{\lambda^{1/\alpha}-\left(\frac{(-\mathrm{i}k)^{
\alpha}K_\alpha}{|\cos{(\alpha\pi/2)}|}\right)^{1/\alpha}}+\left(1-\frac{1}{\alpha}
\right)\frac{1}{\lambda},
\end{eqnarray}
where for the second term we employ the Laplace transform (\ref{eq:LaplMittag}) of
the Mittag-Leffler function. By taking the indefinite integral over $\lambda$ we
obtain
\begin{equation}
q_+(\lambda,k)=\frac{\lambda^{1-1/\alpha}\left(\lambda^{1/\alpha}-\left(\frac{(-
\mathrm{i}k)^{\alpha}K_\alpha}{|\cos{(\alpha\pi/2)}|}\right)^{1/\alpha}\right)}{
\lambda-\frac{(-\mathrm{i}k)^{\alpha}K_\alpha}{|\cos{(\alpha\pi/2)}|}}.
\label{eq:quplus two}
\end{equation}
Then, $dp_+(\lambda,x)/dx$ follows from equation (\ref{eq:quplus two}) by inverse
Fourier transform,
\begin{equation}
\frac{d}{dx}p_+(\lambda,x)=\frac{1}{2\pi}\int\limits_{-\infty}^{\infty}e^{-\mathrm{i}
kx}\left(\frac{1+\mathrm{i}k\left(\frac{K_\alpha}{|\cos{(\alpha\pi/2)}|\lambda}\right)
^{1/\alpha}}{1-(-\mathrm{i}k)^{\alpha}\frac{K_\alpha}{|\cos{(\alpha\pi/2)}|\lambda}}
\right)\mathrm{d}k.
\end{equation}
By defining $s=-\mathrm{i}k$ we have
\begin{eqnarray}
\nonumber
\fl\frac{\mathrm{d}}{\mathrm{d}x}p_+(\lambda,x)&=&\frac{1}{2\pi\mathrm{i}}\int\limits_{
-i\infty}^{i\infty}e^{sx}\frac{1-s\left(\frac{K_\alpha}{|\cos{(\alpha\pi/2)}|\lambda}
\right)^{1/\alpha}}{1-s^{\alpha}\frac{K_\alpha}{|\cos{(\alpha\pi/2)}|\lambda}}\mathrm{d}s\\
\fl&=&\frac{1}{2\pi\mathrm{i}}\int\limits_{-\mathrm{i}\infty}^{\mathrm{i}\infty}e^{sx}
\frac{1}{1-s^{\alpha}\frac{K_\alpha}{|\cos{(\alpha\pi/2)}|\lambda}}\mathrm{d}s-\frac{
1}{2\pi\mathrm{i}}\int\limits_{-\mathrm{i}\infty}^{\mathrm{i}\infty}e^{sx}\frac{s
\left(\frac{K_\alpha}{|\cos{(\alpha\pi/2)}|\lambda}\right)^{1/\alpha}}{1-s^{\alpha}
\frac{K_\alpha}{|\cos{(\alpha\pi/2)}|\lambda}}\mathrm{d}s.
\end{eqnarray}
Using the properties of the Mittag-Leffler function, equations
(\ref{eq:LaplderivMittag}) and (\ref{eq:Lapl2derivMittag}) we can write 
\begin{eqnarray}
\nonumber
\frac{\mathrm{d}}{\mathrm{d}x}p_+(\lambda,x)&=&-\frac{d}{dx}E_\alpha\left(\frac{
|\cos{(\alpha\pi/2)}|\lambda x^\alpha}{K_\alpha}\right)\\
\fl&&+\left(\frac{K_\alpha}{|\cos{(\alpha\pi/2)}|\lambda}\right)^{1/\alpha}\frac{
\mathrm{d}^2}{\mathrm{d}x^2}E_\alpha\left(\frac{|\cos{(\alpha\pi/2)}|\lambda
x^\alpha}{K_\alpha}\right),
\end{eqnarray}
and with the boundary condition $p_+(\lambda,x=0)=0$ we get
\begin{eqnarray}
\nonumber
p_+(\lambda,x)&=&1-E_\alpha\left(\frac{|\cos{(\alpha\pi/2)}|\lambda x^\alpha}{
K_\alpha}\right)\\
&&+\left(\frac{K_\alpha}{|\cos{(\alpha\pi/2)}|\lambda}\right)^{1/
\alpha}\frac{\mathrm{d}}{\mathrm{d}x}E_\alpha\left(\frac{|\cos{(\alpha\pi/2)}|
\lambda x^\alpha}{K_\alpha}\right).
\end{eqnarray}
Thus, for the Laplace transform $\wp(\lambda)=1-p_+(\lambda,d)$ we obtain
\begin{equation}
\fl\wp(\lambda)=E_\alpha\left(\frac{\lambda |\cos{(\alpha\pi/2)}|}{K_\alpha}d^{
\alpha}\right)-\left(\frac{\lambda |\cos{(\alpha\pi/2)}|}{K_\alpha}\right)^{-1/
\alpha}\frac{\mathrm d}{\mathrm d d}E_\alpha\left(\frac{\lambda |\cos{(\alpha
\pi/2)}|}{K_\alpha}d^{\alpha}\right).
\label{eq:Laplacetwo-side_b1}
\end{equation}
By applying the inverse Laplace transform with respect to $\lambda$ and using
the series representation (\ref{eq:seriMittag}) of the Mittag-Leffler function
for the first term on the right side of equation (\ref{eq:Laplacetwo-side_b1})
we have
\begin{eqnarray}
\nonumber
\frac{1}{2\pi\mathrm{i}}\int\limits_{\mathrm{Ha}}e^{\lambda t}E_\alpha\left(\frac{
\lambda|\cos{(\alpha\pi/2)}|}{K_\alpha}d^{\alpha}\right)\mathrm{d}\lambda
\nonumber
&=&\frac{1}{2\pi\mathrm{i}}\int\limits_{\mathrm{Ha}}e^{\lambda t}\displaystyle\sum_{
n=0}^{\infty}\frac{\left(\frac{\lambda |\cos{(\alpha\pi/2)}|}{K_\alpha}d^{\alpha}
\right)^n}{\Gamma(\alpha n+1)}\mathrm{d}\lambda\\
\nonumber
&&\hspace*{-2.2cm}=
\displaystyle\sum_{n=0}^{\infty}\frac{\left(\frac{|\cos{(\alpha\pi/2)}|}{K_
\alpha}d^{\alpha}\right)^n}{\Gamma(\alpha n+1)}\frac{1}{2\pi i}\int\limits_{
\mathrm{Ha}}e^{\lambda t}\lambda^n\mathrm{d}\lambda\\
&&\hspace*{-2.2cm}=
\displaystyle\sum_{n=0}^{\infty}\frac{\left(\frac{|\cos{(\alpha\pi/2)}|}{K_
\alpha}d^{\alpha}\right)^n}{\Gamma(\alpha n+1)\Gamma(-n)}=0,
\end{eqnarray}
where we use $1/\Gamma(-n)=0$ for $n=0,1,2,\ldots$. For the second term on the
right side of equation (\ref{eq:Laplacetwo-side_b1}) we calculate the derivative
with respect to $d$ of the series representation (\ref{eq:seriMittag}) of the
Mittag-Leffler function and get
\begin{eqnarray}
\nonumber
-\left(\frac{\lambda|\cos{(\alpha\pi/2)}|}{K_\alpha}\right)^{-1/\alpha}\frac{
\mathrm d}{\mathrm dd}E_\alpha\left(\frac{\lambda |\cos{(\alpha\pi/2)}|}{K_
\alpha}d^{\alpha}\right)\\
=-\left(\frac{\lambda|\cos{(\alpha\pi/2)}|}{K_\alpha}\right)^{-1/\alpha}\frac{1}{
\mathrm{d}}\displaystyle\sum_{n=1}^{\infty}\frac{\left(\frac{\lambda |\cos{(\alpha
\pi/2)}|}{K_\alpha}d^{\alpha}\right)^{n}}{\Gamma{(\alpha n)}}.
\end{eqnarray}
After inverse Laplace transform and using the integral form
(\ref{eq:intGamma}) of the Gamma function, we obtain
\begin{eqnarray}
\nonumber
-\frac{1}{2\pi\mathrm{i}}&\int\limits_{\mathrm{Ha}}&e^{\lambda t}\left(\frac{\lambda
|\cos{(\alpha\pi/2)}|d^\alpha}{K_\alpha}\right)^{-1/\alpha}\displaystyle\sum_{
n=1}^{\infty}\frac{\left(\frac{\lambda|\cos{(\alpha\pi/2)}|d^{\alpha}}{K_\alpha}
\right)^n}{\Gamma{(\alpha n)}}\mathrm{d}\lambda\\
\nonumber
&=&-\displaystyle\sum_{n=1}^{\infty}\frac{\left(\frac{|\cos{(\alpha\pi/2)}|
d^\alpha}{K_\alpha}\right)^{n-1/\alpha}}{\Gamma(\alpha n)}
\frac{1}{2\pi\mathrm{i}}\int\limits_{\mathrm{Ha}}e^{\lambda t}\lambda^{n-1/\alpha}
\mathrm{d}\lambda\\
&=&-\displaystyle\sum_{n=1}^{\infty}\frac{\left(\frac{|\cos{(\alpha\pi/2)}|
d^{\alpha}}{K_\alpha}\right)^{n-1/\alpha}t^{-n-1+1/\alpha}}{\Gamma(\alpha n)
\Gamma(1/\alpha-n)}.
\end{eqnarray}
Rewriting this expression and using the relation $-\Gamma(\alpha n)\Gamma(1/\alpha
-n)=\alpha\Gamma(\alpha n-1)\Gamma(1+1/\alpha-n)$ yields
\begin{equation}
\wp(t)=\frac{t^{-2+1/\alpha}d^{\alpha-1}}{\alpha\left(K_{\alpha}/|\cos{(\alpha
\pi/2)}|\right)^{1-1/\alpha}}\displaystyle\sum_{n=1}^{\infty}\frac{(|\cos{(\alpha
\pi/2)}|d^\alpha/{K_\alpha t})^{n-1}}{\Gamma{(\alpha n-1)}\Gamma{(1+1/\alpha-n)}}.
\label{eq:fptPDFtwobp}
\end{equation}

\textcolor{black}{
To obtain a closed-form solution by help of equation (\ref{xidef}) we rewrite
equation (\ref{eq:fptPDFtwobp}) as
\begin{eqnarray}
\wp(t)&=&\frac{t^{-2+1/\alpha}d^{\alpha-1}}{\alpha \xi^{1-1/\alpha}}\displaystyle
\sum_{n=1}^{\infty}\frac{(d^{\alpha}/\xi t)^{n-1}}{\Gamma{(\alpha n-1)}\Gamma{(
1+1/\alpha-n)}}\nonumber\\
&=&\frac{t^{-2+1/\alpha}d^{\alpha-1}}{\alpha \xi^{1-1/\alpha}}\displaystyle\sum_{
n=0}^{\infty}\frac{(d^{\alpha}/\xi t)^{n}}{\Gamma{(\alpha n+\alpha-1)}\Gamma{(-n
+1/\alpha)}}.
\label{eq:fptPDFtwobp1}
\end{eqnarray}
Now, the generalised four-parametric Mittag-Leffler function has the series
representation (page 129, equation (6.1.1) \cite{RGorenflo2014})
\begin{equation}
E_{\alpha_1,\beta_1;\alpha_2,\beta_2}(z)=\displaystyle\sum_{k=0}^{\infty}\frac{z^k
}{\Gamma(\alpha_{1}k+\beta_{1})\Gamma(\Gamma(\alpha_{2}k+\beta_{2})},\,\,\,
z\in\mathbb{C} ,
\label{eq:Mittag-fourpar}
\end{equation}
for $\alpha_1$, $\alpha_2$ $\in$ $\mathbb{R}$ $(\alpha_1^2+\alpha_2^2\neq0)$ and
$\beta_1$, $\beta_2$ $\in\mathbb{C}$. It is an entire function, and if $\alpha_1
+\alpha_2>0$ it has the Mellin-Barnes integral form (page 132, equation (6.1.14)
of \cite{RGorenflo2014})
\begin{equation}
E_{\alpha_1,\beta_1;\alpha_2,\beta_2}(z)=\frac{1}{2\pi i}\int\limits_{\mathcal{L}}
\frac{\Gamma(s)\Gamma(1-s)}{\Gamma(\beta_{1}-\alpha_{1}s)\Gamma(\beta_{2}-\alpha_2
s)}(-z)^{-s}\mathrm{d}s,
\label{eq:int-Mellin-Barnes}
\end{equation}
where $\mathcal{L}=\mathcal{L}_{-\infty}$ is a contour running in a horizontal strip,
from $-\infty+i\phi_1$ to $-\infty+i\phi_2$, with $-\infty<\phi_1<0<\phi_2<\infty$.
This contour separates the poles of the Gamma functions $\Gamma(s)$ and $\Gamma(1-s)$.
The function $E_{\alpha_1,\beta_1;\alpha_2,\beta_2}(z)$ with $\alpha_1+\alpha_2>0$
converges for all $z\neq0$. For real values of the parameters $\alpha_1$, $\alpha_2$
$\in$ $\mathbb{R}$ and complex values of $\beta_1$, $\beta_2$ $\in \mathbb{C}$ the
four-parametric Mittag-Leffler function $E_{\alpha_1,\beta_1;\alpha_2,\beta_2}$ can
be represented in terms of the generalised Wright function and the Fox $H$-function.
If $\alpha_1>0$, $\alpha_2<0$ and the contour of integration in expression
(\ref{eq:int-Mellin-Barnes}) is chosen as $\mathcal{L}=\mathcal{L}_{-\infty}$ for
$\alpha_1+\alpha_2>0$, by identification with the corresponding Mellin-Barnes
integral definiont of the $H$-function one can obtain the following representation
of $E_{\alpha_1,\beta_1;\alpha_2,\beta_2}$ in terms of the $H$-function
(page 135, equation (6.1.28) \cite{RGorenflo2014})
\begin{equation}
E_{\alpha_1,\beta_1;\alpha_2,\beta_2}(z)=H_{2,2}^{1,1}\left[-z\left|
\begin{array}{l}(0,1),(\beta_2,-\alpha_2)\\(0,1),(1-\beta_1,\alpha_1)
\end{array}\right.\right].
\label{eq:int-H-function}
\end{equation}
From equations (\ref{eq:fptPDFtwobp}) and(\ref{eq:Mittag-fourpar}) by setting
$\alpha_1=\alpha$, $\beta_1=\alpha-1$, $\alpha_2=-1$, $\beta_2=1/\alpha$, and
$z=d^{\alpha}/{\xi t}$, we finally obtain
\begin{equation}
\wp(t)=\frac{t^{-2+1/\alpha}d^{\alpha-1}}{\alpha \xi^{1-1/\alpha}}H_{2,2}^{1,1}
\left[-\frac{d^\alpha}{\xi t}\left|\begin{array}{l}(0,1),(1/\alpha, 1)\\(0,1),(2
-\alpha,\alpha)\end{array}\right.\right].
\label{eq:FPT-twoside-Hfun}
\end{equation}
}

\subsection{Asymptotic of the first-passage time PDF of $\alpha$-stable processes
with $\alpha\in(1,2)$, $ \beta\in[-1, 1]$, or $\alpha\in(0,1)$, $\beta\in(-1, 1)$,
or $\alpha=1$, $\beta=0$}
\label{skoroapp}

We write equation (\ref{eq:skorokhod8}) as
\begin{equation}
\ln{q_+(\lambda,k)}=\int\limits_{\lambda}^{\infty}\int\limits_0^{\infty}e^{-ut}
\int\limits_0^{\infty}(e^{\mathrm{i}kx}-1)f(x,t)\mathrm{d}x\mathrm{d}t\mathrm{d}u,
\label{eq:SkorokhodGenr1}
\end{equation}
and split this expression into two terms,
\begin{eqnarray}
\nonumber
\ln{q_+(\lambda,k)}&=&\int\limits_{\lambda}^{\infty}\int\limits_0^{\infty}e^{-ut}
\int\limits_0^{\infty}\e^{\mathrm{i}kx}f(x,t)\mathrm{d}x\mathrm{d}t\mathrm{d}u\\
&&-\int\limits_{\lambda}^{\infty}\int\limits_0^{\infty}e^{-ut}\int\limits_0^{
\infty}f(x,t)\,\mathrm{d}x\mathrm{d}t\mathrm{d}u.
\label{eq:SkorokhodGenr2}
\end{eqnarray} 
We now employ theorem 4 from \cite{Zolotarev1957}, which says that the Laplace
transform with respect to $x$ of an $\alpha$-stable law in the Z-form of the
characteristic function has the form\footnote{Except for $\alpha\in(0,1)$,
$\beta=1,-1$}
\begin{equation}
\fl\ell_{\alpha,\rho}^Z(s,t)=\int\limits_0^{\infty}e^{-sx}\ell_{\alpha,\rho}^Z
(x,t)\mathrm{d}x=\frac{\sin{(\pi\rho)}}{\pi}\int\limits_0^{\infty}\frac{\exp(-t
K_{\alpha}^Z(sx)^{\alpha})}{x^2+2x\cos{(\pi\rho)}+1}\mathrm{d}x.
\label{eq:LaplaceTraGeneral}
\end{equation}
In the first term in the right hand side of (\ref{eq:SkorokhodGenr2}) we use
equation (\ref{eq:LaplaceTraGeneral}) with $s\to-\mathrm{i}k$, while for the second
term $s\rightarrow0$. Then we get 
\begin{eqnarray}
\nonumber
\fl\ln{q_+(\lambda,k)}&=&\int\limits_{\lambda}^{\infty}\int\limits_0^{\infty}e^{-ut}
\frac{\sin{(\pi\rho)}}{\pi}\int\limits_0^{\infty}\frac{\exp(-tK_{\alpha}^Z(-
\mathrm{i}kx)^{
\alpha})}{x^2+2x\cos{(\pi\rho)}+1}\mathrm{d}x\mathrm{d}t\mathrm{d}u\\ 
\fl&&-\int\limits_{\lambda}^{\infty}\int\limits_0^{\infty}e^{-ut}\frac{\sin{(\pi
\rho)}}{\pi}\int\limits_0^{\infty}\frac{1}{x^2+2x\cos{(\pi\rho)}+1}\,\mathrm{d}x
\mathrm{d}t\mathrm{d}u.
\end{eqnarray}
In this expression we change the order of integration and first take the integrals
over $t$,
\begin{eqnarray}
\nonumber
\fl\ln{q_+(\lambda,k)}&=&\frac{\sin{(\pi\rho)}}{\pi}\int\limits_{\lambda}^{\infty}
\int\limits_0^{\infty}\frac{1}{x^2+2x\cos{(\pi\rho)}+1}\int\limits_0^{\infty}\left(
\e^{-tu-tK_{\alpha}^Z(-\mathrm{i}kx)^{\alpha}}-\e^{-tu}\right)\mathrm{d}t\mathrm{d}
x\mathrm{d}u\\
\fl&=&\frac{\sin{(\pi\rho)}}{\pi}\int\limits_{\lambda}^{\infty}\int\limits_0^{
\infty}\frac{(u+K_{\alpha}^Z(-\mathrm{i}kx)^{\alpha})^{-1}-(u)^{-1}}{x^2+2x\cos{
(\pi\rho)}+1}\mathrm{d}x\mathrm{d}u.
\end{eqnarray}
In the next step we again change the order of integration,
\begin{eqnarray}
\nonumber
\fl\ln{q_{+}(\lambda,k)}&=&-\frac{\sin{(\pi\rho)}}{\pi}\int\limits_0^{\infty}
\frac{K_{\alpha}^Z(-\mathrm{i}kx)^{\alpha}}{x^2+2x\cos{(\pi\rho)}+1}\int_{\lambda}
^{\infty}\frac{1}{u(u+K_{\alpha}^Z(-\mathrm{i}kx)^{\alpha})}\mathrm{d}u\mathrm{d}x\\
\fl&=&\frac{\sin{(\pi\rho)}}{\pi}\int\limits_0^{\infty}\frac{\ln{\lambda}-\ln{(
\lambda+K_{\alpha}^{Z}(-\mathrm{i}kx)^{\alpha})}}{x^2+2x\cos{(\pi\rho)}+1}\mathrm{d}x
\end{eqnarray}
Now, we split the above equation into two terms,
\begin{eqnarray}
\nonumber
\ln{q_{+}(\lambda,k)}&=&\frac{\sin{(\pi\rho)}}{\pi}\int\limits_0^{\infty}\frac{
\ln{\lambda}}{x^2+2x\cos{(\pi\rho)}+1}\mathrm{d}x\\
&&-\frac{\sin{(\pi\rho)}}{\pi}\int\limits_0^{\infty}\frac{\ln{(\lambda+K_{\alpha}
^Z(-\mathrm{i}kx)^{\alpha})}}{x^2+2x\cos{(\pi\rho)}+1}\mathrm{d}x.
\label{eq:SkorokhodGenral7}
\end{eqnarray}
By defining the first term in the right hand side of equation
(\ref{eq:SkorokhodGenral7}) as
\begin{equation}
\ln{r(\lambda)}=\frac{\sin{(\pi\rho)}}{\pi}\int\limits_0^{\infty}\frac{\ln{
\lambda}}{x^2+2x\cos{(\pi\rho)}+1}\mathrm{d}x
\end{equation}
and using
\begin{equation}
\int\limits_0^{\infty}\frac{1}{x^2+2x\cos{(\pi\rho)}+1}\mathrm{d}x=\frac{\pi
\rho}{\sin{(\pi\rho)}},
\label{eq:HelpInteg}
\end{equation}
we get
\begin{equation}
\ln{r(\lambda)}=\rho\ln\lambda .
\label{eq:SkorokhodGenral10}
\end{equation}
For the second integral on the right side of equation (\ref{eq:SkorokhodGenral7})
we set $\lambda\to0$ and then use equation (\ref{eq:HelpInteg}) and the integral
\begin{equation}
\int\limits_0^{\infty}\frac{\ln{x}}{x^2+2x\cos{(\pi\rho)}+1}\mathrm{d}x=0.
\end{equation}
The result is
\begin{equation}
\frac{\sin{(\pi\rho)}}{\pi}\int\limits_0^{\infty}\frac{\ln{(K_{\alpha}^Z(-\mathrm{i}
kx)^{\alpha})}}{x^2+2x\cos{(\pi\rho)}+1}\mathrm{d}x=\rho\ln{(K_{\alpha}^Z(-\mathrm{i}
k)^{\alpha})}.
\label{eq:SkorokhodGenral12}
\end{equation}
Combining equations (\ref{eq:SkorokhodGenral10}) and (\ref{eq:SkorokhodGenral12})
we obtain the asymptotic of $\ln{q_+(\lambda,k)}$ at small $\lambda$ in the form
\begin{equation}
\ln{q_+(\lambda,k)}\approx\rho\ln{\lambda}-\rho\ln{(K_{\alpha}^Z(-\mathrm{i}k)^{
\alpha})},
\label{eq:SkorokhodGenr18}
\end{equation}
and thus 
\begin{equation}
q_+(\lambda,k)\approx\frac{\lambda^\rho}{(K_{\alpha}^Z(-\mathrm{i}k)^{\alpha})^\rho},
\quad\lambda\to0.
\label{eq:SkorokhodGenr19}
\end{equation}
Going back to equation (\ref{eq:skorokhod8}) by inverse Fourier transform we
find ($x>0$)
\begin{equation}
\frac{d}{dx}p_+(\lambda,x)\approx\frac{1}{2\pi}\int\limits_{-\infty}^{\infty}e^{-
\mathrm{i}kx}\left(\frac{\lambda^\rho}{(K_{\alpha}^Z(-\mathrm{i}k)^{\alpha})^\rho}
\right)\mathrm{d}k=\frac{\lambda^\rho x^{\alpha\rho-1}}{(K_{\alpha}^Z)^{\rho}
\Gamma{(\alpha\rho)}}.
\end{equation}
With the boundary condition $p_+(\lambda,x=0)=0$ we arrive at
\begin{equation}
p_+(\lambda,x)=\frac{\lambda^\rho x^{\alpha\rho}}{(K_{\alpha}^Z)^{\rho}\Gamma{
(1+\alpha\rho)}}.
\end{equation}
Following equation (\ref{eq:LaplFPTandp}), 
\begin{equation}
\wp(\lambda)\sim1-\frac{d^{\alpha\rho}}{(K_{\alpha}^{Z})^{\rho}\Gamma{(1+
\alpha\rho)}}\lambda^\rho.
\end{equation}
Finally, with the help of the Tauberian theorem \cite{Feller1971}, see equation
(\ref{eq:Tauberian1}), the long time asymptotic for the cases $\alpha\in(1,2]$,
$\beta\in[-1,1]$, or $\alpha\in(0,1)$, $\beta\in(-1,1)$, and $\alpha=1$, $\beta
=0$ has the form  
\begin{equation}
\wp(t)\sim\frac{\rho(K_{\alpha}^Z)^{-\rho}d^{\alpha\rho}}{\Gamma(1-\rho)\Gamma(
1+\alpha\rho)}t^{-\rho-1}.
\label{eq:Skorokhodgeneralasym}
\end{equation}
In this expression the exponent of $t$ was obtained in \cite{Bertoin1996},
Proposition VIII.1.2, p.~219, while the prefactor was derived by another method
in \cite{bingham1}, Corollary, p.~564, and \cite{bingham2}, Theorem 3b, p.~285. 

To represent the above equation in the A-form of the characteristic function we
need to use relation (\ref{eq:Prcm1}). Thus, we arrive at the desired result
(\ref{eq:generalasym}).

\section{Some properties of the Mittag-Leffler and the Wright functions}
\label{Appendix Spec-func}

The (one-parameter) Mittag-Leffler function is defined by the following series
representation, which is convergent in the whole complex plane
\cite{Podlubny1999,FMainardi2010}
\begin{equation}
E_\alpha(z)=\displaystyle\sum_{n=0}^{\infty}\frac{z^n}{\Gamma(\alpha n+1)},\quad
\alpha>0,\quad z\in\mathbb{C}.
\label{eq:seriMittag}
\end{equation}
Its integral representation is
\begin{equation}
E_\alpha(z)=\frac{1}{2\pi\mathrm{i}}\int\limits_{\mathrm{Ha}}\frac{\zeta^{\alpha-1}
e^{\zeta}}{\zeta^{\alpha}-z}\mathrm{d}\zeta,\quad\alpha>0,\quad z\in\mathbb{C},
\label{eq:intMittag}
\end{equation}
where the Hankel integration path is a loop which starts and ends at $-\infty$
and follows the circular disk $|\zeta|\leq|z|^{1/\alpha}$ in the positive sense,
$-\pi\leq\arg\zeta\leq\pi$ on Ha. The equivalence between the series and integral
representations can be proven by using the Hankel formula for the Gamma function
\begin{equation}
\frac{1}{\Gamma{(z)}}=\frac{1}{2\pi\mathrm{i}}\int\limits_{\mathrm{Ha}}\e^{\zeta}
\zeta^{-z}\,\mathrm{d}\zeta,\quad z\in \mathbb{C}.
\label{eq:intGamma}
\end{equation}
The Mittag-Leffler function is completely monotonous on the negative real axis
($z<0$) if $0<\alpha\leq1$. The Mittag-Leffler function is connected to the
Laplace integral through the identity \cite{FMainardi2010}
\begin{equation}
\fl E_\alpha(-\lambda x^{\alpha})\div\mathscr{L}\{E_\alpha(-\lambda x^{\alpha});s\}
=\int\limits_0^{\infty}e^{-sx}E_\alpha(-\lambda x^{\alpha})\mathrm{d}x=\frac{s^{
\alpha-1}}{s^{\alpha}+\lambda},
\label{eq:LaplMittag}
\end{equation}
for $\mathrm{Re}(s)>|\lambda|^{1/\alpha}$. From here we easily get two useful formula, see
also equations (E.52), (E.54), and (E.55)) in \cite{FMainardi2010},
\begin{equation}
-\frac{1}{\lambda}\frac{d}{d x}E_{\alpha}(-\lambda x^{\alpha})\div\frac{1}{s^{
\alpha}+\lambda},
\label{eq:LaplderivMittag}
\end{equation}
and
\begin{equation}
-\frac{1}{\lambda}\frac{d^2}{d x^2}E_{\alpha}(-\lambda x^{\alpha})\div\frac{s}{
s^{\alpha}+\lambda},
\label{eq:Lapl2derivMittag}
\end{equation}
where $\alpha>0$ and $\mathrm{Re}(s)>|\lambda|^{1/\alpha}$.

The Wright $W$ function has the series representation \cite{FMainardi2010}
(convergent in the whole complex plane)
\begin{equation}
W_{\alpha,\beta}(z)=\displaystyle\sum_{n=0}^{\infty}\frac{(z)^n}{n!\Gamma(\alpha
n+\beta)},\quad\alpha>-1,\quad\beta\in\mathbb{C}.
\label{eq:seriWfun}
\end{equation}
The integral representation of this function is
\begin{equation}
W_{\alpha,\beta}(z)=\frac{1}{2\pi\mathrm{i}}\int\limits_{\mathrm{Ha}}e^{\sigma+z
\sigma^{-\alpha}}\frac{\mathrm{d}\sigma}{\sigma^{\beta}},\quad\alpha>-1.
\label{eq:intWfun}
\end{equation}
For $\alpha=0$ we get $W_{0,\beta}(z)=e^z/\Gamma(\beta)$.

The Wright $M$ function has the series representation \cite{FMainardi2010}
\begin{equation}
M_\alpha(z)=\displaystyle\sum_{n=0}^{\infty}\frac{(-z)^n}{n!\Gamma(1-\alpha-\alpha
n)}=\frac{1}{\pi}\displaystyle\sum_{n=1}^{\infty}\frac{(-z)^{n-1}}{(n-1)!}\Gamma{
(\alpha n)}\sin{(\alpha\pi n)},
\label{eq:seriMfun}
\end{equation}
where $0<\alpha<1$. We note that $M_\alpha(0)=1/\Gamma{(1-\alpha)}$. The radius of
convergence of the power series is infinite for $0<\alpha<1$. The integral
representation of the $M$-function is
\begin{equation}
M_\alpha(z)=\frac{1}{2\pi\mathrm{i}}\int\limits_{\mathrm{Ha}}e^{\sigma-z\sigma^{
\alpha}}
\frac{\mathrm{d}\sigma}{\sigma^{1-\alpha}},\quad z\in\mathbb{C},\quad0<\alpha<1.
\label{eq:intMfun}
\end{equation}
Since the $M$-function is entire in $z$ the exchange between the
series and the integral in the calculations is legitimate. For the special case
$\alpha=1/2$ the $M$-function can be expressed in terms of the known functions
$M_{1/2}(z)=\exp(-z^2/4)/\sqrt{\pi}$. Another important property of the
$M$-function is the asymptotic representation $M_\alpha(x)$ as $x\to+\infty$.
By a saddle-point approximation it is shown in \cite{FMainardi1995} that
\begin{equation}
M_\alpha(x/\alpha)\sim\frac{x^{(\alpha-1/2)/(1-\alpha)}}{\sqrt{2\pi(1-\alpha)}}
\exp{\left(-\frac{1-\alpha}{\alpha}x^{1/(1-\alpha)}\right)}.
\label{eq:asymMfun}
\end{equation}
Recalling the integral representation for large argument of the Mittag-Leffler
function (\ref{eq:intMittag}), for the Laplace transform of the $M_\alpha(r)$
one can write 
\begin{equation}
M_\alpha(r)\div E_\alpha(-s),\quad0<\alpha<1.
\label{eq:LaplMfun}
\end{equation}
The relevant Laplace transform pair related to the $M_\alpha(r^{-\alpha})$
function is \cite{FMainardi2010}
\begin{equation}
\frac{\lambda\alpha}{r^{\alpha+1}}M_\alpha(\lambda r^{-\alpha})\div e^{-\lambda
s^{\alpha}},\quad0<\alpha<1,\lambda>0.
\label{eq:LaplPair1}
\end{equation}
The M-function is non-negative, integrable, and normalised on the positive
semi-axis \cite{FMainardi2010}
\begin{equation}
\int\limits_0^{\infty}M_{\alpha}(r)\mathrm{d}r=1,\quad0<\alpha<1,
\label{eq:M-normalization}
\end{equation}
and also
\begin{equation}
\int\limits_0^{\infty}\alpha r^{-1-\alpha}M_{\alpha}(r^{-\alpha})\mathrm{d}r=1,
\quad0<\alpha<1.
\label{eq:M-normalization2}
\end{equation}

\section*{References}

\bibliographystyle{iopart-num}

\end{document}